\documentclass[a4paper,11pt]{article}
\pdfoutput=1
\usepackage{jheppub}
\usepackage{subfig}
\usepackage{amsmath,amssymb,bm,graphicx,epsf,colordvi}
\usepackage{physics}
\usepackage{slashed}
\def\bea#1\eea{\begin{align}#1\end{align}} 
\usepackage{comment}
\usepackage{diagbox}
\usepackage{mathtools}
\usepackage{lipsum}
\usepackage{soul}
\usepackage{enumitem}
\usepackage{braket}
\usepackage{bm}
\allowdisplaybreaks 
\renewcommand{\arraystretch}{1.2}
\usepackage{color}
\usepackage[dvipsnames]{xcolor}

\usepackage{mathrsfs}
\usepackage{appendix}
\usepackage{bm}
\usepackage{lipsum}

\newcommand{\bef}{\begin{figure}[hbt]\centering}
\newcommand{\eef}{\end{figure}}

\usepackage{mathtools}
\usepackage{booktabs}
\usepackage{graphicx}
\graphicspath{ {./figure/} }
\usepackage{caption}
\usepackage{lipsum}

\newcommand{\cf}{C_F}
\newcommand{\ca}{C_A}

\newcommand{\beq}{\begin{equation}}
\newcommand{\eeq}{\end{equation}}
\def\bea#1\eea{\begin{align}#1\end{align}}

\def \be  {\begin{equation}}
\def \ee  {\end{equation}}
\def \ba  {\begin{eqnarray}}
\def \ea  {\end{eqnarray}}
\newcommand{\df}{\mathrm{d}}

\allowdisplaybreaks

\usepackage{graphicx}
\usepackage{float}

\usepackage{arydshln}

\usepackage{multicol}
\usepackage[dvipsnames]{xcolor}

\usepackage [autostyle, english = american]{csquotes}
\MakeOuterQuote{"}

\newcommand{\rmd}{{\rm d}}

\def\ca{C_{\rm A}}
\def\cf{C_{\rm F}}
\def\da{d_{\rm A}}
\def\df{d_{\rm F}}
\def\lra{\leftrightarrow}

\newcommand{\ZH}[1]{{\color{red} \bf ZH: #1}}

\newcommand{\arjun}[1]{{\color{olive} \bf ASK: #1}}

\def \[{\left[}
\def \]{\right]}
\def \({\left(}
\def \){\right)}

\def \vp {\bm{p}}
\def \vk {\bm{k}}
\def \vq {\bm{q}}

\title{\boldmath Sensitivity of Jet Observables to Moli\`ere Scattering Off Quasiparticles in Quark-Gluon Plasma}

\author[a]{Zachary Hulcher,}
\author[b]{Arjun Srinivasan Kudinoor,}
\author[c,d]{Daniel Pablos,}
\author[b]{Krishna Rajagopal}

\affiliation[a]{SLAC National Accelerator Laboratory, Menlo Park, CA 94025, USA}
\affiliation[b]{Center for Theoretical Physics --- a Leinweber Institute, Massachusetts Institute of Technology,\\
77 Massachusetts Ave, Cambridge, MA 02139, USA}
\affiliation[c]{Departamento de F\'isica, Universidad de Oviedo, Avda. Federico Garc\'ia Lorca 18,\\
33007 Oviedo, Spain}
\affiliation[d]{Instituto Universitario de Ciencias y Tecnolog\'ias Espaciales de Asturias (ICTEA),\\
Calle de la Independencia 13, 33004 Oviedo, Spain}

\emailAdd{zhulcher@stanford.edu}
\emailAdd{kudinoor@mit.edu}
\emailAdd{pablosdaniel@uniovi.es}
\emailAdd{krishna@mit.edu}

\preprint{MIT-CTP/6012}

\abstract{
Quark-gluon plasma (QGP) behaves as a strongly coupled liquid when viewed at length scales of order the inverse of its temperature and longer. However, when it is probed at sufficiently short length scales, with sufficiently high momentum-exchange, the asymptotic freedom of QCD mandates the presence of quark- and gluon-like quasiparticles. High-energy partons in jets can trigger such perturbative, high momentum-exchange $2 \rightarrow 2$ Moli\`ere scatterings off quasiparticles in the medium, making jets useful probes with which to study the microscopic structure of QGP.
Prior to this work, as appropriate for a strongly coupled liquid, nonperturbative soft momentum-exchanges between jet partons and the hydrodynamic droplet of QGP produced in a heavy-ion collision, as well as the wakes that jets excite in the droplet, had been accounted for in the hybrid strong/weak coupling model of jet quenching.
Here, we present a full calculation of Moli\`ere scattering and describe how it is implemented in the Hybrid Model.
Moli\`ere scattering off a quasiparticle in the medium results in the deflection of the jet parton and the excitation of a parton from the thermal medium that recoils after being kicked. 
The scattered jet parton and the recoil parton continue to propagate through the QGP, lose energy and momentum, excite wakes, and may further re-scatter.
Using the Hybrid Model, we study how Moli\`ere scatterings impact a variety of observables including jet shapes and  fragmentation functions, the Soft Drop splitting angle $R_g$, jet girth $g$, and observables that focus on the number and angular distribution of subjets within jets.
We demonstrate that photon-tagged jets provide a particularly sensitive probe: selecting events by the photon energy mitigates the selection bias inherent in inclusive jet measurements and enhances sensitivity to rare large-angle scatterings. We find that Moli\`ere scatterings broaden both the $R_g$ and $g$ distributions when jets significantly softer than the photon are included. Our results point the way towards
distinctive model-independent experimental signatures of hard scattering 
of jet partons off quasiparticles in QGP.

}

\begin{document} 
\maketitle
\flushbottom

\section{Introduction}
\label{sec:intro}

When viewed at length scales of order the inverse of its temperature $T$ and longer, quark-gluon plasma (QGP) behaves as a strongly coupled liquid. It has no quasiparticles with mean free paths that are long compared to $1/T$ and flows as a hydrodynamic liquid with a lower specific shear viscosity than that of any other known liquid. Relativistic heavy-ion collisions at RHIC and the LHC produce droplets of this high-temperature phase of QCD matter and have shown it is in fact a strongly coupled liquid~\cite{PHENIX:2004vcz,BRAHMS:2004adc,PHOBOS:2004zne,STAR:2005gfr,Gyulassy:2004zy}.
However, when it is probed at sufficiently short length scales with sufficiently high momentum transfer, the asymptotic freedom of QCD mandates the presence of quark-like and gluon-like quasiparticles. This raises a guiding question: How does this strongly coupled liquid emerge from an asymptotically free gauge theory?

A natural method of answering this question is to perform scattering experiments with sufficiently large momentum transfer so that the microscopic, particulate structure of QGP is resolved. High-energy partons within jets produced in a heavy-ion collision can trigger such high momentum-exchanges with the quark- and gluon-like quasiparticles in the droplet of QGP formed in the same collision. This makes jets a particularly compelling probe to access detailed information about the microscopic structure of QGP.

Hard scatterings between partons in the incident nuclei produce two or more quarks, antiquarks or gluons with high virtuality. In the absence of a droplet of QGP (aka in vacuum), each high-virtuality parton relaxes its virtuality down to the QCD confinement scale via successive splittings. This produces a shower of partons within some irregular cone in momentum space around the direction of the initial high-virtuality parton. This shower of partons is called a jet. The production and subsequent evolution of a jet in vacuum is theoretically well described by the high-virtuality, perturbative DGLAP evolution equations. However, the evolution of a jet through a droplet of expanding, flowing, and cooling QGP formed in a heavy-ion collision is more complicated. As the high-energy partons within a jet plow through the droplet of QGP, they undergo strong interactions with the 
strongly coupled colored liquid or, possibly, with the colored quasiparticles that can be resolved within it at short length scales.
Such interactions modify the parton shower, and consequently the energy and structure of the measured jet. The modifications to the energy and structure of jets as a result of their interactions with the droplet of QGP is colloquially referred to as \textit{jet quenching}~\cite{dEnterria:2009xfs,Wiedemann:2009sh,Majumder:2010qh,Jacak:2012dx,Muller:2012zq,Mehtar-Tani:2013pia,Connors:2017ptx,Busza:2018rrf,Cao:2020wlm,Cunqueiro:2021wls,Apolinario:2022vzg,Wang:2025lct}. 
In this paper, we shall focus on one aspect of the interaction between a parton shower and QGP:
$2 \rightarrow 2$ elastic scatterings between high-energy jet-partons and thermal quasiparticles in the medium.  Elastic scattering with sufficiently high momentum transfer can be described perturbatively.
Later in this Introduction, and at greater length in Section~\ref{sec:implementing-in-hybrid}, we
shall also describe the jet quenching dynamics that results from strongly coupled 
interactions between jet partons and the strongly coupled liquid QGP through which they propagate, as a result of which the jet partons lose energy, accumulate soft transverse momentum and excite wakes in the droplet of QGP. 

The short-distance, particulate structure of QGP can be resolved when a high-energy jet-parton scatters off a thermal parton in a droplet of it. 
If QGP were a liquid at all length scales with no particulate constituents at any length scales, then the probability that an energetic parton plowing through it is deflected with some momentum $q_\perp$ perpendicular to its initial direction of motion is a Gaussian distribution in $q_\perp$~\cite{Liu:2006ug, Liu:2006he, DEramo:2010wup, DEramo:2012uzl}. 
So, at sufficiently small values of $q_\perp$ the distribution of transverse momentum broadening experienced by a jet-parton will be Gaussian. 
This soft, Gaussian, strongly coupled transverse momentum broadening falls of rapidly at large $q_\perp$.  
However, because QCD is asymptotically free this cannot be the whole story.
QGP is not a strongly coupled liquid at all scales: it must be particulate at short length-scales. 
Thus, large angle, high momentum-transfer $2 \rightarrow 2$ scatterings of jet-partons with thermal partons in the medium may be rare, but 
they will not be Gaussianly rare. 
Experimental measurements of such large angle, high momentum deflections of jet-partons --- referred to as \textit{Moli\`ere scatterings}~\cite{Kurkela:2014tla} 
after the person who
first discussed the QED analogue~\cite{Moliere:1947zza, Moliere:1948zz, Moliere:1955zz} and a focus in the present context at least since the work of Refs.~\cite{DEramo:2012uzl,Kurkela:2014tla}
--- are the first steps to studying the microscopic structure of QGP and to understanding how a strongly coupled QCD liquid emerges from an asymptotically free gauge theory.

In this paper, we develop a concrete framework for incorporating Moli\`ere scatterings into a realistic model of jet quenching and explore how their presence modifies a broad set of jet observables. Our goal is twofold: first, to embed perturbative elastic $2 \rightarrow 2$ scattering processes consistently within the hybrid strong/weak coupling model (or simply the Hybrid Model) of jet quenching~\cite{Casalderrey-Solana:2014bpa,Casalderrey-Solana:2015vaa,Casalderrey-Solana:2016jvj,Hulcher:2017cpt,Casalderrey-Solana:2018wrw,Casalderrey-Solana:2019ubu,Hulcher:2022kmn,Bossi:2024qho, Kudinoor:2025ilx,Kudinoor:2025gao,Beraudo:2025nvq}, which already captures nonperturbative energy loss,
soft momentum broadening, and the effects of jet wakes that all arise from  strongly coupled interactions between jet partons and QGP; and, second, to identify jet observables that are sensitive to Moli\`ere scatterings off thermal partons in the medium. By doing so, we aim to assess how current and future measurements of jets and jet substructure can meaningfully constrain the short-distance, quasiparticle structure of QGP. 
We shall assess the impact of Moli\`ere scattering on jet observables by turning it off and on --- something that is of course impossible to do in experimental data. 

The Hybrid Model is particularly well suited for the purpose of turning hard scattering of jet partons off QGP quasiparticles off and on because before the present work this model included no high momentum transfer processes.
Elastic scattering of jet partons off QGP quasiparticles, which yields 
recoiling partons, is built into 
other extant Monte Carlo treatments of jet quenching that are based upon the assumption that jet partons interact 
perturbatively with the QGP~\cite{Zapp:2008gi,Zapp:2013vla,He:2015pra,Milhano:2017nzm,Park:2018acg,Ke:2020clc,Dai:2020rlu,Tachibana:2025rcx,Jing:2025bwi,Soudi:2025lei,Boguslavski:2025ylx}. However, these models include such perturbatively calculated processes with momentum transfers that can be soft or hard. This makes it more difficult to use these models to investigate the observable consequences of hard scattering via turning it off. In the Hybrid Model, the 
soft interactions between jet partons and the strongly coupled QGP produced at colliders 
are described as is appropriate at strong coupling (where partons lose energy to the QGP, creating wakes in the QGP, and exchange soft transverse momentum with the QGP)
and at strong coupling there is no distinction in principle or in practice between elastic and inelastic energy loss. In this work, we add Moli\`ere scattering to the Hybrid Model as this is the simplest example of interactions between jet partons and QGP quasiparticles with high momentum transfer, which of course should be treated perturbatively.
Our purpose here is to use the resulting ability to turn hard scattering off and on in the Hybrid Model to identify which jet substructure observables are sensitive to hard scattering of jet partons off QGP quasiparticles.
Clear evidence for such processes in future experimental measurements of such observables
would motivate adding additional weakly coupled 
hard scattering processes to the model.

We shall implement Moli\`ere scatterings in the 
Hybrid Model, which
is a theoretical framework for jet quenching designed to describe the multi-scale processes of jet production and evolution through strongly coupled plasma. Processes that can be described by weakly coupled QCD are described as such, while processes that are intrinsically strongly coupled are also described appropriately.
It treats jet production and hard evolution perturbatively while modeling the soft momentum exchanges between partons in a jet shower and the droplet of QGP 
using insights from holographic calculations derived rigorously in a strongly coupled gauge theory. Each parton in a jet shower loses energy to the plasma via a holographically derived formula~\cite{Chesler:2014jva,Chesler:2015nqz} implemented in Refs.~\cite{Casalderrey-Solana:2014bpa,Casalderrey-Solana:2015vaa,Casalderrey-Solana:2016jvj,Hulcher:2017cpt,Casalderrey-Solana:2018wrw,Casalderrey-Solana:2019ubu,Hulcher:2022kmn,Bossi:2024qho,Kudinoor:2025ilx,Kudinoor:2025gao,Beraudo:2025nvq}. The interaction strength is governed by the dimensionless parameter $\kappa_{\rm sc}$; for illustration, a parton with initial energy $E_{\rm in}$ thermalizes over a distance $x_{\rm stop}=E_{\rm in}^{1/3}/(2\kappa_{\rm sc}T^{4/3})$ if it does not split first. 
$\kappa_{\rm sc}$ can be calculated in the strongly coupled gauge theory where the energy loss formula was derived;  fitting Hybrid Model predictions to data~\cite{Casalderrey-Solana:2018wrw} shows that $\kappa_{\rm sc}$ is smaller by a factor of 3 to 4 in QCD.
Jet-induced wakes are computed by perturbing the stress–energy tensor of the hydrodynamic droplet 
and converting the perturbation into soft hadrons via the Cooper--Frye prescription, assuming a hydrodynamic wake and a small perturbation to the hadron spectra~\cite{Casalderrey-Solana:2016jvj,Casalderrey-Solana:2020rsj,Bossi:2024qho,Kudinoor:2025ilx}. New observables have been proposed to isolate measurable  signatures of jet wakes~\cite{Bossi:2024qho,Kudinoor:2025ilx}.
%
The Hybrid Model incorporates soft, Gaussian-distributed transfers of momentum transverse to the jet parton’s direction of motion, describing the frequent (continuous in the strong coupling limit) back-and-forth exchanges between the jet parton and the medium~\cite{Casalderrey-Solana:2016jvj}.
However, 
prior to this study the Hybrid Model has not included rare, perturbative, high momentum transfer Moli\`ere $2 \rightarrow 2$ scattering between jet partons and weakly coupled quasiparticles resolved in QGP at short length-scales. An early and incomplete version of this study was reported in Ref.~\cite{Hulcher:2022kmn}.

We begin in Section~\ref{sec:moliere} by reviewing perturbative calculations of Moli\`ere scattering probabilities for energetic partons traversing a thermal medium, following and adapting earlier work in Refs.~\cite{DEramo:2012uzl,DEramo:2018eoy}. In particular, we review a calculation~\cite{DEramo:2018eoy} of the probability of finding an outgoing hard parton with energy $p$ and angle $\theta$ relative to the direction of the incident hard jet-parton after a $2 \rightarrow 2$ scattering has occurred. While the calculation of this probability in Ref.~\cite{DEramo:2018eoy} assumed that only one parton in the final state is observed, we extend it to a formulation in which both outgoing partons are explicitly tracked, as is required for a realistic implementation within a description of jet quenching since after the scattering both outgoing partons should be considered jet partons, as both will subsequently lose energy, excite wakes in the QGP, and may scatter again. 

Next, in Section~\ref{sec:phase-space-constraints-and-sampling}, we specify the region of phase space over which a perturbative description of Moli\`ere scatterings is reasonably valid. In particular, we require the momentum transfer squared of a scattering $cd \rightarrow ab$ (for partons $a$, $b$, $c$, and $d$) to be greater than a multiple $a$ (we shall typically chose $a=10$)
of the squared Debye mass $m_D^2$, namely $|t|, \, |u| > am_D^2$. We also introduce a set of kinematic variables that will help us to evaluate the probability distributions calculated in Section~\ref{sec:moliere} subject to the constraints $|t|, \, |u| > am_D^2$. We find that, depending on the value of $a$, certain ranges of outgoing momentum $p$ and scattering angle $\theta$ are disallowed by the constraint $|t| > am_D^2$ and find that the constraint $|u|>a m_D^2$ reduces the probability of finding scattered partons in certain other regions of $p$ and $\theta$. To conclude Section~\ref{sec:phase-space-constraints-and-sampling}, we address the practical task of sampling the kinematics of each scattering process $cd \rightarrow ab$ from the probability distributions that we have computed so that Moli\`ere scatterings may be included within the Hybrid Model of jet quenching.

In Section~\ref{sec:implementing-in-hybrid}, we describe how we incorporate the Moli\`ere scattering processes evaluated in Sections~\ref{sec:moliere} and~\ref{sec:phase-space-constraints-and-sampling} into the Hybrid Model. At each timestep of the in-medium evolution of a parton shower, we compute the probability that a hard scattering occurs and, when it does, probabilistically select which subprocess $cd \rightarrow ab$ takes place and choose the corresponding final-state kinematics by sampling from the appropriate probability distributions. The calculation of the total probability with which a Moli\`ere scattering occurs, as well as the process by which we sample the kinematics of each scattering, is specified by a set of four integrals, given in Section~\ref{sec:sampling}. We perform two of the integrations analytically (as described in explicit detail in Appendix~\ref{app:integrals}). 
Although doing so is laborious,
the analytic evaluation of two of the four types of integrals provides us with a significant computational advantage, as it means that we only need to evaluate a double integral numerically, not a quadruple integral.

Momentum transfers below the hard-scattering threshold $|t|, \,|u|<a m_D^2$ are not computed perturbatively; instead, their cumulative effect is treated as Gaussian transverse momentum broadening in the Hybrid Model~\cite{Casalderrey-Solana:2016jvj}, since the transverse momentum distribution of energetic partons propagating through a strongly coupled 
liquid are Gaussian-distributed~\cite{Liu:2006ug,DEramo:2010wup,DEramo:2012uzl}.  We choose the magnitude of
the soft transverse momentum broadening so as to
ensure a smooth matching between its Gaussian distribution and the probability for rare hard Moli\`ere scatterings that we have calculated. These Moli\`ere scattering and Gaussian transverse momentum broadening effects are embedded in the Hybrid Model alongside its nonperturbative description of strongly coupled energy loss.
At each time-step, each jet parton loses energy to, and exchanges soft momentum with, the droplet of QGP and may, with a probability that we compute in this paper, experience a Moli\`ere scattering.
After a Moli\`ere scattering, the two outgoing partons lose energy to the droplet of QGP like any other partons in the jet.
That is, we track the energy and momentum deposited 
by all the jet partons and each
recoiling thermal parton after a Moli\`ere scattering, and convert all of this ``lost'' energy and momentum into a hydrodynamic wake 
in the droplet of QGP which, via the Cooper-Frye prescription at freezeout, becomes soft hadrons in the final state.  
Including soft jet wakes in the analysis is important to elucidating the consequences of hard Moli\`ere scattering for observables because
the scattered partons lose energy and 
produce wakes, in many cases to such a degree that the most significant consequence of a Moli\`ere scattering 
arises from the modification of the shape of the wake of the jet.

In our implementation, we do not include the effects of a finite nonzero medium resolution length $L_{\rm res}$, defined such that two partons separated by a distance less than $L_{\rm res}$ will interact with the plasma as if they were a single color-charged object, and will interact with the plasma independently if and only if they are separated by a distance greater than $L_{\rm res}$. The effects of a finite nonzero $L_{\rm res}$ have been implemented and studied in the Hybrid Model in Refs.~\cite{Hulcher:2017cpt, Kudinoor:2025gao}. In this paper, we treat all partons as if they interact independently with the medium at all times, and leave the analysis of Moli\`ere scatterings between multiple unresolved jet-partons and thermal medium-partons~\cite{Pablos:2024muu} to future studies.

In Section~\ref{sec:results}, we present a systematic study of our implementation of Moli\`ere scatterings in the Hybrid Model
and how their inclusion impacts a variety of jet observables in PbPb collisions, including the suppression $R_{\rm AA}$ of charged-hadrons and jets, jet shapes and fragmentation functions, the jet girth $g$, groomed jet observables like the Soft Drop angle $R_g$ and 
leading $k_T$, and observables that focus on the number and angular distribution of (sub)jets within jets.
We find that, compared to when elastic Moli\`ere scatterings are excluded from the Hybrid Model, the presence of elastic scatterings broadens the radial distribution of hadronic energy within jets (i.e. broadens the jet shape and girth) and it broadens the Soft Drop angle $R_g$ of photon-tagged and inclusive jets in PbPb collisions.   
We confirm that the selection bias intrinsic to selecting jets based upon their own energy can be partially mitigated by selecting gamma-jet events based on the photon energy. In particular, we find that the distributions of both the Soft Drop angle $R_g$ and the girth $g$ of gamma-jets in PbPb collisions broaden due to Moli\`ere scatterings as long as jets which are sufficiently softer than the photon are included.

Our analysis in Section ~\ref{sec:results} of Hybrid Model calculations of
both the Soft Drop splitting angle $R_g$ and the jet girth $g$ in photon-tagged $R=0.2$ jets and $R=0.6 $ jets, in each case
considering both samples selected with $x_J\equiv p_T^{\rm jet}/p_T^\gamma>0.2$ and $x_J>0.8$
provides a compelling case study.
We find that in samples of $R=0.2$ photon–tagged jets with $x_J>0.2$ recoiling against photons with $p_T^\gamma>150$~GeV, the broadening of the $R_g$ and $g$ distributions induced by Moli\`ere scattering in PbPb collisions relative to those in pp collisions 
can overcome the narrowing associated with jet selection bias and show that jet wakes make a negligible contribution to these observables in such narrow jets.  Comparing the PbPb/pp ratios of these distributions in samples selected with $x_J>0.8$ and $x_J>0.2$ serves to quantify the effects of selection bias due to energy loss, and comparing these distributions for jets with $R=0.6$ and $R=0.2$ reveals a fascinating interplay between 
the effects of jet wakes and Moli\`ere scattering.
Jet wakes make a dominant contribution to the PbPb/pp ratio for both $R_g$ and $g$ in a sample of $R=0.6$ jet with $x_J>0.2$, but only to $g$ in a sample of these wide jets with 
the more restrictive selection $x_J>0.8$.
This provides a roadmap for how experimentalists can use their ability to dial two knobs --- the radius $R$ of the jets they reconstruct  and 
the $x_J$ selection criterion for which jets they include in their sample --- so as to 
separately focus on and highlight
the effects of selection bias due to energy loss, jet wakes, and Moli\`ere scattering.
With regards to Moli\`ere scattering in particular,
we explain how our study, in combination with the existing tantalizing CMS measurements~\cite{CMS:2024zjn} of $R_g$ and $g$ for $R=0.2$ photon–tagged jets with $p_T^\gamma>100$~GeV and $x_J>0.4$, indicates that  measurements with a modestly higher photon-$p_T$ 
that allows for jets with a somewhat lower $x_J$ to be included in the analysis
can reveal an enhancement of photon–tagged jets with large $R_g$ and large $g$ in PbPb collisions relative to pp collisions. This would be a distinctive model-independent experimental signature of hard scattering 
of jet partons off quasiparticles in QGP, and would open the door to studying the microscopic particulate nature of QGP and seeing how a strongly coupled QCD liquid emerges from an asymptotically free gauge theory.
We conclude and look ahead, in particular towards the importance of making such measurements in oxygen-oxygen collisions, in Section~\ref{sec:conclusions}.


\section{Moli\`ere scattering}
\label{sec:moliere}

Motivated by the tantalizing possibility that large angle, large momentum-transfer, elastic scatterings between jet partons and thermal quasiparticles can reveal the short-distance structure of QGP, we now turn to a quantitative description of these scattering processes. In this section, we review and adapt earlier work in Ref.~\cite{DEramo:2018eoy} on the perturbative framework needed to compute the differential probabilities with which such Moli\`ere scatterings occur. Our goal here is to establish  notation and define the kinematic variables, and probability distributions that will later allow us to consistently incorporate elastic $2 \rightarrow 2$ scatterings into the Hybrid Model. 

\subsection{Review of a Previous Calculation}

\begin{figure}[t]
    \centering
    \includegraphics[width=0.7\textwidth]{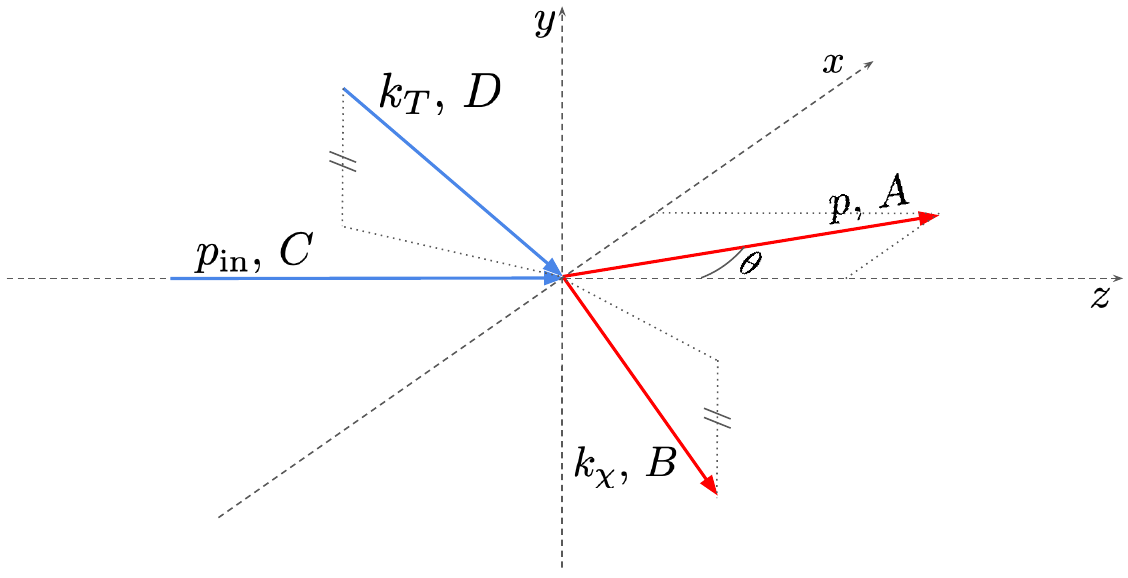}
    \caption{Schematic of a $2\rightarrow 2$ Moli\`ere scattering process, where an incoming particle of type $C$ with momentum $p_{\rm in}$ in the $+\hat z$ direction scatters off a thermal particle of type $D$ with momentum $k_T$.  The outgoing particle of type $A$ has momentum $p$ and scatters
    in the $(z,x)$ plane at an angle $\theta$ relative to the direction of propagation of the incoming hard parton. The other outgoing parton of type $B$ has momentum $k_\chi$ and scatters at an angle which can be determined using energy-momentum conservation. For example, momentum conservation means that $k_\chi$ has the same $\hat y$ component as $k_T$, as indicated in the diagram.  
    $A$, $B$, $C$, and $D$ can each be a gluon, quark, or antiquark.} 
    \label{fig:eventexample}
\end{figure}

When viewed at length scales of order the inverse of its temperature $T$ and longer, quark-gluon plasma (QGP) behaves as a strongly coupled liquid. It has no quasiparticles with mean free paths that are long compared to $1/T$ and flows as hydrodynamic liquid with a lower specific shear viscosity than that
of any other known liquid. 
However, when it is probed at shorter length scales with sufficiently high momentum transfer, the asymptotic freedom of QCD mandates the presence of quark-like and gluon-like quasiparticles. 
High energy partons within jets can trigger such exchanges, revealing the particulate structure of QGP. Such large-angle, high-momentum transfer, $2 \rightarrow 2$ elastic scatterings between high energy partons in a jet and quasiparticles in the medium are rare but not exponentially rare. Fig.~\ref{fig:eventexample} shows a schematic of what a Moli\`ere scattering process looks like. 
It is a high momentum transfer scattering between an incident parton with momentum $p_{\rm in}$ and a target thermal parton from the QGP medium with momentum $k_T$ (where the subscript $T$ refers to the word ``thermal'', not ``transverse'') that produces two outgoing partons with momenta $p$ and $k_\chi$. Each of the partons can be a gluon, quark, or antiquark. We shall treat the quarks and antiquarks as massless.
Note that the two outgoing partons can each be either the deflected incident parton or the recoiling QGP parton after it has been kicked. The calculation of the probability that an incident parton with momentum $p_{\rm in}$ experiences a Moli\`ere scattering that yields an outgoing parton with momentum $p$  at an angle $\theta$ was set up in Refs.~\cite{DEramo:2012uzl,DEramo:2018eoy}. In this Section, we review the results of Ref.~\cite{DEramo:2018eoy}.

We adopt the notation used in Ref.~\cite{DEramo:2018eoy} to denote different parton species with lower case letters (i.e. $a = u, \overline{u}, d, \overline{d}, s, \overline{s}, g$) and to denote different types of partons, like gluons, quarks, and antiquarks using uppercase indices (i.e. $A = G, Q, \overline{Q}$). For a $2 \rightarrow 2$ Moli\`ere scattering like that shown in Fig.~\ref{fig:eventexample}, $F^{C \rightarrow A}(p, \theta; p_{\rm in})$ denotes the probability of finding an outgoing parton of type $A$ with energy $p$ at angle $\theta$ with respect to the direction of an incoming parton of type $C$ with energy $p_{\rm in}$. The leading order perturbative QCD calculation of $F^{C \rightarrow A}(p, \theta; p_{\rm in})$ is introduced and described in Ref.~\cite{DEramo:2018eoy}.

To implement Moli\`ere scatterings in the Hybrid Model, two steps are required. First, at each time step in the evolution of a parton shower, for each parton in the shower at that time, we need to calculate the probability that a Moli\`ere scattering occurs. This can be done by integrating $F^{C \rightarrow A}(p, \theta; p_{\rm in})$ over the variables $p$ and $\theta$. Second, if any Moli\`ere scatterings occur, we then 
need to determine which type of $2 \rightarrow 2$ scattering process takes place (e.g. $qq \rightarrow qq$, $q\overline{q} \rightarrow gg$, etc.) and what the momenta are of the two outgoing partons. In Section~\ref{sec:phase-space-constraints-and-sampling}, we shall describe how we accomplish these two steps and then in Section~\ref{sec:implementing-in-hybrid} we shall implement them in the Hybrid Model.

Before we move forward with any computations, it is useful to review some key details about the calculation of $F^{C \rightarrow A}(p, \theta; p_{\rm in})$ as presented in Ref.~\cite{DEramo:2018eoy}; this is our goal in this Section. Suppose that $f_a(p, \theta, t)$ denotes the probability of finding a parton of species $a$ in a phase-space cell at angle $\theta$ (relative to the incident parton) with energy $p$ at time $t$. So, if a parton enters a static brick of QGP at time $t = t_I$ with energy $p_I$, then $f_a(p, \theta, t_I) = 0$ for all values of $p \neq p_I$ and $\theta \neq 0$. Assume that this incident parton propagates through the plasma for a time $\Delta t$. Then, $f_a(p, \theta, t_I + \Delta t)$ is a phase space distribution function for an outgoing parton of species $a$ that can be nonzero for $p \neq p_I$ and $\theta \neq 0$.


One can then write the probability distribution function $F^{C \rightarrow A}(p, \theta; p_{\rm in})$ as a sum over all possible processes with $C$ in the initial state and $A$ in the final state
\begin{equation}
    F^{C \rightarrow A}(p, \theta; p_{\rm in}) = V \frac{p^2 \text{sin}(\theta)}{(2\pi)^2} \sum_{a \in A} \nu_a f_a(p, \theta, t_I + \Delta t),
\end{equation}
where the prefactor in front is the Jacobian of the phase space integration and $\nu_a$ is a degeneracy factor that accounts for helicity and color configurations. In particular, $\nu_a = 2(N_c^2 - 1)$ if $a$ is a gluon and $\nu_a = 2(N_c - 1)$ if $a$ is a quark or antiquark. The probability of finding a parton of any type with momentum $p$ and angle $\theta$ in the final state is then obtained by summing over all possible $A$:
\begin{equation}
    F^{C \to \text{all}}(p, \theta) = F^{C \to G}(p, \theta) + F^{C \to Q}(p, \theta) + F^{C \to \overline{Q}}(p, \theta).
\end{equation}

Suppose that we wish to consider an elastic $2 \rightarrow 2$ scattering process $cd \rightarrow ab$.  We further assume that both the partons $a$ and $b$ are detected in the final state; this differs from the assumption in Ref.~\cite{DEramo:2018eoy}, where only the parton $a$ is detected in the final state and the parton $b$ is undetected. After examining the form of $f_a(p, \theta, t_I + \Delta t)$ in Eq. (2.4) of Ref.~\cite{DEramo:2018eoy} and noting the difference described in the preceding sentence, $F^{C \to A}(p, \theta; p_{\text{\rm in}})$ is given by the expression 
\begin{equation} \label{eq:probfinal}
\begin{aligned}
F^{C \to A}(p, \theta; p_{\text{\rm in}}) &= V \, \frac{\kappa}{T} \frac{p^2 \sin \theta}{(2\pi)^2} \sum_n w_C^{(n)} \, \frac{\tilde{\delta}_{a,A}}{1 + \delta_{cd}} 
\int_{\bm{p}', \bm{k}', \bm{k}} \frac{|\mathcal{M}_{ab \leftrightarrow cd}^{(n)}|^2}{g_s^4} \\
&\quad \times \frac{1}{\nu_C} \left[ \tilde{\delta}_{d,C} \, f_I(\bm{k}') \, n_c(\bm{p}') + \tilde{\delta}_{c,C} \, f_I(\bm{p}') \, n_d(\bm{k}') \right]\\
&\quad \times {\left[ 1 \pm n_a(\bm{p}) \right]} \left[ 1 \pm n_b(\bm{k}) \right],
\end{aligned}
\end{equation}
where $\kappa \equiv g_s^4 T \Delta t$ is a dimensionless parameter that depends on the thickness $\Delta t$ of the QGP brick and the strength of the QCD coupling $g_s$,
\begin{equation}
    f_I(\vp) \equiv \nu_a f_a(\vp,t_I) = \frac{1}{V}\,\frac{4\pi^{2}}{p_{\rm in}^{2}}\,
\delta(p - p_{\rm in})\,\delta(\cos\theta - \cos\theta_{\rm in}),
\end{equation}
where we have already chosen $\theta_{\rm in} = 0$, $\tilde{\delta}_{a, A}$ is a generalized Kronecker delta which equals 1 if $a \in A$ and is 0 otherwise, 
$n_a$ is the generalized medium ``soft” distribution function
\begin{equation}
    n_a(p) = \frac{\tilde{\delta}_{a, G}}{e^{p/T} - 1} + \frac{\tilde{\delta}_{a, Q} + \tilde{\delta}_{a, \overline{Q}}}{e^{p/T} + 1}, 
    \label{eq:ndefn}
\end{equation}
$n_b$, $n_c$ and $n_d$ are defined analogously, and
$|\mathcal{M}_{ab \leftrightarrow cd}^{(n)}|^2/g_s^4$ is the leading order expression for the corresponding QCD squared matrix element for the different binary collision processes $ab \leftrightarrow cd$ of type $(n)$
in vacuum, summed over initial and final polarizations and colors. The different binary collision processes, these matrix elements, and their associated degeneracy factors $w_C^{(n)}$ are listed in Table~\ref{tab:QCDprocesses}.

\begin{table}
\begin{center}
\renewcommand{\arraystretch}{1.3}
\begin{tabular}{|c | c | c | c | c | c|}
\hline 
$(n)$ & Process & $\left|{\cal M}^{(n)}\right|^2 / g_s^4$ & $w^{(n)}_Q$ & $w^{(n)}_{\bar Q}$ & $w^{(n)}_G$ \\ \hline  \hline
$1$ & $q q \lra q q$ & $8\,  \frac{\df^2 \, \cf^2} {\da} \left( \frac{s^2+u^2}{t^2} + \frac{s^2+t^2}{u^2} \right) + 16\, \df \, \cf \left( \cf {-} \frac{\ca}{2} \right) \frac{s^2}{tu}$ &  
$1$ & $0$ & $0$ \\
$2$ & $\bar{q} \bar{q} \lra \bar{q} \bar{q}$ & $\left|{\cal M}^{(1)}\right|^2 / g_s^4$ &  $0$ & $1$ & $0$\\ \hdashline
$3$ & $q \bar{q} \lra q \bar{q}$ & $8\,  \frac{\df^2 \, \cf^2}{\da}   \left( \frac{s^2+u^2}{t^2} + \frac{t^2+u^2}{s^2} \right)
	    + 16\, \df \, \cf \left( \cf {-} \frac{\ca}{2} \right) \frac{u^2}{st}$ &  $1$ & $1$& $0$\\ \hdashline
$4$ & $q q^\prime \lra q q^\prime$ & $8\,  \frac{\df^2 \, \cf^2}{\da} \left( \frac{s^2+u^2}{t^2} \right)$ & $N_f - 1$ & $0$ & $0$ \\ 
$5$ & $\bar{q} \bar{q}^\prime \lra \bar{q} \bar{q}^\prime$ & $\left|{\cal M}^{(4)}\right|^2 / g_s^4$ &  $0$ & $N_f - 1$ & $0$ \\
$6$ & $q \bar{q}^\prime \lra q \bar{q}^\prime$ & $\left|{\cal M}^{(4)}\right|^2 / g_s^4$ &  $N_f - 1$ & $N_f - 1$ & $0$ \\ \hdashline
$7$ & $q \bar{q} \lra q^\prime \bar{q}^\prime$ & $8\, \frac{\df^2 \, \cf^2}{\da} \left( \frac{t^2 + u^2}{s^2} \right)$ &  $N_f - 1$ & $N_f - 1 $& $0$ \\ \hdashline
$8$ & $q \bar{q} \lra g g$ & $8\, \df \, \cf^2 \left( \frac{t^2 + u^2}{t u}  \right) - 8\, \df \, \cf \, \ca \left( \frac{t^2+u^2}{s^2} \right)$ &  $1$ & $1$ & $N_f$ \\ \hdashline
$9$ & $q g \lra q g$ & $-8\, \df \, \cf^2 \left( \frac{u}{s}  +  \frac{s}{u} \right) + 8\, \df \, \cf \, \ca \left( \frac{s^2 + u^2}{t^2} \right)$ &  $1$ & $0$ & $N_f$\\
$10$ & $\bar{q} g \lra \bar{q} g$ & $\left|{\cal M}^{(9)}\right|^2 / g_s^4$ &  $0$ & $1$ & $N_f$\\  \hdashline
$11$ & $g g \lra g g$ & $ 16\, \da \, \ca^2 \left( 3 - \frac{su}{t^2} - \frac{st}{u^2} - \frac{tu}{s^2} \right)
$ &  $0$ & $0$& $1$ \\
\hline
\end{tabular}
\end{center}
\caption{List of the binary collision processes labeled by $(n)$ that can produce a hard parton in the final state with large transverse momentum with respect to the incoming probe. 
Here, $q$ and $q'$ are quarks of distinct flavors, $\bar q$ and $\bar{q}'$ the associated antiquarks, and $g$ is a gluon. The third column lists explicit leading order perturbative QCD expressions for the corresponding squared matrix elements, in vacuum, summed over initial {\it and final} polarizations and colors, as a function of the Mandelstam variables $t=-2(p' p-\vp'\cdot\vp)$, $u=-2(p' k -\vp'\cdot\vk)$ and $s = -t -u$. (See Ref.~\cite{Arnold:2002zm}.) In an SU($N_{c}$) theory with fermions in the fundamental representation, we have for the dimensions of the representations and the Casimir factors $\df= \ca = N_{c}$, $\cf=(N^2_{c}-1)/(2N_{c})$, and $\da = 2\, \df \cf=N^2_{c}-1$. For SU(3) (i.e. QCD), $\df = \ca = 3$, $\cf = 4/3$, and $\da = 8$. Finally, we give the degeneracy factors $w^{(n)}_C$ appearing in Eq.~\eqref{eq:probfinal}. Here, $N_f$ is the number of light flavors; we take $N_f=3$ throughout.
}
\label{tab:QCDprocesses}
\end{table}

Throughout this work, we shall assume that the momentum distributions of partons from the QGP medium are given
simply by Eq.~\eqref{eq:ndefn} and the matrix elements are given simply by the leading order perturbative QCD expressions in Table~\ref{tab:QCDprocesses}. The first of these assumptions 
must be questioned, as QGP is strongly coupled, but it suffices for our purposes of identifying experimental observables that are well-suited for revealing and, ideally, isolating the effects of Moli\`ere scattering.  We leave investigating more sophisticated forms for $n_a(p)$ to future work. Ideally, future experimental measurements that succeed in isolating effects of Moli`ere scattering may one day be used to infer constraints on the form of 
$n_a(p)$ in strongly coupled QGP.  The second of these assumptions may also be questioned, but we are less concerned about it since we shall only consider elastic scattering processes in which the squared momentum transfer is larger than the squared Debye mass 
by at least a factor of 10, which provides some justification for assuming weak coupling.

The expression on the right-hand side of Eq.~\eqref{eq:probfinal} for a scattering process $cd \rightarrow ab$, can be understood as follows. As in Ref.~\cite{DEramo:2018eoy}, our convention is that the outgoing parton $a$ in the final state comes from the parton $c$ in the initial state, and the outgoing parton $b$ in the final state comes from the parton $d$ in the initial state. 
So, the term $\tilde{\delta}_{c,C} \, f_I(\bm{p}') \, n_d(\bm{k}')$ corresponds to the case where the outgoing parton $a$ came from the incident parton $c$, and the outgoing parton $b$ came from the thermal medium parton $d$ after it was kicked by the incident parton $c$, as illustrated in Fig.~\ref{fig:eventexample}.  
Conversely, the term $\tilde{\delta}_{d,C} \, f_I(\bm{k}') \, n_c(\bm{p}')$ corresponds to the case when the outgoing hard parton $a$ originates from the thermal medium parton $c$ after having been kicked by the incident parton $d$, which is deflected and sources the outgoing parton $b$. 
The Kronecker delta $\delta_{c, d}$ accounts for when $c$ and $d$ are identical. The modified Kronecker delta $\tilde{\delta}_{a, A}$ ensures that only final state partons $a$ of type $A$ contribute to the total probability $F^{C \to A}$. Finally, the sum runs over all possible binary collision processes $ cd \rightarrow ab $, with $\bm{p}', \bm{k}', \bm{p}, \bm{k}$ representing the momenta of $c, d, a, b$, respectively. 
The phase space integral is written in the compact form
\begin{equation}
\begin{aligned}
\int_{\bm{p}', \bm{k}', \bm{k}} &\equiv \frac{1}{2p} 
\int \frac{d^3\bm{k}}{2k (2\pi)^3} 
\int \frac{d^3\bm{p}'}{2p' (2\pi)^3} 
\int \frac{d^3\bm{k}'}{2k' (2\pi)^3} \\
&\quad \times (2\pi)^4 \, \delta^{(3)}\left( \bm{p} + \bm{k} - \bm{p}' - \bm{k}' \right) 
\, \delta\left( p + k - p' - k' \right) .
\end{aligned}
\end{equation}

\subsection{Decomposition of the Total Probability}
\label{sec:decomposition}

The sum over $n$ in Eq.~\eqref{eq:probfinal} runs over all 11 processes in Table~\ref{tab:QCDprocesses}. $F^{C \to A}(p, \theta; p_{\text{\rm in}})$ may therefore be decomposed into the partial contributions from each of the 11 binary collision processes for a scattering $cd \rightarrow ab$:
\begin{equation}
F^{C \to A}(p, \theta; p_{\text{\rm in}}) \equiv \sum_n F^{C \to A}_{(n)}(p, \theta; p_{\text{\rm in}})\ .
\end{equation}
Following Section 2.5 of Ref.~\cite{DEramo:2018eoy}, each partial contribution $F^{C \to A}_{(n)}(p, \theta; p_{\text{\rm in}})$ is given by a linear combination of two phase space integrals:
\begin{equation} \label{eq:helper1}
\langle (n) \rangle_{D,B} \equiv \frac{1}{T} \frac{{\left[ 1 \pm n_A(p) \right]} \, p^2 \sin \theta}{(2\pi)^2} 
\int_{\bm{p}', \bm{k}', \bm{k}} \frac{|\mathcal{M}^{(n)}|^2}{g_s^4} 
f_I(\bm{p}') \, n_D(\bm{k}') \left[ 1 \pm n_B(k) \right],
\end{equation}

\begin{equation} \label{eq:helper2}
\langle (\tilde{n}) \rangle_{D,B} \equiv \frac{1}{T} \frac{{\left[ 1 \pm n_A(p) \right]} \, p^2 \sin \theta}{(2\pi)^2} 
\int_{\bm{p}', \bm{k}', \bm{k}} \frac{|\mathcal{M}^{(\tilde{n})}_{{ t \leftrightarrow u}}|^2}{g_s^4} 
f_I(\bm{k}') \, n_D(\bm{p}') \left[ 1 \pm n_B(k) \right],
\end{equation}
where uppercase indices $A, B, C, D$ denote whether the outgoing partons of type $A$ and $B$ and the incoming partons of type $C$ and $D$ are quarks, antiquarks, or gluons {and where we shall explain the $t\leftrightarrow u$ notation further below}.
The factors ${[1 \pm n_A]}$ and $[1 \pm n_B]$ (the sign is $+$ if $A$ or $B$ is a boson and $-$ if $A$ or $B$ is a fermion) describe Bose enhancement or Pauli blocking and depend on the occupation of the mode in which the outgoing particles $A$ or $B$ in the final state are produced. Note that in the limit of high outgoing momentum $p$ for final state particle $a$, the Pauli blocking factor ${[1 \pm n_A]} \approx 1$ and we recover 
Eqs.~(2.15a) and (2.15b) of Ref.~\cite{DEramo:2018eoy}. 
In the present work, we restrict to 
hard scattering processes with high momentum {\it transfer}, but do not require the momenta of any particles to be high.
One can use Eqs.~\eqref{eq:helper1} and \eqref{eq:helper2} and 
Table~\ref{tab:QCDprocesses} to write expressions for all possible $F^{C \to A}(p, \theta; p_{\text{\rm in}})$:

\begin{align}
F^{Q\to Q}(p,\theta; p_{\rm in})
&=
\frac{\kappa}{\nu_q}
\Bigg\{
\frac{1}{2} \langle(1)\rangle_{Q,Q}
+
\frac{1}{2} \langle(\tilde 1)\rangle_{Q,Q}
+
\langle(3)\rangle_{Q,Q}
+
\langle(9)\rangle_{G,G}
\nonumber\\
&\qquad
+
(N_f - 1)
\Big[
\langle(4)\rangle_{Q,Q}
+
\langle(\tilde{4})\rangle_{Q,Q}
+
\langle (6) \rangle_{Q,Q}
+
\langle(7)\rangle_{\bar Q,\bar Q}
\Big]
\Bigg\}\  , \label{eq:fqq} \\
F^{Q\to G}(p,\theta; p_{\rm in}) &= \frac{\kappa}{\nu_q}
\Bigg\{ \langle(8)\rangle_{Q,G} + \langle(\tilde 8)\rangle_{Q,G} + \langle(\tilde 9)\rangle_{G,Q} \Bigg\}\  , \label{eq:fqg} \\
F^{Q\to \bar{Q}}(p,\theta; p_{\rm in})
&=
\frac{\kappa}{\nu_q}
\Bigg\{ \langle(\tilde 3)\rangle_{Q,Q}
+ (N_f - 1) \Big[ \langle(\tilde 6)\rangle_{Q,Q} + \langle(\tilde{7})\rangle_{Q,Q}
\Big]
\Bigg\}\  , \label{eq:fqqbar} \\
F^{G\to Q}(p,\theta; p_{\rm in}) &= \frac{\kappa}{\nu_g} N_f
\Bigg\{ \langle(8)\rangle_{G, Q} + \langle(\tilde 9)\rangle_{Q,G} \Bigg\}\  , \label{eq:fgq} \\
F^{G\to G}(p,\theta; p_{\rm in}) &= \frac{\kappa}{\nu_g}
\Bigg\{ N_f \Big[ \langle(9)\rangle_{Q, Q} + \langle(10)\rangle_{Q,Q} \Big] + \frac{1}{2} \langle (11) \rangle_{G,G} + \frac{1}{2} \langle (\tilde{11}) \rangle_{G,G} \Bigg\}\  , \label{eq:fgg} \\
F^{G\to \bar{Q}}(p,\theta; p_{\rm in}) &= \frac{\kappa}{\nu_g} N_f
\Bigg\{ \langle(8)\rangle_{G, Q} + \langle(\tilde{10})\rangle_{Q,G} \Bigg\}. \label{eq:fgqbar} 
\end{align}

Next, we recast Eqs.~\eqref{eq:helper1} and \eqref{eq:helper2} 
into forms suitable for analytic and numerical evaluation. To this end, let $\omega \equiv p_{\rm in} - p$ and $\vq \equiv \vp - \vp_{\rm in}$ denote the energy and momentum difference, respectively, between the incident parton and the outgoing parton $a$. Then, the momentum transfer is $Q^\mu = (\omega, \vq)$.
Then, upon defining 
\begin{equation}
q \equiv |\vq | = \sqrt{p_{\rm in}^2 + p^2 - 2p\, p_{\rm in}  \cos \theta } \ , 
\label{eq:qdefn}
\end{equation}
energy–momentum conservation requires the incoming thermal parton to have an energy of at least $k_T^{\rm min} = (q - \omega)/2$. 
Furthermore,   we denote the 4-momentum of the thermal parton from the medium involved in the elastic scattering
by $(k_T, \vk_T)$ and note that the 
four-momentum of the outgoing parton $b$ is given by $(k_\chi,\vk_\chi)$ with $k_\chi = k_T + \omega$ and $\vk_\chi=\vk_T + \vq$. 
(In our phase space integration, both $\vk'=\vk_T$ and $\vp'=\vk_T$ will contribute; similarly, both $\vp=\vk_\chi$ and $\vk=\vk_\chi$ will contribute.)

Given all this, one can reformulate Eq.~\eqref{eq:helper1} as
\begin{equation}
\langle (n) \rangle_{D,B} = \frac{{\left[ 1 \pm n_A(p) \right]}}{16 (2\pi)^3} \left( \frac{p \sin \theta}{p_{\text{\rm in}} q T} \right) 
\int_{k_{\text{min}}}^{\infty} dk_T \, n_D(k_T) \left[ 1 \pm n_B(k_\chi) \right]
\int_0^{2\pi} \frac{d\phi}{2\pi} \, \frac{|\mathcal{M}^{(n)}|^2}{g_s^4},
\end{equation}
where $\phi$ is the angle between the two planes identified by the pair of vectors $(\bm{p}, \bm{q})$ and $(\bm{q}, \bm{k}_T)$, and $\mathcal{M}^{(n)}$ are the QCD matrix elements that appear in Table~\ref{tab:QCDprocesses}. These matrix elements are functions of the Mandelstam variables $t$ and $u$. Following Ref.~\cite{DEramo:2018eoy}, we define the quantities 
{$t$, $u$ and $s$ as}
\begin{eqnarray} \label{eq:mandelstahm}
{t}\  & \equiv & \ \omega^2 - q^2 = -2p \, p_{\rm in}(1 - \textrm{cos}(\theta)), \nonumber\\
{u}\  & \equiv & \ -{s} - {t},\nonumber\\
{s}\  & \equiv &\  \left( -\frac{{t}}{2q^2} \right)
\left\{ \left[ (p_{\text{\rm in}} + p)(k_T + k_\chi) + q^2 \right]
- \sqrt{(4p_{\text{\rm in}}p + {t})(4k_T k_\chi + {t})} \cos \phi \right\}.
\end{eqnarray}
{These are the Mandelstam variables in the matrix elements 
$\mathcal{M}^{(n)}$ in Eq.~\eqref{eq:helper1}, but in the 
matrix elements $\mathcal{M}^{(n)}_{t\leftrightarrow u}$ in Eq.~\eqref{eq:helper2} the quantity $t$ ($u$) defined in Eq.~\eqref{eq:mandelstahm} plays the role of the Mandelstam variable $u$ ($t$) --- hence the subscript $t\leftrightarrow u$ in Eq.~\eqref{eq:helper2}.}  
%
Thus,
\begin{equation}
\begin{aligned}
\langle (\tilde{n}) \rangle_{D,B} 
&= \frac{{\left[ 1 \pm n_A(p) \right]}}{16 (2\pi)^3} \left( \frac{p \sin \theta}{p_{\text{\rm in}} q T} \right)
\int_{k_{\text{min}}}^{\infty} dk_T \, n_D(k_T) \left[ 1 \pm n_B(k_\chi) \right]
\int_0^{2\pi} \frac{d\phi}{2\pi} \frac{|\mathcal{M}^{(n)}_{{t\leftrightarrow u}}|^2
}{g_s^4}\ .
\end{aligned}
\end{equation}
The phase space integrals $\langle (n) \rangle_{D, B}$ 
and $\langle (\tilde n)\rangle_{D,B}$
for $n = 1, ... , 11$ are functions of $\theta$ and $p$. 

Recall that implementing elastic $2 \rightarrow 2$ scatterings within a framework of jet quenching requires two steps: determining whether or not such a scattering has occurred, and then determining which process $cd \rightarrow ab$ occurred with what momenta for the partons $a$ and $b$ after the elastic scattering. The first of these two steps requires that we integrate our probability $F^{C \to A}(p, \theta; p_{\text{\rm in}})$ over all allowed values of angle $\theta$ and outgoing momentum $p$. This amounts to performing a total of four integrals, over $\phi$, $k_T$, $\theta$ and $p$. 

Finally, following Ref.~\cite{DEramo:2018eoy} we observe that $|\mathcal{M}^{(n)}({t}, {u})|^2$ and $|\mathcal{M}^{(\tilde{n})}({u}, {t})|^2$ can be decomposed into a linear combination of 7 terms $m_i$:
\begin{equation}
    \frac{1}{g_s^4} \left| \mathcal{M}^{(n)}({t}, {u}) \right|^2 
    = \sum_i c_i^{(n)} m_i({t}, {u}), \qquad \frac{1}{g_s^4} \left| \mathcal{M}^{(\tilde{n})}({u}, {t}) \right|^2 
    = \sum_i \tilde{c}_i^{(\tilde{n})} m_i({t}, {u}),
\label{eq:lin-comb-7}
\end{equation}
where the 7 terms $m_i$ in question are
\begin{equation}
\begin{split}
m_1 = \left( \frac{{s}}{{t}} \right)^2, \quad
m_2 = -\left( \frac{{s}}{{t}} \right), \quad
m_3 = 1, \quad
m_4 = -\left( \frac{{t}}{{s}} \right), \quad
m_5 = \left( \frac{{t}}{{s}} \right)^2, \\
m_6 = -\left( \frac{{t}}{{s} + {t}} \right) 
= \frac{{t}}{{u}}, \quad
m_7 = \left( \frac{{t}}{{s} + {t}} \right)^2 
= \left( \frac{{t}}{{u}} \right)^2.
\end{split}
\label{eq:seven-ms}
\end{equation}
The coefficients $c_i^{(n)}$ and $\tilde{c}_i^{(\tilde{n})}$ 
can be determined from Table~\ref{tab:QCDprocesses} and are given explicitly in Appendix~\ref{app:processes}. 
Since the coefficients $c_i^{(n)}$ and $\tilde{c}_i^{(\tilde{n})}$ only depend on $N_c$, we may write $\langle (n) \rangle_{D,B}$ as
\begin{equation} \label{eq:lincomb}
    \langle (n) \rangle_{D,B} = \sum_{i = 1}^7 c_i^{(n)} \langle m_i({t}, {u}) \rangle,
\end{equation}
where $\langle m_i({t}, {u}) \rangle$ is defined as
\begin{equation} \label{eq:mi}
    \langle m_i({t}, {u}) \rangle \equiv \frac{{\left[ 1 \pm n_A(p) \right]}}{16 (2\pi)^3} \left( \frac{p \sin \theta}{p_{\text{\rm in}} q T} \right) 
\int_{k_{\text{min}}}^{\infty} dk_T \, n_D(k_T) \left[ 1 \pm n_B(k_\chi) \right]
\int_0^{2\pi} \frac{d\phi}{2\pi} \, m_i({t}, {u}).
\end{equation}

This concludes our review of the calculation as formulated in Ref.~\cite{DEramo:2018eoy}, with small adaptations for the purposes of our work that we have described along the way.

\section{Phase-Space Constraints and Sampling Kinematic Variables}
\label{sec:phase-space-constraints-and-sampling}

Before we can embed the $2 \rightarrow 2$ elastic scattering processes described in the previous Section into a Monte Carlo model of jet showers in heavy ion collisions, we must define the portion of phase space where a perturbative description of elastic scatterings with \textit{high momentum exchange} is trustworthy and, conversely, excise the soft sector where 
the shower partons and the QGP are strongly coupled.
In the Hybrid Model, making this excision can also be understood as avoiding double counting.  
The Hybrid Model implementation of energy loss is based upon a strongly coupled calculation in which there is no distinction between elastic and radiative energy loss processes.  
By fitting the Hybrid Model parameter $\kappa_{\rm sc}$ that controls the energy loss of jet partons to experimental data we are already incorporating effects that might alternatively be described as elastic scattering with soft momentum exchange.
Here, therefore, we need to ensure that what we are adding to the Hybrid Model is only elastic scattering where the exchanged momentum is above a threshold that we must choose.  We shall require that the exchanged momentum, squared, is greater than $a m_D^2$, where we shall take the 
Debye mass squared to
be $m_D^2=\frac{1}{3}g_s^2 T^2 (N_c+N_f/2) = \frac{3}{2}g_s^2 T^2$.
Throughout this paper, we shall choose $a=10$ unless otherwise noted.  The QCD coupling $g_s$ that sets the magnitude of the matrix elements in Table~\ref{tab:QCDprocesses} as well as the Debye mass and the threshold parameter $a$ that defines how large the momentum transfer must be in elastic scattering processes that we are adding to the Hybrid Model are the two new Hybrid Model parameters that we introduce in this study.   

In Section \ref{sec:phasespace} below, we provide an explicit description of how we impose the phase-space constraints corresponding to keeping only those elastic scattering processes with exchanged momentum squared greater than $a m_D^2$.
Having identified the kinematic constraints that isolate hard, perturbative, elastic scatterings, we turn to the practical question of how to 
integrate over the allowed phase space in order to determine the probability that such an elastic scattering occurs 
as well as how to
sample the allowed phase space with the appropriate probability distribution. In Sec.~\ref{sec:sampling}, we describe a Monte Carlo procedure for generating elastic scattering events according to the differential probabilities derived in Sec.~\ref{sec:moliere}, while consistently enforcing the phase-space restrictions discussed in Sec.~\ref{sec:phasespace}. This step is essential for embedding Moli\`ere scatterings into a dynamical jet-quenching model such as the Hybrid Model.

\subsection{Phase-Space Constraints}
\label{sec:phasespace}

Following Ref.~\cite{DEramo:2018eoy}, 
we shall require that the square of the four momentum difference
between the incident parton and each of the two outgoing partons must be larger than $a m_D^2$. That is, we shall require that
\begin{equation} \label{eq:constraint1}
|\tilde t|=2p_{\rm in}p(1-\cos\theta)>a m_D^2\ ,
\end{equation}
and
\begin{equation} \label{eq:constraint2}
|\tilde u|=2(p_{\rm in}k_{\chi} - \bm p_{\rm in}\!\cdot\!\bm k_\chi)>a m_D^2\ .
\end{equation}
The two constraints in Eqs.~\eqref{eq:constraint1} and~\eqref{eq:constraint2} serve together to eliminate the phase space corresponding to low momentum exchange processes.

The constraint on $|t|$ in Eq.~\eqref{eq:constraint1} 
is easy to impose as it is directly expressed as a constraint 
on $p$ and $\theta$, cutting off the region where either 
is too close to zero. In Fig.~\ref{fig:bounds} (presented 
and described in Section~\ref{sec:sampling}), this excision corresponds to the black excluded regions along the $p$ and $\theta$ axes.

The constraint on $|u|$ in Eq.~\eqref{eq:constraint2} requires
further examination, because it is not phrased as a constraint on $p$ and $\theta$.  
For any choice of $p$ and $\theta$, there are some values of the other kinematic variables $\phi$ and $k_T$ for which $|u|$ satisfies Eq.~\eqref{eq:constraint2} and other values of these kinematic variables that must be excluded because $|u|$ is too small.
Two illustrative limits highlight how Eq.~\eqref{eq:constraint2}
(together with Eq.~\eqref{eq:constraint1})
constrains the regime of kinematic variables that we must include/exclude.
First, we examine the limit in which the momentum $k_T$ of the thermal parton from the QGP involved in the elastic scattering is very small.
In Appendix~\ref{app:uatktmin}, we evaluate $|u|$ at the minimum 
value of $k_T$ allowed by energy and momentum conservation for given specified values of $p_{\rm in}$, $p$ and $\theta$, namely $k_T = k_T^{\rm min} = (q-\omega)/2$,
obtaining
\begin{equation}
    |u\bigl(k_T = k_T^{\min}\bigr)| =\frac{|t|\, (p - k_T^{\min})}{q} = \frac{2p_{\rm in}p(1-\cos(\theta))\, (p - k_T^{\min})}{q}\ .
    \label{eq:uKmin}
\end{equation}
We then see that if $k_T=k_T^{\rm min}$ both requirements $|t|>a m_D^2$ and $|u|>a m_D^2$ are fulfilled at sufficiently 
high $p$ and $\theta$ whereas at small enough $p$ and/or $\theta$ 
one or both can be violated.
The second illustrative limit that we examine is a collinear limit where we suppose $k_T=p$. Then,
\begin{equation}
    |u\bigl(k_T=p\bigr)| =\left(\frac{2p_{\rm in}p\,\sin(\phi/2)\sin\theta}{q}\right)^2,
\end{equation}
which can become small when $\phi \to 0$ or $\theta \to \pi$, as well as when either $\theta$ or $p$ is small as before.  Hence, for $k_T$ in this regime additional restrictions on $\phi$ and $\theta$ are needed in order to satisfy Eq.~\eqref{eq:constraint2}.

Elastic scattering in the regimes where $|t|$ and $|u|$ are small are exactly the regimes of phase space with collinear divergences that would need need to be treated more carefully 
if we wanted to include such processes. 
This would entail identifying the regions in the
$(\phi,k_T,p,\theta)$ phase space where $|u|$ and $|t|$ are large enough vs. where they are not, and modifying the amplitudes in
the latter region 
%
%
according to the prescription from Ref.~\cite{Arnold:2002zm}. 
As we have already noted, though, doing this would 
not make sense
in the Hybrid Model where we have already incorporated a description of the strongly coupled, soft, exchanges between 
jet partons and the strongly coupled QGP, which we should not double count.
%
%
For our purposes, in the regions in $(\phi,k_T,p,\theta)$ phase space where $|u|$ and $|t|$ are large that describe the large momentum transfer elastic scattering events that are intrinsically
weakly coupled that we seek to add to the Hybrid Model 
we need not modify the scattering amplitudes, and we need
to exclude the regions where $|u|$ and/or $|t|$ are $<a m_D^2$.

We now describe how we implement the conditions $|u|>a m_D^2$ and $|t|>a m_D^2$ in the $(\phi,k_T,p,\theta)$ phase space.
To save notation, we rewrite Eqs.~\eqref{eq:mandelstahm} 
as 
\begin{equation}
    \tilde s = -\tilde t(C - D\cos\phi), \quad \tilde u= -\tilde s -\tilde t = \tilde t(C_u-D\cos\phi),
\label{eq:rewriting-stu}
\end{equation}
where
\begin{align}
    2q^2 C(p_{\text{in}}, p, q, k_T) &\equiv (p_{\text{in}} + p)\,(k_T + k_\chi) + q^2, \label{eq:c} \\
C_u(p_{\text{in}}, p, q, k_T) &\equiv C(p_{\text{in}}, p, q, k_T) - 1, \label{eq:cu} \\
2q^2 D(p_{\text{in}}, p, q, k_T) &\equiv \sqrt{\bigl(4p_{\text{in}}\,p + t\bigr)\,\bigl(4k_T k_\chi + t\bigr)}\ . \label{eq:d}
\end{align}
We also define $k_{\chi}^{\rm min}$ as the minimum
value of $k_\chi$ allowed by energy and momentum conservation
given specified values of $p_{\rm in}$, $p$ and $\theta$, which can be written as
\begin{equation}
k_\chi^{\rm min} = \frac{|t|}{4 k_T^{\rm min}}=\frac{|t|}{2(q-\omega)}\ .
\label{eq:kchi-min}
\end{equation}
%
And, finally, we define the dimensionless quantities 
$x\equiv a m_D^2/|t|$ and $\bar t \equiv t/a m_D^2 = -1/x$
%
and introduce the convention that a tilde over a momentum or energy variable makes it dimensionless via dividing by $T$: 
$\tilde p_{\rm in}=p_{\rm in}/T$, $\tilde p = p/T$, 
$\tilde q=q/T$, $\tilde \omega=\omega/T$, $\tilde k_T \equiv k_T/T$, $\tilde k_\chi = k_\chi/T$ $\ldots$ 
With these definitions, instead of using the variables $(\phi,k_T,p,\theta)$ we can equivalently describe our phase space using the new variables
$\tilde{k}\equiv \tilde{k}_T-\tilde{k}_{T}^{\rm min}$,
$\tilde{k}_{\chi}^{\rm min}$ and $x$ as well as $\phi$. 
(Note that the new variable $\tilde k$ that we have defined here and that we will use throughout the following is {\it not} given by $k/T$, where $k$ is the integration variable in Eq.~\eqref{eq:probfinal}.) 
We have defined new variables in the way that we have because the constraints \eqref{eq:constraint1} and \eqref{eq:constraint2} can be expressed explicitly in terms of these variables, as we now explain.
We have defined $x$ in such a way that satisfying the 
constraint~\eqref{eq:constraint1}, namely $|t|>am_D^2$, can be accomplished by requiring that $x<1$.
In order to satisfy the constraint~\eqref{eq:constraint2}, we need $\cos\phi < (C_u - x)/D$. This condition follows from writing $u = t \left( C_u - D\cos(\phi) \right)$ as in Eq.~\eqref{eq:rewriting-stu}, which makes clear that imposing $|u| > am_D^2$ yields $C_u - D \cos(\phi) > am_D^2/|t| = x$. If $|C_u - x| > D$, this is automatically satisfied for all $\phi$. If $|C_u - x| \leq D$, then $\phi$ is restricted by $\cos\phi < (C_u - x)/D$.
Furthermore, $x$, $\bar{t}$ and  $\tilde{k}_\chi^{\rm min}$ 
must be nonzero, with the additional constraint that $ k_\chi^{\rm min} < p_{\rm in}$, as we now explain.
Since we work only in the hard-scattering regime where $|t|>a m_D^2 > 0$, it is apparent that $x = a m_D^2/|t|$ and $\bar{t} = t/(a m_D^2)$ must be nonzero and from 
Eq.~\eqref{eq:kchi-min} we see that $\tilde k_\chi^{\rm min}$ is also nonzero. 
%
Turning now to showing that $k_\chi^{\min}<p_{\rm in}$, from
Eq.~\eqref{eq:kchi-min} we see that this is equivalent to showing that $|t|<2p_{\rm in}(q-\omega)$.
With $\omega=p_{\rm in}-p$ and 
$q=\sqrt{p_{\rm in}^2+p^2-2p_{\rm in}p\cos\theta}$, since $|t|=2p_{\rm in}p(1-\cos\theta)$ it follows that
\begin{equation} \label{eq:simpler-inequality}
    |t|<2p_{\rm in}(q-\omega) \iff p(1-\cos\theta) < q+p-p_{\rm in} \iff q>p_{\rm in}-p\cos\theta.
\end{equation}
Therefore it suffices to show that $q>p_{\rm in}-p\cos\theta$.
From the
explicit form of $q$ given in Eq.~\eqref{eq:qdefn}, we note that
\begin{equation}
    q^2-(p_{\rm in}-p\cos\theta)^2=p^2\sin^2\theta\ge0.
\end{equation}
Since $q \geq 0$, $q^2 \geq (p_{\rm in}-p\cos\theta)^2 \implies q \geq p_{\rm in}-p\cos\theta$, with equality only for exactly collinear scattering. Thus, throughout the region of space with $|t|>0$,
$k_\chi^{\min}<p_{\rm in}$.



Recall that our first task is (for each parton in a parton shower, for each time step $\Delta t$) to determine whether or not an elastic scattering with high momentum transfer as we have defined it has occurred, and that this requires integrating our probability distribution $F^{C\rightarrow A}$ over the allowed regions of $\phi$, $k_T$, $\theta$ and $p$.  We shall, instead,
formulate this integral in terms of the new variables 
$\phi$, $\tilde k$, $\tilde k_\chi^{\rm min}$ and $x$. Writing the constraint $|t| > am_D^2$ in terms of the new variables is trivial since $|t| > am_D^2 \implies x < 1$. To make explicit that the constraint $|u|>am_D^2$ can indeed be written in terms of the new variables
$(\phi,\tilde k,\tilde k_\chi^{\rm min},x)$, we begin by observing that since
$u= t\,(C_u-D\cos\phi)$ [cf.\ Eq.~(3.5)],
\begin{equation}
|u|>am_D^2 \quad\iff\quad C_u-D\cos\phi > \frac{am_D^2}{|t|}\equiv x
\quad\iff\quad C_u-x-D\cos\phi>0\,.
\label{eq:u-constraint-CD}
\end{equation}
We now need to show that $C_u$ and $D$, which were defined in Eqs.~\eqref{eq:cu} and~\eqref{eq:d} 
in terms of the variables $\omega$, $q$, $p$, $t$,
$k_T$ and $k_\chi$  (as well as the fixed parameter $p_{\rm in}$) 
can be written in terms of the variables
$(\tilde k,\tilde k_\chi^{\rm min},x)$. 
We do so as follows.
First, we 
note that $t=-a m_D^2/x$ and 
observe that $\omega^2-q^2=t$
and Eq.~\eqref{eq:kchi-min} in the form $q-\omega=|t|/(2k_\chi^{\rm min})$ together imply that
\begin{equation}
\omega=k_\chi^{\rm min}-\frac{|t|}{4k_\chi^{\rm min}},\qquad {\rm and} \qquad
q=k_\chi^{\rm min}+\frac{|t|}{4k_\chi^{\rm min}}\,,
\label{eq:omega-and-q}
\end{equation}
Upon recalling also that $p=p_{\rm in}-\omega$, we see that we have written $\omega$, $q$, $p$ and $t$ in terms of $x$ and $k_\chi^{\rm min}$.
Next, the definition $\tilde k\equiv \tilde k_T-\tilde k_T^{\rm min} = \tilde k_T - (\tilde q - \tilde \omega)/2$ allows us to write $\tilde k_T$
in terms of $\tilde k$ as well as $q$ and $\omega$, which we have already written in terms of $x$ and $k_\chi^{\rm min}$.
Last, we observe that 
\begin{equation}
\tilde k +\tilde k_\chi^{\rm min} = 
\tilde k_T + \tilde k_\chi^{\rm min} - \frac{\tilde q-\tilde \omega}{2} = \tilde k_T +\tilde k_\chi^{\rm min}-\frac{|t|}{4k_\chi^{\rm min}}=\tilde k_T+\tilde \omega = \tilde k_T+ \tilde p_{\rm in} - \tilde p = \tilde k_\chi\ ,
\label{eq:five-equalities}
\end{equation}
where we have used Eqs.~\eqref{eq:kchi-min} and \eqref{eq:omega-and-q} as well as energy conservation to relate $\tilde k_\chi$ to $\tilde k$ and $k_\chi^{\rm min}$.
%
With these substitutions, $C_u$ and $D$ as defined in Eqs.~\eqref{eq:cu} and~\eqref{eq:d} 
are expressed in terms of the variables
$(\tilde k,\tilde k_\chi^{\rm min},x)$ and the fixed parameter $\tilde p_{\rm in}$. This means that, per Eq.~\eqref{eq:u-constraint-CD}, when we 
write the constraint $|u|> a m_D^2$ as 
\begin{equation}
{\cal P}_u=\Theta(|u|-am_D^2)=\Theta(C_u-x-D\cos\phi)\ ,
\label{eq:curly-Pu}
\end{equation}
we have cast this constraint entirely
in terms of the new variables $(\phi,\tilde k,\tilde k_\chi^{\rm min},x)$.
Of course, the constraint $|t|>a m_D^2$ 
can straightforwardly be cast as
\begin{equation}
{\cal P}_t=\Theta(|t|-am_D^2)=\Theta(1-x)\ .
\end{equation}
%
Writing the two constraints in this fashion
enables the phase space integral to be  transformed as follows:
%
%
\begin{equation} \label{eq:changeofvariables}
    \int dp\int d\theta\int dk_T\int d\phi {\cal P}_t {\cal P}_u \rightarrow \int_{0}^{1} dx\int_0^{p_{\rm in}}d\tilde{k}_{\chi}^{\rm min}\int_0^{\infty} d\tilde{k}\int_0^{2\pi} d\phi {\cal P}_u|\mathcal{J}|,
\end{equation}
where $|\mathcal{J}|$ is a Jacobian factor which results from the changes of variables and is given by
%
%
\begin{equation}
    \frac{1}{16(2\pi)^3}\frac{p\sin\theta}{p_{\rm in}qT}|\mathcal{J}|=\frac{T^2}{p_{\rm in}^2(4\pi)^3}\frac{am_D^2}{4k_{\chi}^{\rm min}x}\frac{1}{xT}=\frac{1}{\tilde{p}_{\rm in}^2(4\pi)^3}\frac{k_{T}^{\rm min}}{xT}
    =\frac{\eta \tilde{k}_{T}^{\rm min}}{x}
    \label{eq:jacobianprefactor}
\end{equation}
with $\eta^{-1}\equiv (4\pi)^3 \tilde p_{\rm in}^2$.
By construction, this phase space integral transformation is useful for evaluating the probabilities with which elastic scattering processes occur.  We shall see that it is also useful for determining the kinematics of these scatterings. We shall focus on these two goals in the next subsection.


\subsection{Evaluating the Probability of Elastic Scatterings and Sampling Kinematic Variables} \label{sec:sampling}

Having constrained our region of phase space to that which is allowed by the kinematic constraints that include only hard, perturbative, elastic scatterings, we now turn to the practical tasks of evaluating the probabilities of such scatterings and sampling their associated kinematics. 
We begin by computing the probability that a parton with energy $p_{\rm in}$ undergoes Moli\`ere scattering during a time interval $\Delta t$
by integrating the differential 
probability density $F^{C \rightarrow A}(p, \theta; p_{\rm in})$ over the allowed region of phase space.
Next, we describe how to
choose the momenta of the two partons after the scattering 
(if a Moli\`ere scattering process occurs)
in a way that satisfies energy and momentum conservation 
and the constraints~\eqref{eq:constraint1} and \eqref{eq:constraint2} by sampling the kinematic variables 
of the Moli\`ere scattering over the allowed phase space according
to the appropriate probability distribution.
%
This sampling procedure is required to generate physically consistent final-state momenta for the outgoing partons from a $2 \rightarrow 2$ Moli\`ere scattering.


We begin by computing the total probability with which a hard parton of type $C$ will undergo an elastic scattering with a quasiparticle in the medium, which is given by
\begin{equation} \label{eq:totprob}
    P_{\rm tot} = \int dp \int d\theta \, F^{C \rightarrow \rm all} (p, \theta) = \int dp \int d\theta \, \left( F^{C \to G}(p, \theta) + F^{C \to Q}(p, \theta) + F^{C \to \overline{Q}}(p, \theta) \right)\ .
\end{equation}
And,  per Eqs.~\eqref{eq:fqq}--\eqref{eq:fgqbar},
each $F^{C \rightarrow A} (p, \theta; p_{\rm in})$ can be written as a linear combination of terms $\langle (n) \rangle_{D,B}$ and $\langle (\tilde{n}) \rangle_{D,B}$ corresponding to the eleven different possible $2 \rightarrow 2$ scattering processes labeled by $n$.
According to Eq.~\eqref{eq:lincomb}, each $\langle (n) \rangle_{D,B}$ and $\langle (\tilde{n}) \rangle_{D,B}$ may be written as a linear combination of seven terms $\langle m_i({t}, {u}) \rangle$ defined in Eq.~\eqref{eq:mi} as

\begin{equation}
    \langle m_i({t}, {u}) \rangle  = \int_{k_T^{\rm min}}^{\infty} dk_T \int_0^{2 \pi} d\phi \, m_i^{D,B}({t}, {u}),
\end{equation}
where
\begin{equation} \label{eq:midb}
    m_i^{D,B}({t}, {u}) \equiv \frac{{\left[ 1 \pm n_A(p) \right]}}{16 (2\pi)^4} \left( \frac{p \sin \theta}{p_{\text{\rm in}} q T} \right) \, n_D(k_T) \left[ 1 \pm n_B(k_\chi) \right] \, m_i({t}, {u}).
\end{equation}
Here, $i$ runs from 1 to 7, and the coefficients 
that specify the 22 linear combinations of the $m_i$'s that make up the $\langle (n) \rangle_{D,B}$ and $\langle (\tilde{n}) \rangle_{D,B}$ are given in Appendix~\ref{app:processes}.
With all this in mind, we define
\begin{align}
    P_i & \equiv \int dp \int d\theta \int dk_T \int d\phi \, m_i^{D,B}({t}, {u}) {\cal P}_t {\cal P}_u \\
    & = \int_0^1dx\int_0^{p_{\rm in}}d\tilde{k}_{\chi}^{\rm min}\int_0^{\infty} d\tilde{k}\int_0^{2\pi} d\phi \, |\mathcal{J}| \, m_i^{D,B}({t}, {u})  {\cal P}_u\ ,
    \label{eq:four-integrals-to-be-done}
\end{align}
where in the second equality we have used the change of variables in Eq.~\eqref{eq:changeofvariables} and where $i$ runs from 1 to 7. 
Defining $P^{(n)} \equiv \sum_{i = 1}^7 c_i^{(n)} P_i$ and $\tilde{P}^{(\tilde{n})} \equiv \sum_{i = 1}^7 \tilde{c}_i^{(\tilde{n})} P_i$ (where the coefficients $c_i^{(n)}$ and $\tilde{c}_i^{(\tilde{n})}$ are the same coefficients as in Eq.~\eqref{eq:lin-comb-7}), the total probability $P_{\rm tot}$ in Eq. \eqref{eq:totprob} for the occurrence of an elastic scattering between an incoming parton $c$ of type $C$ with incoming momentum $p_{\rm in}$ and a quasiparticle in the medium is then given by a linear combination of $P^{(n)}$ and $\tilde{P}^{(\tilde{n})}$, analogous to Eqs.~\eqref{eq:fqq}+\eqref{eq:fqg}+\eqref{eq:fqqbar} for an incident quark or antiquark and analogous to Eqs.~\eqref{eq:fgq}+\eqref{eq:fgg}+\eqref{eq:fgqbar} for an incident gluon.
We defer a description of how we compute the integrals in Eq.~\eqref{eq:four-integrals-to-be-done} to Appendix~\ref{app:integrals}.

Note that each $F^{C \rightarrow A}$ term in the total probability \eqref{eq:totprob} is proportional to the chosen timestep $\Delta t$ (c.f. Eq.~\eqref{eq:probfinal}). So, Eq.~\eqref{eq:totprob} is accurate only when $\Delta t$ is small enough that the probability of more than one hard scattering occurring in the interval $\Delta t$ is negligible. We interpret $P_{\rm tot}$ as the first-order expansion of a Poisson process with total rate $P_{\rm tot}/\Delta t$. Then, the probability that at least one elastic scattering with high momentum transfer occurs in the time interval $\Delta t$ is then
\begin{equation}
    P_{\rm step} = 1 - e^{-P_{\rm tot}}.
    \label{eq:Pstep}
\end{equation}
Evaluating $P_{\rm step}$ is our first goal. In Appendix~\ref{app:integrals}, we describe 
how we evaluate the four integrals in Eq.~\eqref{eq:four-integrals-to-be-done}, doing two of them analytically and two of them numerically. 
Note that if $P_{\rm tot}$ is sufficiently small, then $P_{\rm step} \approx P_{\rm tot}$. 
Furthermore, note that the probability of $n$ Moli\`ere scatterings occurring within the time interval $\Delta t$ is $(P_{\rm tot})^n e^{-P_{\rm tot}}$, which is suppressed by powers of $P_{\rm tot} \ll 1$. 
Although in answering the question of whether an elastic scattering with high momentum transfer has occurred we shall evaluate $P_{\rm step}$,
we shall always make sure that $\Delta t$ is small enough that $P_{\rm tot} \ll 1$ and
$P_{\rm step} \approx P_{\rm tot}$ so that we can assume that only one elastic scattering with high momentum transfer 
occurs during the time interval $\Delta t$ with probability $P_{\rm step}$.


\begin{figure}[h]
    \centering
    \includegraphics[width=.49\textwidth]{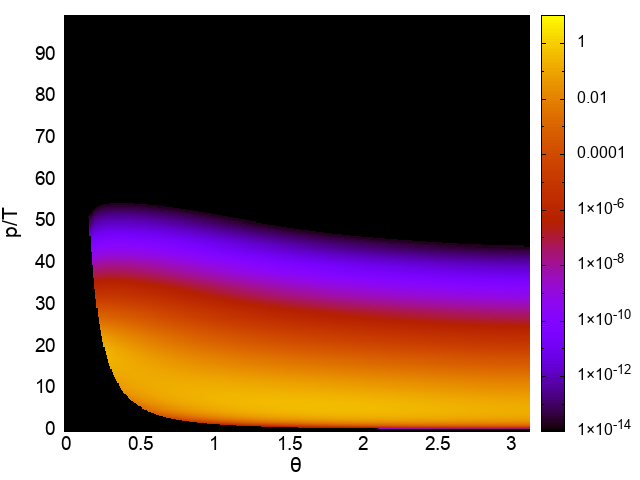}
    \includegraphics[width=.49\textwidth]{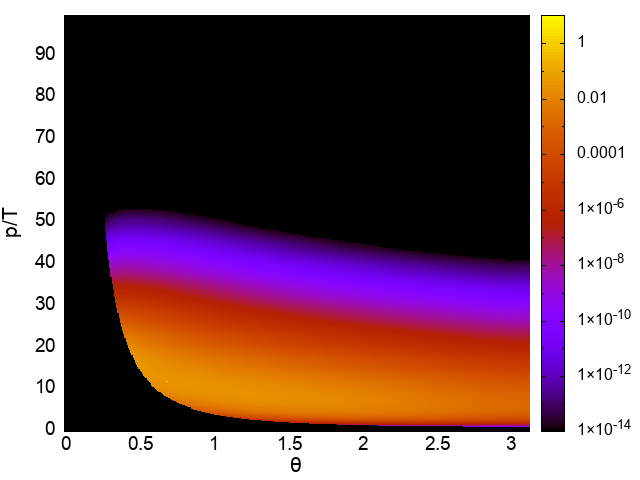}

     \includegraphics[width=.49\textwidth]{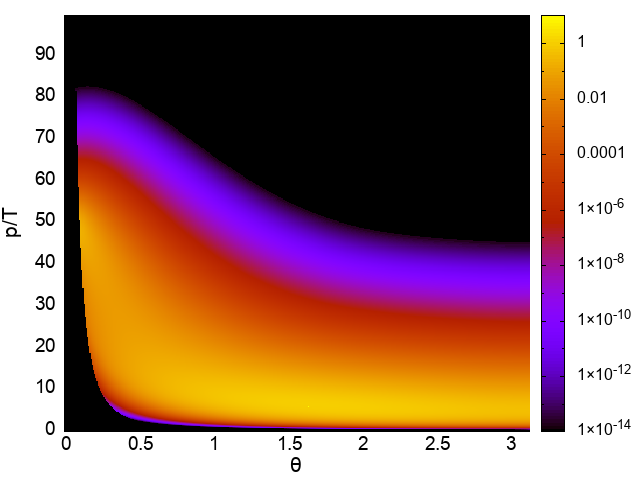}
    \includegraphics[width=.49\textwidth]{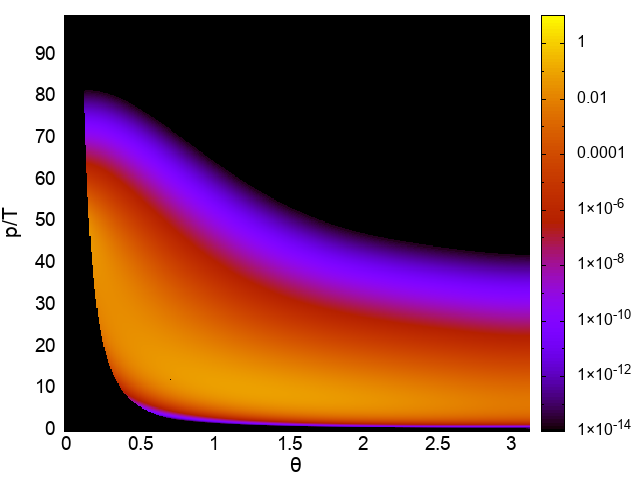}

     \includegraphics[width=.49\textwidth]{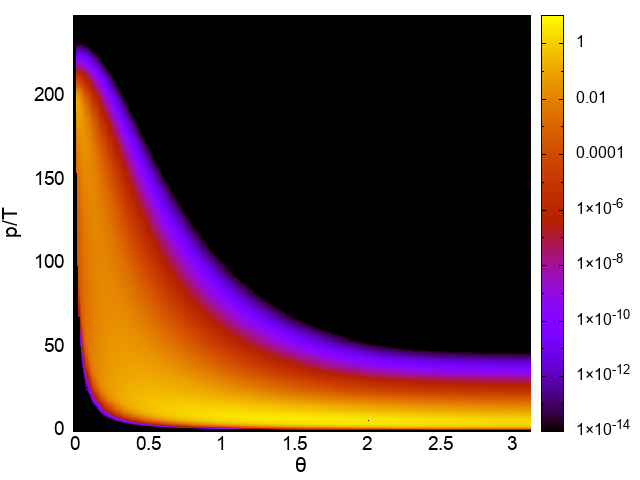}
    \includegraphics[width=.49\textwidth]{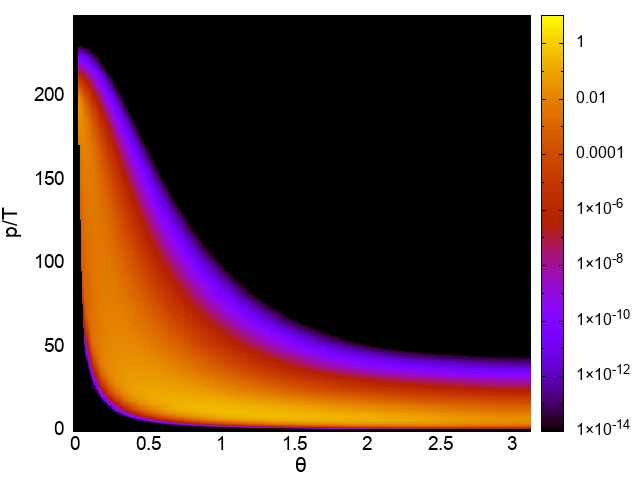}
    
    \caption{Probability distributions $F^{G \rightarrow \rm all}(p, \theta; p_{\rm in})$ for finding a scattered parton with energy and angle $(p,\theta)$ from
    the Moli\`ere scattering of an incoming gluon with $p_{\rm in} = 20T$ (top panels), $50T$ (middle panels), and $200T$ (bottom panels) that has traveled for a time $\Delta t=6/T$ through a ``brick'' of QGP with a constant temperature $T$. We have chosen the two parameters that specify the elastic scattering probabilities (see text for their definition and description) to be $g_s=2.25$ and $a=4$ in the left panels, and $g_s=2.25$  and $a=10$ in the right panels.
    }
    \label{fig:bounds}
\end{figure}

Before continuing onward to describing how we sample the kinematic variables of the Moli\`ere scattering process so as to choose the momenta of the two partons after a scattering from the appropriate probability distribution, it will be helpful to visualize that probability distribution.  Visualizing the full
four-dimensional distribution, either over the variables $(p,\theta,k_T,\phi)$ or over the variables $(x,\tilde k_\chi^{\rm min},\tilde k,\phi)$, would be a challenge and is not necessary.
It will suffice to visualize the probability distribution
over the variables $(p,\theta)$, which we do by plotting 
$F^{G\rightarrow {\rm all}}(p,\theta;p_{\rm in})$ in Fig.~\ref{fig:bounds}.  Recall that this is the probability density
for finding an outgoing particle of any type with an energy $p$ at an angle $\theta$ with respect to the direction of an incoming gluon with energy $p_{\rm in}$.
In plotting Fig.~\ref{fig:bounds}, we have evaluated $F^{G\rightarrow {\rm all}}(p,\theta;p_{\rm in})$ at a grid of values of $(\theta,p)$ where the constraint~\eqref{eq:constraint1} is satisfied.
At each point in the grid, we evaluate $k_T^{\rm min}=(q-p_{\rm in}+p)/2$ with $q$ defined in Eq.~\eqref{eq:qdefn}
and then perform the phase space integrations
over $k_T$ from $k_T^{\rm min}$ to infinity and
over $\phi$ 
that are needed to evaluate the seven functions
$\langle m_i(t,u)\rangle$ defined in 
Eq.~\eqref{eq:mi}
numerically, with the Heaviside $\Theta$ function that defines the constraint \eqref{eq:constraint2} included in the integrand.  
We then take the appropriate linear combinations of these functions for each possible $2\rightarrow 2$ scattering process, see Eq.~\eqref{eq:lincomb}, 
and then, along the lines of Eqs.~\eqref{eq:fqq}--\eqref{eq:fgqbar},
we sum over all the $2\rightarrow 2$ processes
that contribute to $F^{G\rightarrow {\rm all}}(p,\theta;p_{\rm in})$.
 

In Fig.~\ref{fig:bounds}, we show probability distribution $F^{G\to {\rm all}}(p,\theta; p_{\rm in})$ for the Moli\`ere scattering of a relativistic gluon with $p_{\rm in} = 20 T$ (top panels), $50T$ (middle panels), and $200T$ (bottom panels)  traveling for a time interval $\Delta t = 6/T$
through a ``brick'' of QGP with a constant temperature $T$ 
and a thickness of at least $6/T$.
In the Hybrid Model calculations that will be our focus throughout most of this paper, $\Delta t$ will be a small timestep, but for visualizing the probability distribution and so as to make 
comparisons to the results of Ref.~\cite{DEramo:2018eoy} possible, it is helpful here to consider a brick of QGP.
In all panels of Fig.~\ref{fig:bounds} we have chosen the value 
$g_s=2.25$ (corresponding to $\alpha_s=0.4$) for the QCD coupling constant that sets the magnitude of the elastic scattering matrix elements and the Debye mass $m_D$; we have displayed results obtained with two different choices of the parameter $a$ (recall that we include only those elastic scattering processes with momentum transfer $>a m_D^2$),
with $a=4$ in the left panels of the Fig.~\ref{fig:bounds} and $a=10$ in the right panels.  (Throughout most of this paper, we shall choose $a=10$.)
Fig.~\ref{fig:bounds} illustrates how the phase-space constraints derived in Section~\ref{sec:phasespace} restrict the allowed values of momentum $p$ and scattering angle $\theta$ for the outgoing hard partons after the scattering. 
The black region at small $p$ or/and $\theta$ is excluded by requiring  $|t|>a m_D^2$, which is a constraint on $p$ and $\theta$, see Eq.~\eqref{eq:constraint1}.  
Enforcing the constraint $|u|>a m_D^2$ is responsible for reducing the probability for obtaining scattered particles with small $p$ or/and $\theta$ in the region of $(p,\theta)$
just outside the black excluded region.  If we had only enforced the constraint on $t$, this region would have been yellow; in fact, it is red/purple. By comparing the right panels to the left panels, we see that increasing $a$ expands the 
regime of low-momentum transfer elastic scattering that we are excluding, which is as expected.  Increasing $a$ has little effect on the 
probability for the larger momentum transfer elastic scattering
processes that we shall include in the Hybrid Model.

By comparing the upper, middle and lower panels of Fig.~\ref{fig:bounds} at around, say, $\theta=0.5$ radians, we see that 
increasing the energy of the incident parton shifts the probability distribution for the scattered parton energy
to larger $p$ and squeezes the probability distribution for the scattering angle $\theta$ to smaller angles, for high energy scattered partons.  These scattered partons with larger values of $p$ are most likely the scattered incident partons, but note that these are not all that is plotted in Fig.~\ref{fig:bounds}.  We are plotting the probability distribution for finding 
a scattered parton with $(p,\theta)$, but this scattered parton may well be a parton from the QGP that was kicked by the incident gluon.  Indeed, the yellowest regions in the probability distributions plotted in Fig.~\ref{fig:bounds} that correspond to scattered partons with energies $\sim 10 T$ found at all angles including very large angles are most likely partons from the QGP.
This intuition, visualized in Fig.~\ref{fig:bounds} will be important in understanding results from our Hybrid Model calculations that we shall present later.


Having visualized the 
probability distribution for the energy and angle of 
partons resulting from Moli\`ere scattering processes, 
we now turn to the calculations that we need in order to be able to implement Moli\`ere scattering in the Hybrid Model.  Earlier in this Section, we have set up the computation of $P_{\rm step}$, defined in Eq.~\eqref{eq:Pstep}, namely the probability that a Moli\`ere scattering occurs during a small timestep $\Delta t$.
We now describe how we choose the identities, and momenta,
of the two partons that result from such an elastic $2\rightarrow 2$ scattering process.
Although in Fig.~\ref{fig:bounds} we only asked about the probability for finding a single scattered parton thus and so, in the Hybrid Model we will need to specify both scattered partons,
with their identities and momenta sampled from
the appropriate probability distribution.

During each timestep, we choose a random number between 0 and 1 that, first, determines whether or not the incoming parton of type $C$ undergoes a Moli\`ere scattering $cd \rightarrow ab$, noting that the probability that an 
elastic scattering has occurred
is $P_{\rm step}$.
If the Moli\`ere scattering has occurred, we 
calculate the three individual probabilities
for the cases where $A=Q$ or $A=\bar Q$ or $A=G$. Based upon
the value of another random number, we then decide whether $A=Q$ or $\bar Q$ or $G$
.
We can then determine which of the eleven processes in Table~\ref{tab:QCDprocesses} occurred by calculating the conditional probability with which each process may occur, given that the incoming parton is $c$, and the outgoing parton is of type $A$ as determined in the previous step
. 
The flavor of each (anti-)quark involved in the $2 \rightarrow 2$ scattering is then chosen uniformly from ($\bar{u}, \bar{d}, \bar{s}$) $u, d, s$.


Once we have decided (probabalistically, as above) which of the eleven $2 \rightarrow 2$ scattering processes occurred, we know the identities of the two scattered partons. To specify their momenta, we need to choose the values of the kinematic variables 
$(x,\tilde k_\chi^{\rm min},\tilde k,\phi)$ 
for the Moli\`ere scattering process from the appropriate
probability distribution. We do so as follows.
%
First, we define conditional probability distribution for the
variable $x$:
\begin{equation} \label{eq:x-basis}
    P_i(x | cd \rightarrow ab) \propto \int_0^{p_{\rm in}}d\tilde{k}_{\chi}^{\rm min}\int_0^{\infty} d\tilde{k}\int_0^{2\pi} d\phi \, m_i^{D,B}({t}, {u}) |\mathcal{J}| {\cal P}_u \ .
    \end{equation}
We need to choose $x$ by sampling this distribution.
We accomplish this via first computing the cumulative
distribution functions $\Phi_i(x)=\int_0^x dx' P_i(x')$
corresponding to each of the seven probability distribution functions $P_i(x)$ and adding them in the same way that we described in the paragraph following Eq.~\eqref{eq:four-integrals-to-be-done}
%
%
so as to obtain the
total cumulative distribution function $\Phi(x)$. Then, we choose a random number $u$ with a uniform distribution between 0 and 1. Finally, we solve for $x=\Phi^{-1}(u)$ using the bisection method. 
In this way, we have picked a value of $x$ by sampling 
the conditional probability distribution~\eqref{eq:x-basis}.
Having chosen $x$, we move on to choosing $\tilde{k}_{\chi}^{\rm min}$ according to the sum of the seven conditional probability distributions
\begin{equation} \label{eq:kchimin-basis}
    P_i(\tilde{k}_{\chi}^{\rm min}|x) \propto \int_0^{\infty} d\tilde{k}\int_0^{2\pi} d\phi \, m_i^{D,B}({t}, {u}) |\mathcal{J}| {\cal P}_u\ ,
\end{equation} 
which again we accomplish via first defining a cumulative distribution function, then picking a random number between 0 and 1, and choosing $\tilde{k}_{\chi}^{\rm min}$ to be the value at which the cumulative distribution function matches that random number.
Having found $\tilde{k}_{\chi}^{\rm min}$ and $x$, we then sample $\tilde{k}$ using the cumulative distribution functions of
\begin{equation} \label{eq:k-basis}
    P_i(\tilde{k} | \tilde{k}_{\chi}^{\rm min}, x) \propto \int_0^{2\pi} d\phi \, m_i^{D,B}({t}, {u}) |\mathcal{J}| {\cal P}_u.
\end{equation}
Finally, having found $\tilde{k}$, $\tilde{k}_{\chi}^{\rm min}$, and $x$, we sample $\phi$ using the cumulative distribution functions of
\begin{equation} \label{eq:phi-basis}
    P_i(\phi |\tilde{k}, \tilde{k}_{\chi}^{\rm min}, x) \propto m_i^{D,B}({t}, {u}) |\mathcal{J}| {\cal P}_u.
\end{equation}
All that remains is for us to describe how 
we compute the conditional probabilities on the right-hand sides of Eqs.~\eqref{eq:phi-basis}, \eqref{eq:k-basis}, \eqref{eq:kchimin-basis} and \eqref{eq:x-basis}.
Eqs.~\eqref{eq:phi-basis} and \eqref{eq:k-basis} are explicitly 
calculated in Appendices~\ref{app:phi-integrals} and ~\ref{app:k-integrals}, respectively. The conditional probability distributions in Eqs.~\eqref{eq:kchimin-basis} and ~\eqref{eq:x-basis} are evaluated using numerical integration, as described in Appendix~\ref{app:numerical-integrals}. These integrals are complicated by the constraint ${\cal P}_u$, which takes different forms in different regimes of $k$. These complications are made precise and handled in Appendix~\ref{app:integrals}.

The sampling procedure described above yields values for the set of kinematic variables $(x,\tilde k_\chi^{\rm min},\tilde k,\phi)$ that fully specify the kinematics of a Moli\`ere scattering process. Around Eqs.~\eqref{eq:omega-and-q} and \eqref{eq:five-equalities}, we have described how to obtain the variables $\omega$, $q$, 
$p$, $t$, $k_T$ and $k_\chi$
from $(x,\tilde k_\chi^{\rm min},\tilde k,\phi)$
and $\theta$ can then be obtained via 
$|t|=2 p_{\rm in} p (1-\cos\theta)$.
Consequently, as we make explicit in Appendix~\ref{app:kinematics}, the variables $(x,\tilde k_\chi^{\rm min},\tilde k,\phi)$
specify the energies and angles
of the two scattered partons that result from the elastic $2 \rightarrow 2$ scattering process --- which is what we will need in the Hybrid Model.  
The Hybrid Model calculation of a parton shower developing within a droplet of QGP specifies the 
energy and direction of an incident jet parton and the temperature
of the QGP it finds itself in during each time interval $\Delta t$; if a Moli\`ere scattering event occurs during a time interval, the kinematic variables 
$(p,\theta,k_T,\phi)$
that we obtain as we have described specify the 
energy and direction of the QGP parton that this jet parton
scattered from, as well as the energies and directions of the
two partons after the elastic scattering.  From that time onward, these two partons are treated as a part of the jet shower and evolved accordingly 
by the Hybrid Model.  They will lose energy and momentum to, and create wakes in, the droplet of QGP.  And, although this is unlikely, they may scatter elastically again.

\section{Incorporating Moli\`ere Scatterings in the Hybrid Model}
\label{sec:implementing-in-hybrid}

The hybrid strong/weak coupling model, or more simply the Hybrid Model, is a theoretical framework designed to describe the multi-scale processes of jet production and evolution through  a droplet of strongly coupled QGP whose expansion and cooling is described via hydrodynamics. The weakly coupled processes of jet production 
in an initial hard scattering 
and 
the subsequent fragmentation into, and evolution 
of, parton showers
are described perturbatively and are implemented in the model using 
PYTHIA 8~\cite{Sjostrand:2014zea} with the EPS09 nuclear parton distribution functions (nPDFs)~\cite{Eskola:2009uj}.
Before this study, the Hybrid Model 
incorporated interactions between jet partons and
the strongly coupled QGP in a way that would be appropriate if these interactions are strongly coupled.
This is motivated by the fact that the most common momentum transfers $q$ are dominated by the characteristic scales of the medium --- such as the temperature $T$ or the Debye mass $m_D$ --- 
at which the strong coupling constant is not small.
However, although they are rarer, processes with larger momentum transfer between a jet parton and the QGP can of course occur, and these should be described perturbatively. Elastic Moli\`ere scattering is the simplest such process --- it must occur although it will not be the only perturbative process via which large momentum transfer occurs.   
For large-enough momentum-transfers, one can calculate the probability of Moli\`ere scatterings perturbatively, as we have done in Sections~\ref{sec:moliere} and \ref{sec:phase-space-constraints-and-sampling}. By implementing the Moli\`ere scattering processes described in Table~\ref{tab:QCDprocesses}, we incorporate 
weakly coupled jet-medium processes with large momentum transfer in the Hybrid Model model for the first time. 
The purpose of our study is to identify  jet quenching observables that are sensitive to the presence of Moli\`ere scatterings off quasiparticles in the medium. By doing so, we can learn how best to identify and isolate signatures of weakly coupled large momentum transfer processes involving a jet parton and a quasiparticle from the QGP in experimental data.
We look forward to the day when this has been done --- as at that point if there are quantitative discrepancies between experimental measurements of suitably sensitive observables to the predictions we make this would motivate including other weakly coupled large momentum tranfer processes in addition to Moli\`ere scattering, which is the simplest example of such a process and which must be present.


In this Section, after 
reviewing aspects of the Hybrid Model in detail, 
we review how we incorporate 
softer Gaussian-distributed transverse momentum broadening, as appropriate in a strongly coupled plasma~\cite{Casalderrey-Solana:2016jvj}
and then describe in detail how we incorporate hard Moli\`ere scatterings in the Hybrid Model.

\subsection{Hybrid Model description of 
strongly coupled jet-medium interactions}

An energetic parton produced in a hard QCD process initiated by two partons from the incident nuclei in a heavy ion collision starts out with a virtuality $Q$ of the order of its large transverse momentum $p_T$.
Via a series of splittings that 
are well described via perturbative QCD, specifically via the DGLAP evolution equations, the hard parton branches into a parton shower. 
In vacuum, perturbative splitting proceeds until the partons in the shower have virtualities of
order $\Lambda_{\rm QCD}$ at which point they
hadronize.
The processes of jet production in a hard process, parton shower evolution as described by the DGLAP evolution equations, and hadronization are implemented in the Hybrid Model using PYTHIA 8~\cite{Sjostrand:2014zea} with nPDFs from EPS09~\cite{Eskola:2009uj}.

When a parton shower develops within the QGP formed in a heavy ion collision,
the early high-virtuality splittings in the shower 
are not significantly modified relative to 
how they would occur in vacuum because of the
large separation between the scale $Q$ and the temperature of the medium $T$.
As these vacuum-like splittings take place, however, interactions between the partons in the developing shower 
and the QGP are nevertheless occurring. These interactions are dominated by very common, soft, momentum exchanges that are strongly coupled --- that have been the focus of the Hybrid Model prior to this study as we review in this subsection --- as well as occasional larger-than-average momentum exchanges that can be described at weak coupling that will be our focus in this paper.

In the Hybrid Model, we model the common, strongly coupled, soft exchanges between an energetic parton and the 
expanding, cooling, droplet of QGP through which it propagates  
that result in the loss of energy and momentum in the direction of the parton's direction of propagation based upon what we know from holographic calculations that can be performed in the limit of strong coupling.
We assume that
an energetic quark traversing the droplet of QGP loses energy at a rate which takes the same form as that for a light quark or gluon 
traversing strongly coupled plasma in $\mathcal{N} = 4$ SYM theory in the limit of infinite coupling and large $N_c$. 
A massless parton that began with energy $E_{\rm in}$ and has traveled a distance $x$ through the strongly coupled plasma loses energy to the plasma at a rate~\cite{Chesler:2014jva,Chesler:2015nqz}
\begin{equation}
\label{eq:elossrate}
   \left. \frac{\rmd E}{\rmd x}\right|_{\rm strongly~coupled}= - \frac{4}{\pi} E_{\rm in} \frac{x^2}{x_{\rm stop}^2} \frac{1}{\sqrt{x_{\rm stop}^2-x^2}} \quad ,
\end{equation}
where 
$x_{\rm stop}\equiv  E_{\rm in}^{1/3}/(2{T}^{4/3}\kappa_{\rm sc})$ is the maximum distance the parton can travel before completely thermalizing.   
Here, $\kappa_{\rm sc}$ is a dimensionless parameter which governs the strength of the interaction between the massless parton and the
strongly coupled medium.
$\kappa_{\rm sc}$ is calculable in $\mathcal{N} = 4$ SYM theory and is proportional to $\lambda^{1/6}$, where $\lambda = g^2 N_c$ is the 't Hooft coupling, with a prefactor that depends on details of the theory.  Additionally, note that the stopping distance of a gluon is reduced (meaning that $\kappa_{\rm sc}$ is increased) by a factor $(C_A/C_F)^{1/3}$ compared to that of a quark \cite{Gubser:2008as}, where $C_A$ and $C_F$ are the Casimirs of the adjoint and fundamental representations of the color gauge group, respectively.
In the Hybrid Model, we aim to describe the soft strongly coupled interaction between jet partons and strongly coupled QGP in QCD, 
not in ${\cal N}=4$ SYM theory.
We assume that the rate of energy loss of a quark or anti-quark in the parton shower is governed by Eq.~\eqref{eq:elossrate} with 
$\kappa_{\rm sc}$ treated as a parameter to be determined by fitting 
to experimental data on the suppression of high-$p_T$ charged-hadrons and jets in PbPb collisions at the LHC, and assume that gluons
in the parton shower lose energy as
described by Eq.~\eqref{eq:elossrate} with
\begin{equation} \label{eq:kappagluon}
   \kappa_{\rm gluon} = (9/4)^{1/3} \kappa_{\rm sc}\ .
\end{equation}
Fitting Hybrid Model calculations to data~\cite{Casalderrey-Solana:2018wrw} shows that $\kappa_{\rm sc}$ is smaller (and the stopping distance of massless partons is larger) by a factor of 3 to 4 in QCD than in $\mathcal{N} = 4$ SYM theory.

As the rate of energy loss $dE/dx$ of a parton in the jet depends on the local temperature of the medium at the position and time where the jet parton is found, in any treatment of jet quenching in heavy ion collisions 
one must assign a spacetime picture to the development of the momentum-space DGLAP shower. In the Hybrid Model,
we use a formation time argument, where each parton in the shower propagates for a time $\tau_f\equiv2E/Q^2$ before splitting. During this time, the propagating parton loses energy as specified by Eq.~\eqref{eq:elossrate} as long as the local temperature is above a temperature
which in this work we set to $T_c=145$~MeV.

The energy and momentum that the jet partons lose, as described by
Eq.~\eqref{eq:elossrate},
is deposited into the droplet of QGP.
The resulting perturbations of the stress-energy tensor of the liquid QGP (which must have a net energy and momentum equal to that lost by the jet) are referred to as jet wakes.  In a hydrodynamic fluid, jet wakes are composed of sound modes (compression and rarefaction) and diffusive modes that describe a region of moving fluid behind the jet where the fluid has picked up net momentum in the direction of the jet.
This wake evolves hydrodynamically as the medium expands, flows, and cools until it reaches the freezeout hypersurface and a complete description of this evolution, including the interplay between the jet-induced wake and the background flow of the expanding droplet, requires 3+1-dimensional relativistic hydrodynamic simulations.

In the present paper, we adopt a much simpler description of the observable consequences of jet wakes, based upon a set of simplifying assumptions first employed in Ref.~\cite{Casalderrey-Solana:2016jvj}. 
First, we neglect transverse flow, and assume that the longitudinal expansion of the background fluid is boost invariant, i.e. is a Bjorken flow. Second, we assume that the perturbation stays close in rapidity to the rapidity at which it was deposited. Third, we assume that the jet-induced perturbations to the hydrodynamic stress-energy tensor are small. And, fourth, we assume that 
after freezeout according to the standard
Cooper-Frye prescription at the freezeout hypersurface~\cite{PhysRevD.10.186}, the perturbations to the spectra of the resulting hadrons is small at all transverse momenta $p_T$. 
This fourth assumption, which is only valid 
in the very soft particle limit, 
allows us to integrate the flow and entropy profiles over the whole freezeout hypersurface and express the perturbations to the hadronic spectra after freezeout in terms of the deposited momentum and energy. This (over)simplification 
allows us to ignore the details of the evolution of the perturbed fields and specify the 
distribution of soft particles originating from the wake of a jet entirely
from energy-momentum conservation. The one-body distribution then reads~\cite{Casalderrey-Solana:2016jvj}
\begin{equation}
\label{eq:onebody}
\begin{split}
E\frac{d\Delta N}{d^3p}=&\frac{1}{32 \pi} \, \frac{m_T}{T^5} \, \cosh(y-y_j)  \exp\left[-\frac{m_T}{T}\cosh(y-y_j)\right] \\
 &\times \Bigg\{ p_{\perp} \Delta P_{\perp} \cos (\phi-\phi_j) +\frac{1}{3}m_T \, \Delta M_T \, \cosh(y-y_j) \Bigg\} \, ,
\end{split}
\end{equation}
where $m_T$, $p_\perp$, $\phi$ and $y$ are the transverse mass, transverse momentum, azimuthal angle and rapidity of the hadrons originating from a jet wake and $\Delta M_T$ and $\Delta P_\perp$ are the transverse mass and transverse momentum 
that a jet parton with azimuthal angle $\phi_j$
and rapidity $y_j$ lost per Eq.~\eqref{eq:elossrate} 
and deposited in the hydrodynamic fluid.
The distribution~\eqref{eq:onebody} is oversimplified but it has the advantage of being fully specified analytically, with no new free parameters. 
And, it captures  important key features, including an enhancement of soft particle production around the jet direction and a depletion of soft particles in the direction opposite to the jet direction in 
azimuthal angle $\phi$. Both are direct consequences of the diffusive modes in the wake: the jet ``boosts'' a region of fluid in the direction of the jet, resulting in more soft particles in that direction and fewer soft particles in the opposite direction after freezeout than would have been the case if that region of fluid had not been moving.
The depletion of soft particle production in the direction opposite to the jet has recently been observed in $Z$-jet events by CMS~\cite{CMS:2025dua}, with a magnitude that is well described by Eq.~\eqref{eq:onebody}.
Eq.~\eqref{eq:onebody} does not,
however, account for the interplay between the hydrodynamic evolution of the wake and the transverse flow of the background fluid, among other effects, meaning that the particles generated with Eq.~\eqref{eq:onebody} are too soft and too wide in azimuthal angle~\cite{Casalderrey-Solana:2016jvj}. 
An efficient description of a more complete treatment of the response of the flowing hydrodynamic fluid to the passage of a jet, 
based on the linearized hydrodynamics solutions of Ref.~\cite{Casalderrey-Solana:2020rsj}, is work in progress by these authors.

While Eq.~\eqref{eq:elossrate} describes energy loss, it has long been known that an energetic parton travelling through strongly coupled plasma will in addition accumulate momentum $k_T$ transverse to its direction of propagation via continuous soft momentum exchange, referred to as transverse momentum broadening. In a weakly coupled plasma, transverse momentum broadening occurs via repeated soft scattering. In a strongly coupled liquid, it occurs continuously in time with a Gaussian distribution for $k_T$ whose width
$\langle k_{\perp}^2 \rangle$ grows linearly with the time $\delta t$ for which the parton propagates: $\langle k_{\perp}^2 \rangle = \hat q\, \delta t$, where $\hat q$ is a property of the strongly coupled medium referred to as the 
jet quenching parameter~\cite{Liu:2006ug,Liu:2006he,DEramo:2010wup,DEramo:2012uzl}.
In the strongly coupled plasma of $\mathcal{N}=4$ SYM theory in the limit of large coupling and large $N_c$, the jet quenching parameter is given by~\cite{Liu:2006ug,Liu:2006he,DEramo:2010wup,DEramo:2012uzl}
\begin{equation}
\label{eq:adsqhat}
    \widehat{q}_{\rm AdS/CFT} \approx 7.53 \sqrt{\lambda} T^3 \equiv K T^3 \, ,
\end{equation}
where $\lambda$ is the 't Hooft coupling. 
Note that $\hat q$ in Eq.~\eqref{eq:adsqhat} 
is not proportional to $N_c^2$, which would be the case in a weakly coupled plasma where $\hat q$ would be proportional to the number density of
quasiparticles off which a jet parton could scatter.  
This highlights the fact that soft Gaussian transverse momentum broadening occurs in a strongly coupled plasma
that, as in ${\cal N}=4$ SYM theory, is strongly coupled at all scales and has no quasiparticles.


Transverse momentum 
broadening with $\hat q = K T^3$ was first implemented in the Hybrid Model in Ref.~\cite{Casalderrey-Solana:2016jvj}.
As there, since we are interested in transverse momentum broadening in strongly coupled QGP in QCD not in ${\cal N}=4$ SYM theory, we shall take $K$ as a parameter that should ultimately be fitted to experimental data. The analysis of Ref.~\cite{Casalderrey-Solana:2016jvj} found that the experimental observables available at that time showed little sensitivity to transverse momentum broadening even for $K$ as large as $\sim 40$, whereas Eq.~\eqref{eq:adsqhat} suggests 
$K\sim 29$ in ${\cal N}=4$ SYM theory with $\lambda=4\pi$, meaning $\alpha=0.4$, and
in the strongly coupled QGP of QCD, $K$ will be smaller. If we knew that $K$ in QCD was smaller than $K$ in ${\cal N}=4$ SYM theory by the same factor as for $\kappa_{\rm sc}$, we could use what has been learned by fitting $\kappa_{\rm sc}$ to data to estimate $K\sim 7-10$, but there is no available holographic computation that relates the energy loss and transverse momentum broadening of massless partons, meaning that ultimately the
value of $K$ used in the Hybrid Model should be determined via fitting to experimental data. We shall describe how we choose $K$ in this paper in the next subsection.

\subsection{Adding perturbative hard processes within strongly coupled plasma} \label{sec:choices-of-a-K}

The $2\rightarrow 2$ elastic 
scattering processes with large momentum transfer, namely the Moli\`ere scattering processes that are our object of study in this paper, impart  transverse momentum to jet partons that is {\it not} Gaussian distributed, with a harder distribution at high momentum transfer.  (This goes back to Rutherford scattering, with momentum transfer distributed $\sim k_T^{-4}$. In the present context, the power-law tail of the distribution can even be somewhat harder than this~\cite{Caucal:2021lgf,Caucal:2022fhc}.)
Although the soft low momentum exchange
processes that lead to Gaussian transverse momentum broadening as specified by Eq.~\eqref{eq:adsqhat} are more common than
Moli\`ere scattering (and in fact happen continuously at strong coupling), rare weakly coupled high momentum transfer processes
where a jet parton resolves and scatters off a quasiparticle in the QGP
are important because, uniquely, they give us access to the
microscopic, short lengthscale, 
properties of QGP. 
%

The new element that we add to the Hybrid Model
in this paper is that at every time-step
during the in-medium evolution, for each jet parton in the parton shower, 
we allow for the possibility that 
this parton scatters elastically off
a weakly coupled quasiparticle 
from the QGP with a momentum transfer
that is greater than a minimum threshold
that we specify. We describe such Moli\`ere scattering processes to leading order 
in weak coupling as in Section~\ref{sec:moliere}
and, as described in Section~\ref{sec:phase-space-constraints-and-sampling}, only include processes in which the 
Mandelstam variables $t$ and $u$ both
exceed a threshold expressed as $|t|, \, |u|>a m_D^2$. Here, $m_D^2=g_s^2T^2\bigl(N_c+N_f/2\bigr)/3$ and we shall take $g_s=2.25$, corresponding to $\alpha_s\simeq 0.4$, throughout and shall
of course take
$N_c = N_f = 3$.
We are free to choose the threshold parameter $a$. In this subsection we shall consider $a=4$ and $a=10$. Elsewhere in this paper, we shall take $a=10$ which is more conservative: choosing a larger threshold $a$ reduces the effects of
Moli\`ere scattering in experimental observables,
improves the validity of our assumption that
these elastic scattering processes are weakly coupled and can be treated perturbatively, 
and reduces the risk that we are double counting by extending 
the Moli\`ere scattering calculation down into
the low momentum transfer regime that we
have already included via strongly coupled transverse momentum broadening as specified by the parameter $K$.



\begin{figure}
    \centering
    \includegraphics[width=0.49\columnwidth]{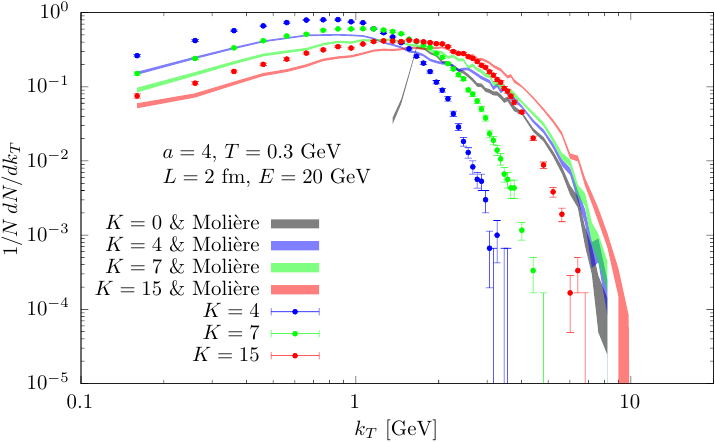}
    \includegraphics[width=0.49\columnwidth]{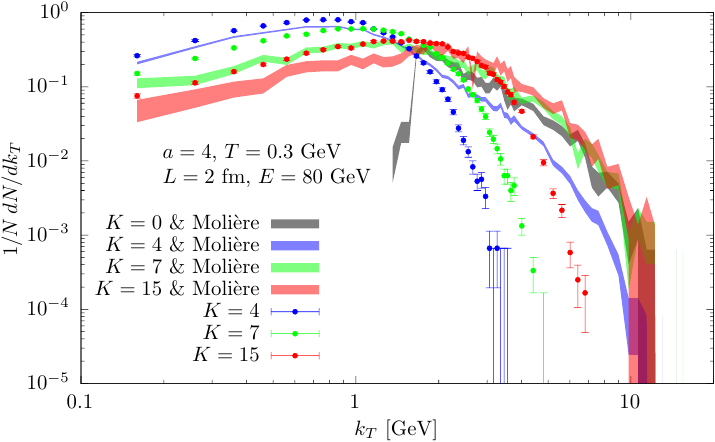}
    \includegraphics[width=0.49\columnwidth]{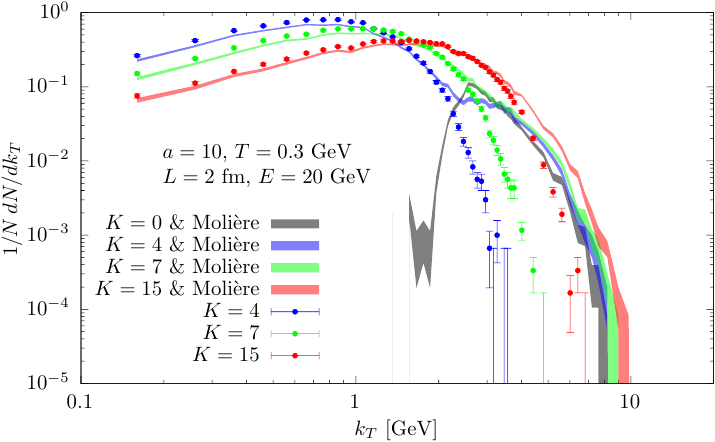}
    \includegraphics[width=0.49\columnwidth]{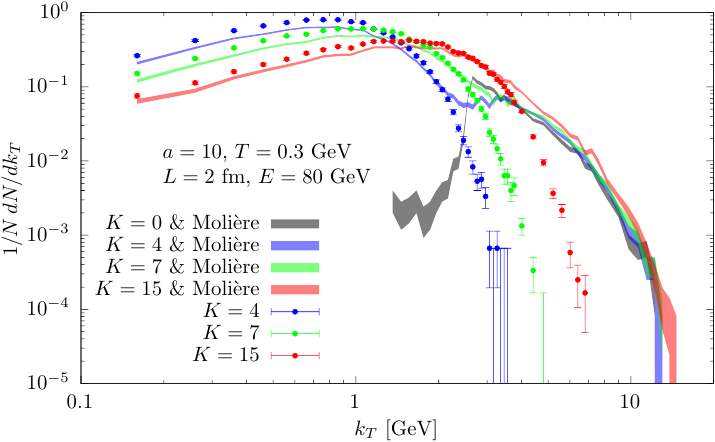}
    \caption{Probability for an incoming quark with energy $E=20$~GeV (left panels) or $E=80$~GeV (right panels) to obtain a transverse momentum $k_T$ after traversing a brick of QGP with length $L=2$~fm and temperature $T=0.3$~GeV. Different colors correspond to different choices of the parameter $K$ that governs the magnitude of the soft Gaussian transverse momentum broadening. The bands include Moli\`ere scattering with squared momentum transfer greater than $a m_D^2$, where we have chosen the threshold parameter 
    as $a=4$ (upper panels) or $a=10$ (lower panels). 
    The points with a given color show the results with that value of $K$ in the absence of Moli\`ere scattering. All probability distributions are normalized. (In the case of the grey bands, where $K=0$, the most probable outcome is $k_T=0$, no Moli\`ere scattering; this is not depicted.)
    }
    \label{fig:ktdists}
\end{figure}

The considerations above suggest that, given a value of $a$, we should adjust the value of $K$
to ensure a smooth matching between the probability distribution for $k_T$ coming from rare hard Moli\`ere scatterings and that coming from the strongly coupled soft Gaussian transverse momentum broadening specified by Eq.~\eqref{eq:adsqhat}.
This happens naturally in a weakly coupled treatment in which the soft Gaussian-distributed transverse broadening arises from multiple soft scattering treated in the same calculation as the rare hard Moli\`ere scattering, for example as in Refs.~\cite{Barata:2020rdn,Barata:2021wuf}.  As we are treating the soft Gaussian transverse momentum broadening as strongly coupled, we will ultimately need to fit the value of $K$ to data and for the present need to adjust its value by hand.
We illustrate how we do this in Fig.~\ref{fig:ktdists}, 
which shows (for different choices of $a$ and $K$) the probability with which an energetic light quark picks up momentum
$k_T$ transverse to its direction of propagation after traversing a distance $L=2$~fm through a static brick of QGP with 
temperature $T=0.3$~GeV.
(We have chosen $L=2$~fm because in the Hybrid Model the mean distance, in the fluid rest frame, that a parton in the shower propagates 
from the splitting at which it originates 
to the point where it either splits in two or leaves the medium is $\sim 2$~fm in a central PbPb collision.)
The calculation of the Moli\`ere scattering contribution to this probability distribution proceeds as described around Fig.~\ref{fig:bounds}. For $K\neq 0$, 
the energetic parton picks up an additional
Gaussian-distributed transverse momentum with $\langle k_T^2\rangle = KT^3 L$.
%
%
The energetic particle is a quark with energy $E=20$ GeV in the left panels, while in the right panels we choose $E=80$ GeV. The value of the threshold parameter is $a=4$ in the upper panels and $a=10$ in the lower panels.
All panels show results with only Moli\`ere scattering ($K=0$; black bands) and with 
soft Gaussian transverse momentum broadening with $K=4$, 7 and 15, with and without Moli\`ere scattering.


The first thing that one clearly appreciates from Fig.~\ref{fig:ktdists} is that the large $k_T$ region is indeed dominated by the Moli\`ere scattering, as expected. 
Comparing the left and right panels also makes it evident that while the contribution coming 
from strongly coupled Gaussian transverse momentum broadening does not depend on the energy $E$ of the jet parton, the contribution from Moli\`ere scattering does.
This is also as expected, since there is more phase space where elastic scattering off a thermally distributed medium quasiparticle 
imparts a momentum $k_T$ to the jet parton
when the incident jet parton has a higher $E$.
%
One can appreciate this by noting 
that when $E=20$~GeV the $k_T$ distribution drops off steeply as $k_T$ approaches 10 GeV, roughly half $E$,
whereas when $E=80$~GeV the harder power-law
distribution in $k_T$ is apparent for $k_T\sim 10$~GeV and beyond.


We can now describe how we use the results plotted in Fig.~\ref{fig:ktdists} to motivate our choice of $K$, for a given choice of $a$.  Let us begin with $a=10$. (With $g=2.25$ and $T=0.3$~GeV, this corresponds to requiring $|t|$, $|u|>(2.6~{\rm GeV})^2$. 
We see in both bottom panels that if we were then to choose $K=4$, we obtain a rather unphysical looking probability distribution with
a dip in it.  From the bottom left panel, we see that with $a=10$ if we choose either $K=7$ or $K=15$, we obtain a smooth probability distribution for $k_T$ for jet partons with $E=20$~GeV.  From the bottom right panel, though, we see that this is better achieved with $K=15$ than with $K=7$ for jet partons with $E=80$~GeV.
Since we wish to choose $a=10$ in order to be conservative in the sense described above, 
we shall take $K=15$ throughout the rest of this paper.  Any value in the range $7<K<15$ is not
unreasonable, however, and in future both $K$ and $a$ should be constrained via fitting Hybrid Model calculations of sufficiently many sufficiently differential observables to experimental data.
Note that we see from the upper panels of
Fig.~\ref{fig:ktdists} that if we were to choose the lower, less conservative, value $a=4$ for the threshold parameter, including elastic scattering processes with lower momentum transfers, we could accommodate a lower value of $K$.

We close this subsection by stressing that Moli\`ere scattering is only the simplest example of a weakly coupled process where a jet parton scatters off a quasiparticle from the QGP 
at high momentum transfer; such processes can also be inelastic. Strongly coupled soft inelastic processes are of course already incorporated in the Hybrid Model, but we leave a treatment of weakly coupled inelastic processes with large momentum transfer to future work.
At a qualitative level, though, it seems reasonable to speculate that including such processes while somewhat increasing the threshold $a$ that defines which elastic scatterings we include 
could yield comparable results, both for the probability distribution for $k_T$ in Fig.~\ref{fig:ktdists} and for the experimental observables that we shall investigate using the Hybrid Model.

\subsection{Monte Carlo implementation within the Hybrid Model}

In the previous subsection, we have 
used a toy calculation involving a brick of QGP to
fix our choice of the two parameters that appear in our
weakly coupled description of Moli\`ere scattering at large momentum transfer, namely $g_s=2.25$ and $a=10$, and the one parameter that appears in our
description of strongly coupled soft transverse momentum broadening, namely $K=15$.  We turn now to implementing Moli\`ere scattering in the Hybrid Model, so that in the next Section we can calculate Hybrid Model predictions for experimental observables with and without including Moli\`ere scattering.

At each time step in the Hybrid Model
description of an event, we have 
jet partons propagating through a droplet of
QGP whose expansion and cooling is described 
via a hydrodynamic profile.
At each time step in the laboratory frame $\Delta t_{\rm LAB}$, a given parton in the shower will be located in a fluid cell with some temperature $T$ and fluid velocity $v^i$, in the lab frame.
The temperature of the QGP is defined in the local fluid rest frame, however. 
So, before computing the energy loss according to Eq.~\eqref{eq:elossrate},
Gaussian transverse momentum broadening
with $\langle k_T^2\rangle\propto K T^3$, and (with some probability) Moli\`ere scattering
we must first boost from the lab frame to the local fluid rest frame.
%
%
After doing these computations, we then boost back to the lab frame and, in the lab frame, determine where the jet parton (or two jet partons if a Moli\`ere scattering has occurred) are located at the next time step and what the fluid velocity and fluid temperature are at that new spacetime 
point.

We note here as an aside that because we boost to the local fluid rest frame, compute Moli\`ere scattering there, and boost back to the lab frame (which is the same as the collision center of mass frame), and because we make no eikonal approximation, our calculation has encoded within it the way that the flowing medium impacts the shape of jets --- at least  via high-momentum transfer elastic scattering. The manner in which the hydrodynamic flow of a droplet of QGP can give jets 
in a fluid flowing transverse to the jet direction
a shape that is asymmetric around the jet axis  has been of long-standing interest~\cite{Armesto:2004pt} and has been the subject of much recent work~\cite{He:2020iow,Barata:2022utc,Fu:2022idl,Sadofyev:2021ohn,Barata:2022krd,Barata:2023zqg,Kuzmin:2023hko,Kuzmin:2024smy,Bahder:2024jpa}.
Although the observables that we shall investigate in this study do not do so, in future work one could use the same simulations that we have performed to investigate the contribution of Moli\`ere scattering to observables that may be sensitive to jet distortion.
The Hybrid Model calculations that we perform in this study do not include distortions of jet wakes arising from fluid flow transverse to the jet, but by building upon 
the work of Ref.~\cite{Casalderrey-Solana:2020rsj} 
this could be added too.

Returning to describing our implementation, we note following Ref.~\cite{Casalderrey-Solana:2015vaa} that
in order to apply the strongly coupled energy loss rate from Eq.~\eqref{eq:elossrate}, one observes that $dE_{L}(x_L)/dx_L=dE_{F}(x_F)/dx_F$,
where the subscripts refer to the lab frame and the local fluid rest frame.
This means that the amount of energy lost in the lab frame can be computed using Eq.~\eqref{eq:elossrate} in the local fluid rest frame, where all we need from the local fluid rest frame is the 
value of $T$.
%
The computation of the Gaussian transverse momentum broadening must be performed in the local fluid rest frame, where we impose that the kick received by the parton is transverse to the parton velocity in that frame, that it does not modify its virtuality, and that its energy remains the same in that frame~\cite{Casalderrey-Solana:2016jvj}. Both these processes --- longitudinal energy loss and Gaussian transverse momentum broadening --- modify the kinematics of a given parton at each time step in the in-medium evolution.

The new element in this paper is
that at each time step, for each parton in the shower, after boosting to the local fluid rest frame we apply the results that we have obtained in Section~\ref{sec:phase-space-constraints-and-sampling}, using expressions given explicitly in the Appendices, to compute the probability that this parton, whose ``incident'' momentum in the local fluid rest frame we refer to as $p_{\rm in}$,
scatters off a quasiparticle from the QGP with momentum $k$, assumed to be in thermal equilibrium in the local fluid rest frame.
(Note that in order to apply the formulae that we have presented in the Appendices, when we boost to the local fluid rest frame we must also rotate such that $p_{\rm in}$ lies along the $z$-axis, and when we later wish to boost back to the lab frame we must first undo this rotation. Note also that in order to compute the probability of Moli\`ere scattering during a specified time step,
in the expressions from Section~\ref{sec:phase-space-constraints-and-sampling} and the Appendices we must employ
the time step $\Delta t_F$ in the local fluid rest frame.)
%
%
If an elastic $2\rightarrow 2$ scattering occurs,
the outcome of this collision will be two partons,
whose momenta in the local fluid rest frame are
$p$ and $k_\chi$, specified as described in Appendix~\ref{app:kinematics}. (One of these outgoing partons, most often the one with higher energy, is the scattered incident parton; the other is the scattered parton that originated from the medium.)
%
%
It is important that in our Hybrid Model
implementation of this physics we treat both outgoing partons fully dynamically.  At all subsequent time steps, both are treated as jet partons meaning that both lose energy and momentum,
both contribute to the creation of the jet wake,
both experience strongly coupled soft Gaussian momentum broadening, and in principle either or both may experience a subsequent Moli\`ere scattering at a later time step.
%
%
%
We also need to take into account that we have 
removed four-momentum $k$ from the droplet of QGP 
by removing a thermal quasiparticle
with this four-momentum and, after it scatters, counting it as a parton in the jet.
This means that after freezeout the background of soft particles that is uncorrelated with the jet has $k$ less four-momentum than it would have had absent the Moli\`ere scattering.
Since we do not simulate the uncorrelated background, in order to preserve energy-momentum conservation
%
we need to include a ``hole'' or ``negative particle'' with four-momentum $k$ (boosted back to the lab frame) when we compute experimental observables.

There are two remaining subtleties that need to be addressed in order to complete the Monte Carlo implementation of Moli\`ere scattering within the Hybrid Model.  First, suppose that the parton in the shower that initiates a Moli\`ere scattering would have split later in the DGLAP evolution of the parton shower.  After the Moli\`ere scattering, this parton has become two partons, both of which we are treating as elements of the parton shower; do these partons split later in the shower evolution?  
We have chosen the following prescription: (i) we make the scattered parton with the higher momentum split as and when would have happened absent the 
Moli\`ere scattering, with this component of the developing parton shower now pointing in the new direction set by the higher-energy parton emerging from the Moli\`ere scattering; and, (ii) we assume that the scattered parton with lower momentum (which is most likely the parton from the medium) does not split.
Improving upon these prescriptions is a goal for future work, since in reality the Moli\`ere scattering will modify the subsequent DGLAP evolution of both scattered partons.
Other possible prescriptions have been
investigated in the literature~\cite{Armesto:2009fj,Zapp:2008gi,JETSCAPE:2017eso}.

The second subtlety that remains to be discussed relates to the hadronization of the parton shower in the Hybrid Model.  In the Hybrid Model, some partons in the shower lose all (and all partons in the shower lose some) of their energy and momentum to the fluid.  The energy and momentum from the parton shower that ends up in the jet wake is hadronized according to Eq.~\eqref{eq:onebody}.  
Note that in many cases the softer of the two scattered partons coming out of a Moli\`ere scattering process subsequently loses all of its energy and momentum to the fluid, meaning that an important consequence of Moli\`ere scattering is the modification of the wake of the jet.
What about the partons that remain in the shower after the Hybrid Model evolution, which is to say after the temperature of the medium in which they are propagating has dropped below 145~MeV?
In the Hybrid Model before the present work, 
jet partons that do not lose all of their energy are hadronized
using the Lund string model present in PYTHIA 8~\cite{Sjostrand:2014zea}, upon assuming for simplicity that these shower partons have retained the same color flow that they would have had in vacuum where a colorless configuration is needed between string endpoints (quarks) in order to produce the hadrons. 
With the addition of scattered partons coming 
from Moli\`ere scattering, 
we can no longer use this simplified assumption. In order to hadronize the surviving shower partons, in this work we use the ``colorless hadronization prescription'', introduced by the JETSCAPE collaboration~\cite{JETSCAPE:2019udz}. In essence, it consists of building up colorless strings based on a minimization procedure such that consecutive links are built between particles whose distance is minimal in $(\eta,\phi)$-space. This procedure does not require knowledge of the original color labels of the shower partons, making it
more consistent with the fact that interactions with the medium will randomly rotate the color of  partons in the shower, rearranging their initial color flow configuration. For this reason, changing from hadronizing the surving shower partons using the Lund string model from PYTHIA 8 to using the colorless hadronization prescription would be a small improvement to the Hybrid Model even without Moli\`ere scattering. Finally, we note that any ``hole partons'' (representing the removal of partons from the QGP by Moli\`ere scattering) 
are hadronized separately, so as to be able
to identify the contribution to the hadronic final state that needs to be subtracted relative to
the background  (hadrons that are uncorrelated with
the jet). This also follows a prescription 
first introduced by the JETSCAPE collaboration~\cite{JETSCAPE:2024nkj}.


Having described in full how we calculate
the probability for and kinematics of weakly coupled large momentum transfer $2\rightarrow 2$ elastic scattering (in previous Sections) and the
way in which we implement these Moli\`ere scattering processes, which are sensitive to the microscopic particulate structure of QGP, in the Hybrid Model, we turn now to the study of their
observable consequences.

\section{Results}
\label{sec:results}

In this Section, we examine a suite of jet observables that have been studied extensively in the past, some groomed and some ungroomed, with the goal of identifying characteristic signatures of $2\rightarrow 2$ Moli\`ere scatterings between jet partons and thermal QGP quasiparticles. Our purpose here is not to perform detailed comparisons with experimental data, but rather to compare Hybrid Model calculations of these observables with and without the inclusion of Moli\`ere scattering.
In this way we can
identify jet observables that are particularly sensitive to
Moli\`ere scattering in heavy-ion collisions, and more generally that are particularly sensitive to
weakly coupled scattering of jet partons off quark- and gluon-quasiparticles in QGP 
resolved at high momentum transfer. 

In all twenty panels in the Figures in this Section, we show four curves corresponding to different implementations of the Hybrid Model in which Moli\`ere scatterings and jet wakes are switched on or off: ``No Moli\`ere Scatterings \& No Wake'', in black; ``No Moli\`ere Scatterings \& With Wake'', in blue; ``With Moli\`ere Scatterings \& No Wake'', in green; and ``With Moli\`ere Scatterings \& With Wake'', in red. This controlled Monte Carlo setup allows us to identify which observables and which regions of phase space are most sensitive to the physics of interest.

The first aspect we need to address is the choice of the parameters of the model. In the present setup, there are four parameters: (i) $\kappa_{\rm sc}$, which enters via Eq.~\eqref{eq:elossrate} controls the magnitude of the rate of strongly coupled energy loss experienced by partons in jet showers; (ii)
$K$, which is defined in Eq.~\eqref{eq:adsqhat},
controls the magnitude of the strongly coupled
Gaussian-distributed soft transverse momentum
broadening; (iii) the strong coupling constant $g_s$ that controls the magnitude of the perturbative QCD matrix elements for Moli\`ere scattering; and (iv) the threshold parameter $a$
which sets the minimum squared momentum transfer
$a m_D^2$ for the $2\rightarrow 2$ elastic scatterings that we treat as weakly coupled, add to the Hybrid Model, and refer to as Moli\`ere scatterings.\footnote{The QCD coupling arising within the Debye mass 
(recall that $m_D^2=\frac{3}{2}g^2 T^2$ in QCD with $N_c=N_f=3$) 
could be taken to have a different value 
than that the $g_s$ arising in the Moli\`ere scattering matrix elements, since Moli\`ere scattering involves momentum transfers that are larger than thermal.  Doing this would not represent adding an additional parameter, though, as it is equivalent to changing the value of $a$.}
Throughout this study, we choose $g_s=2.25$ (corresponding to $\alpha_s\approx 0.4$), 
$a=10$ and $K=15$,
as described in Section~\ref{sec:choices-of-a-K}, as these are reasonable values for our purposes here, namely for identifying jet observables that are sensitive to Moli\`ere scattering.
%
We are thus left with fixing $\kappa_{\rm sc}$. 
In previous work~\cite{Casalderrey-Solana:2018wrw},
we fixed $\kappa_{\rm sc}$ in the Hybrid Model without Moli\`ere scattering and with $K=0$ by
fitting to the experimental data on the suppression of jets and single charged hadrons
in PbPb collisions relative to pp collisions
available at that time, 
finding $\kappa_{\rm sc}^{\rm no~Moli\grave{e}re}=0.404$.
%
In future work, the values of $\kappa_{\rm sc}$,
$K$, $g_s$ and $a$ should be constrained
together via a Bayesian uncertainty quantification
that takes advantage of the many and varied 
experimental data sets available today.
For the present study, upon choosing $K=15$ and including Moli\`ere scattering with $g_s=2.25$ and $a=10$, 
we shall set $\kappa_{\rm sc}=0.37$. 
We do so because, as we shall see
in Section~\ref{sec:jet-and-hadron-suppression},
with this choice
the suppression $R_{\rm AA}$ of jets and single
hadrons is approximately the same
as in the absence of Moli\`ere scattering with $\kappa_{\rm sc}^{\rm no~Moli\grave{e}re}=0.404$, the value obtained by fitting to data in Ref.~\cite{Casalderrey-Solana:2018wrw}.




\subsection{Jet and single hadron suppression, and 
introducing results to follow}
\label{sec:jet-and-hadron-suppression}

\begin{figure}[t!]
    \centering
    \includegraphics[width=0.49\columnwidth]{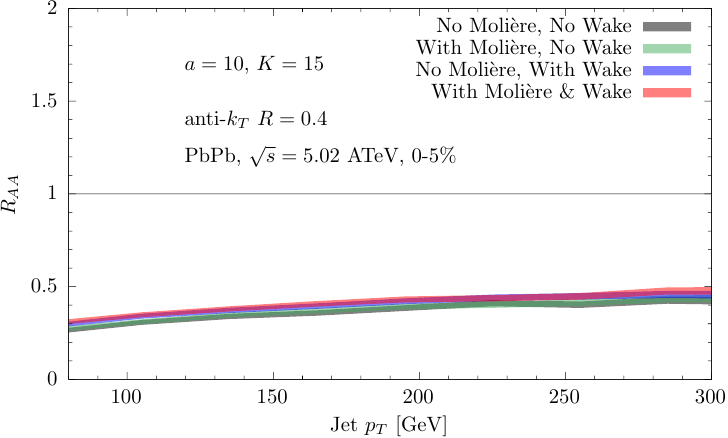}
    \includegraphics[width=0.49\columnwidth]{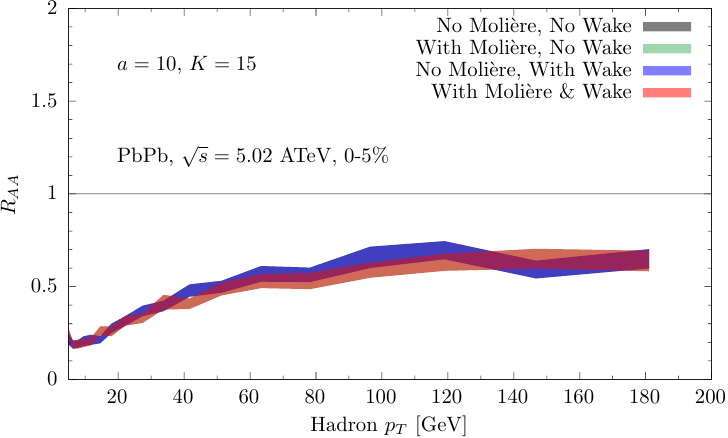}
    \caption{Left: Hybrid Model calculations of $R_{\rm AA}$ of inclusive anti-$k_t$ jets with $R=0.4$ as a function of jet $p_T$. Right: Hybrid Model calculations of $R_{\rm AA}$ of single inclusive charged hadrons with transverse momentum $p_T$. In both panels, $R_{\rm AA}$ is the suppression in PbPb collisions with 0-5\% centrality and $\sqrt{s_{\rm NN}}=5.02$~TeV relative to pp collisions.  The four colored bands show results from Hybrid Model calculations with and without Moli\`ere scattering with $g_s=2.25$ and $a=10$, and with and without the soft hadrons originating from jet wakes. 
    Soft Gaussian transverse momentum broadening with $K=15$ is included in all calculations. Calculations with Moli\`ere scattering (red and green bands) have the parameter that controls the magnitude of strongly coupled energy loss set to $\kappa_{\rm sc}=0.37$ so as to match prior calculations for jet and hadron suppression without Moli\`ere scattering (blue and black bands) calculated with $\kappa_{\rm sc}=0.404$, the value obtained by fitting Hybrid model calculations including jet wakes but without Moli\`ere scattering to data in Ref.~\cite{Casalderrey-Solana:2018wrw}.
    } 
    \label{fig:raa}
\end{figure}

In the left panel of Fig.~\ref{fig:raa} we show Hybrid Model calculations of jet $R_{\rm AA}$ for jets reconstructed with the anti-$k_t$ algorithm~\cite{Cacciari:2008gp, Cacciari:2011ma} and radius parameter $R=0.4$. 
In the right panel, we show Hybrid Model calculations of the $R_{\rm AA}$ for charged hadrons. In the calculations without Moli\`ere scattering (black and blue curves) we have 
chosen $\kappa_{\rm sc}=0.404$. We have adjusted
the value of $\kappa_{\rm sc}$ in the calculations
that include Moli\`ere scattering (green and red)
such that the jet $R_{\rm AA}$ and hadron $R_{\rm AA}$ are very similar when soft hadrons from
jet wakes are included (red curves on top of blue curves) or are not included (green curves on top of
black curves).  This corresponds to choosing $\kappa_{\rm sc}=0.37$ when we include Moli\`ere scattering.
%
%

Since Moli\`ere scatterings impart rare, but sizeable kicks to jet partons, they can be found by looking for large-angle scatterings of individual partons within a jet~\cite{Kurkela:2014tla}. With this goal in mind, in the following subsections we will investigate a series of ungroomed and groomed jet and jet substructure observables, keeping all parameters fixed as described here.  Whenever we see an observable where the red and blue Hybrid Model calculations are similar to each other and well separated from the green and black Hybrid Model calculations, we will conclude that such an observable is more sensitive to jet wakes (a classic strongly coupled, soft, modification
to jets that does not involve adding extra prongs) than to Moli\`ere scattering.
And, whenever we see an observable where the red and green Hybrid Model calculations are similar
to each other and well separated from the blue and black Hybrid Model calculations, we will conclude that such an observable is more sensitive
to Moli\`ere scattering (our classic
example of a weakly coupled, hard, modification 
to jets that does involve adding extra prongs).

Perhaps this goes without saying, but the black, green and blue curves in the Figures should not be compared to experimental data since  it is impossible to turn either Moli\`ere scattering or jet wakes off in an experiment! 
The red curves may be compared to experimental data, ideally upon using measurements of many observables like those in the following subsections to fix the values of the four model parameters via a Bayesian uncertainty quantification analysis.




\subsection{Two ungroomed observables}

\begin{figure}[t!]
    \centering
    \includegraphics[width=0.49\columnwidth]{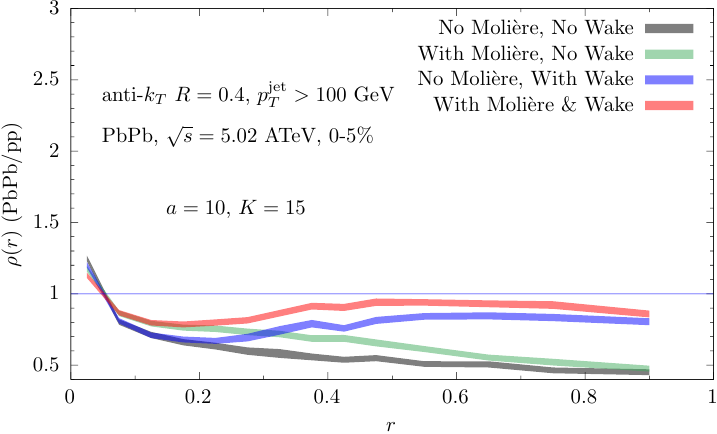}
     \includegraphics[width=0.49\columnwidth]{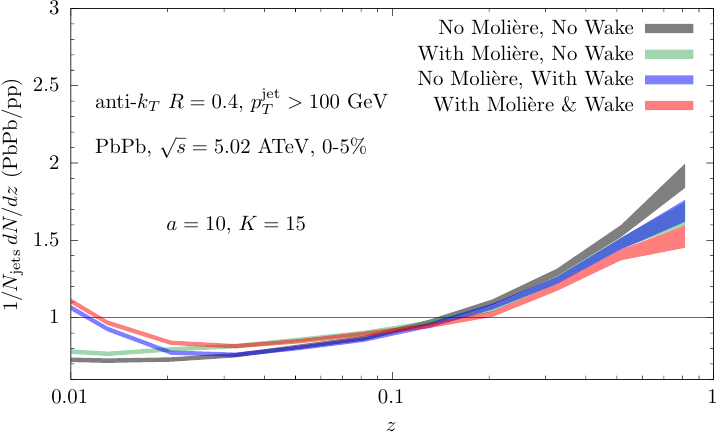}
    \caption{Hybrid Model results for the ratios between the jet shapes (left panel) and jet fragmentation functions (right panel) in PbPb collisions with $\sqrt{s_{\rm NN}}=5.02$~TeV and 0-5\% centrality and pp collisions for inclusive samples of anti-$k_t$ jets with $p_T^{\rm jet}>100$~GeV and $R=0.4$.}
    \label{fig:ffshap}
\end{figure}

Having fixed the parameters in the model, we now begin by examining Hybrid Model results with and without Moli\`ere scattering for two of the most well-studied jet observables, namely the jet shape and the 
jet fragmentation function.   Hybrid Model calculations of these observables were first presented in Ref.~\cite{Casalderrey-Solana:2016jvj},
where the (significant) observable consequences of jet wakes
and the (small) consequences of soft Gaussian transverse momentum broadening were analyzed.

In the left panel of Fig.~\ref{fig:ffshap}, we present the ratio between distributions of the jet shape in PbPb and pp collisions. The jet shape is an observable which measures the distribution of hadronic transverse momentum relative to the jet axis. It is defined as $\rho(r)\equiv\langle \sum_i p_T^i(r)/p_T^{\rm jet} \rangle$, where in our analysis we take the average $\langle \cdot \rangle$ among all jets that satisfy $p_T^{\rm jet}>100$ GeV. $r$ is the distance in the $(\eta,\phi)$ plane between a charged particle $i$ with $p_T^i>1$ GeV and the jet axis, and is defined as $r=\sqrt{\Delta \eta^2+\Delta \phi^2}$. Note that the sum over $i$ need not be limited to those particles contained within the jet radius $R=0.4$. 
The first thing that we observe is the marked narrowing near the jet core for all four curves. This is due to the selection bias 
in PbPb collisions
towards those jets with a given $p_T$ that lost the least energy, sometimes referred to as the survivor bias.
Because the jet production cross-section falls steeply with $p_T$, the jets found with a given PbPb are more likely to be surviving jets which lost little energy than jets which would have had much more energy absent jet quenching.
Since jets that have a narrower shape, and fewer harder fragments, lose less energy, we see
that the ensemble of jets reconstructed in PbPb collisions are narrower (left panel
of Fig.~\ref{fig:ffshap}) and harder (right panel of Fig.~\ref{fig:ffshap}, described below)
than in pp collisions, as has been seen
in many previous 
analyses~\cite{Milhano:2015mng,Rajagopal:2016uip,Casalderrey-Solana:2016jvj,Brewer:2017fqy,Hulcher:2017cpt,
Mehtar-Tani:2017web,Casalderrey-Solana:2018wrw,Casalderrey-Solana:2019ubu,Caucal:2019uvr,
Du:2020pmp,
Caucal:2021cfb,
Brewer:2021hmh,Pablos:2022mrx,Hulcher:2022kmn,Kudinoor:2025gao}. 

Here, in the left panel of Fig.~\ref{fig:ffshap} we observe that including Moli\`ere scatterings, which can be seen by comparing the green curve to the black curve or by comparing the red curve to the blue curve, broadens the jet shape distribution in a way that partially compensates for its narrowing caused by the 
selection bias due to energy loss.  It is clear that the narrowing of the shapes of jets found in PbPb collisions due to selection bias is a much larger effect than the broadening of these jets by Moli\`ere scattering.
Furthermore, the effects of Moli\`ere scattering are also dwarfed by the effects of jet wakes (compare red and blue curves to black and green curves).  Since the soft hadrons originating from jet wakes have a broad angular distribution, including them broadens the jet shape~\cite{Casalderrey-Solana:2016jvj}, and in an ungroomed observable like this one this 
effect is much larger than that of Moli\`ere scattering. We conclude that although Moli\`ere scattering does modify the jet shape distribution, this observable is not well-suited for identifying Moli\`ere scattering unambiguously in experimental data.


It is nevertheless interesting to observe that the effect of elastic scatterings is exhausted after $r\gtrsim 0.6$~radians, which gives us a ballpark estimate of the angle out to which
the scattered partons from Moli\`ere scattering 
are to be found.
The soft hadrons from jet wakes extend out to
even larger angles, raising the large angle tail
of the jet shape via the addition of soft hadrons coming from the region of QGP behind the jet that 
has picked up momentum in the direction of the jet.
Their effect is largest where the hard particles originating  from the (quenched) parton shower are absent, away from the jet core.
We see that the effects of jet wakes, and to a lesser degree of Moli\`ere scattering, persist out to angles beyond the jet radius parameter, here $R=0.4$, employed in reconstructing the sample of jets.

We obtain a similar story in the right panel of Fig.~\ref{fig:ffshap}, where we compute PbPb/pp ratios of the jet fragmentation functions.
The jet fragmentation function describes how
the momentum of the jet is distributed among the momentum (in the direction of the jet) carried by each charged hadron in the jet.
It
is defined as the distribution of charged hadrons within the jet with $p_T>1$~GeV
with a given momentum 
fraction $z\equiv (p_T \cos r)/p_T^{\rm jet}$, where $r$ is defined as in the jet shape. 
For each charged hadron, $p_T \cos r$ is the component of its momentum in the jet direction.
In this analysis, we average the fragmentation function
over all jets with $p_T^{\rm jet}>100$ GeV, and by definition it is normalized to the average number of charged tracks per jet.  

In the right panel of Fig.~\ref{fig:ffshap} 
we see in jet fragmentation functions the manifestation of the same physical effects that we 
saw affecting jet shapes in the left panel.
Selection bias towards jets that have lost less energy favors jets with a harder fragmentation function in PbPb collisions, pushing the PbPb/pp 
fragmentation function ratio up at large $z$ and down at small $z$. This is the biggest effect.
Jet wakes add soft hadrons, with small $z$.
This has a significant effect on the fragmentation function at very low $z$.  (Note that for a 100 GeV jet, $z\sim 0.01-0.02$ corresponds to hadrons with $p_T\sim 1-2$~GeV originating from jet wakes.)
Because of the way that the fragmentation functions are defined, pushing the fragmentation function up at low $z$ also results in pushing it slightly downwards at large $z$. 
As for the jet shape, effects of Moli\`ere scattering are visible in the fragmentation function ratio by comparing red to blue or green to black: a small decrease at large $z$ and a small increase at lower $z$.  We can observe that Moli\`ere scattering pushes the fragmentation function up all the way out to $z\sim 0.1$, meaning that the hadrons that result from these scatterings are  harder than those originating from jet wakes.  However, the effects of Moli\`ere scattering in this observable are so much smaller than the effects of jet wakes and selection bias
that, here too, this ungroomed observable 
is not well-suited for identifying Moli\`ere scattering in experimental data.

From the study of these two important
ungroomed jet observables we can draw several conclusions.  First, Moli\`ere scattering
does modify the shapes and fragmentation functions of jets, broadening the former and softening the latter.  However, selection bias pushes the shapes and fragmentation functions of the samples of jets reconstructed in PbPb collisions more strongly in the opposite directions.  These will not be the only observables where we will see effects of Moli\`ere scattering partially compensating
for effects of selection bias due to energy loss.
Second, in ungroomed observables where soft hadrons within the jets make a significant contribution, the soft hadrons from jet wakes 
modify jets in PbPb collisions relative to jets in pp collisions.  This is of significant interest in its own right, but it represents a confounding effect if we are seeking observables with which to observe unambiguous consequences of Moli\`ere scattering in experimental data. 
This points us toward the strategy that 
we shall follow in the next subsections: analyze groomed observables, including observables that characterize semi-hard subjets within jets.

\subsection{Groomed observables}

In general terms, groomed jet observables refer to observables involving some algorithmic procedure that serves to remove some soft hadrons from the jets.
In this subsection, we analyze 
three groomed jet observables that have been defined so as to
exploit knowledge about the clustering history of a jet.
The jet-finding algorithm that is most robust in the presence of an underlying event is the anti-$k_t$ algorithm~\cite{Cacciari:2008gp}, as it preferentially clusters particles around a high-$p_T$ constituent, which typically lies close to the true partonic jet axis. Once the jet and its constituents have been identified, however, one can probe its internal structure by reclustering the jet using a purely angular-ordered algorithm such as Cambridge--Aachen (C/A)~\cite{Dokshitzer:1997in, Wobisch:1998wt}. The C/A algorithm clusters particles based solely on their angular separation, without reference to their momenta or energies. To leading logarithmic accuracy in the DGLAP evolution variable, vacuum parton showers exhibit strong angular ordering. As a result, when applied in reverse the C/A algorithm approximately reconstructs the splitting history of the jet. Each step in the clustering history can then be characterized by the momentum sharing fraction
\begin{equation}
z_{12} \equiv \frac{\min(p_{T,1}, p_{T,2})}{p_{T,1}+p_{T,2}},
\end{equation}
where $p_{T,1}$ and $p_{T,2}$ are the transverse momenta of the two branches being merged, and by their relative angular separation
\begin{equation}
R_{12} \equiv \sqrt{\Delta \eta_{12}^2 + \Delta \phi_{12}^2}
\end{equation}
in $(\eta,\phi)$ space. The grooming procedures 
we describe were designed so as to reduce
sensitivity to soft, nonperturbative, splittings
within the parton shower by
selecting  a specified subset of these clustering steps.
When applied to jets in heavy ion collisions,
these grooming procedures also serve to remove most of the sort hadrons originating from
jet wakes.

\begin{figure}
    \centering
    \includegraphics[width=0.49\columnwidth]{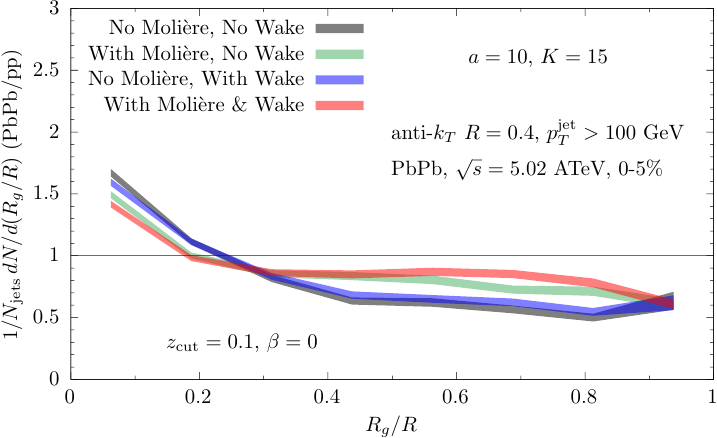}
     \includegraphics[width=0.49\columnwidth]{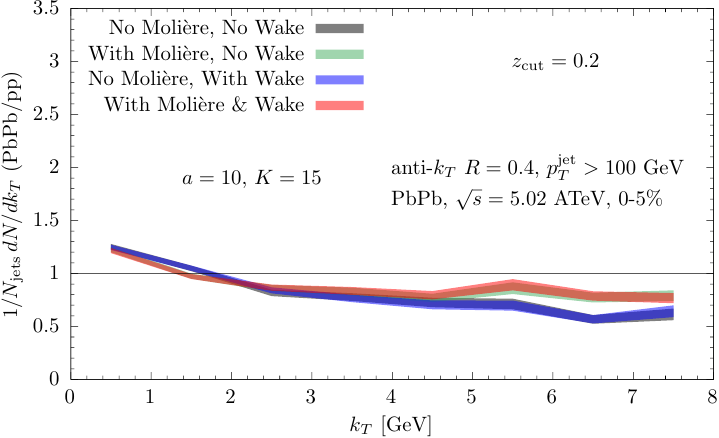}
    \caption{Left: Hybrid Model results
    for the ratios between the distribution $(1/N_{\rm jets})\,dN/d(R_g/R)$ of the scaled Soft Drop angle $R_g/R$ (left panel) 
    and the distribution $(1/N_{\rm jets})\,dN/dk_T$ of the $k_T$ of the largest splitting (right panel) of jets in PbPb 
    collisions with $\sqrt{s_{\rm NN}}=5.02$~TeV
    and 0-5\% centrality and pp collisions for
    inclusive samples of anti-$k_t$ jets with 
    $p_T^{\rm jet}>100$~GeV and $R=0.4$.
    In the left panel, the Soft Drop grooming procedure with $z_{\rm cut}=0.1$ and $\beta=0$ is employed.} 
    \label{fig:rgkt}
\end{figure}

In particular, the Soft Drop grooming algorithm~\cite{Larkoski:2014wba} identifies the first splitting  within an anti-$k_t$ jet of radius $R$ that satisfies the condition
\begin{equation}
z_g > z_{\rm cut} \left(\frac{R_{12}}{R}\right)^\beta,
\end{equation}
where $z_{\rm cut}$ and $\beta$ are parameters that specify the grooming algorithm. 
The angular separation $R_{12}$ of this first splitting is referred to as $R_g$.
The distributions of $z_g$ and $R_g$ for this first accepted splitting have been measured in pp collisions in Ref.~\cite{CMS:2017qlm} and Ref.~\cite{ATLAS:2019mgf}, respectively, 
and have been measured in heavy-ion collisions by CMS~\cite{CMS:2017qlm, CMS:2018fof, CMS:2024zjn}, ALICE~\cite{ALICE:2019ykw, ALargeIonColliderExperiment:2021mqf}, and ATLAS~\cite{ATLAS:2022vii, ATLAS:2025svn}.
%
A previous Hybrid Model analysis (without Moli\`ere scattering) shows almost no modification to the $z_g$ distribution in heavy-ion collisions compared to pp, with $z_{\rm cut}=0.1$~\cite{Casalderrey-Solana:2019ubu}, and we have checked that including Moli\`ere scattering does not modify this conclusion.
(The origin of the soft $z_g$ enhancement seen in experimental data that has not been unfolded~\cite{CMS:2017qlm} can be reproduced by embedding our results in a fluctuating background~\cite{Andrews:2019hvc}.
However, the distribution of the Soft Drop splitting angle, $(1/N)\, dN/dR_g$, was found to be significantly modified~\cite{Casalderrey-Solana:2019ubu}, an effect that is largely
driven by selection bias, as has been discussed in theoretical studies~\cite{Casalderrey-Solana:2019ubu,Caucal:2019uvr,Du:2020pmp,Brewer:2021hmh,Kudinoor:2025gao,Apolinario:2026hff} and in papers reporting experimental measurements~\cite{ALargeIonColliderExperiment:2021mqf, CMS:2024zjn} of the Soft Drop splitting angle. 
As we have seen, because of energy loss the jet sample in PbPb collisions becomes biased toward narrower, less-modified jets that retain a larger fraction of their original energy within the jet cone. 
This bias suppresses the contribution of jets with large opening angles and enhances the relative weight of narrow configurations in the sample of jets selected in PbPb collisions relative to that in pp collisions, thereby modifying the measured $(1/N)\, dN/dR_g$ distribution.  
Jets with larger $R_g$, which can in qualitative terms be thought of as jets with two semi-hard
substructures separated by a larger angle $R_g$,
lose more energy than jets with smaller $R_g$ and so are less numerous among the jets in a given $p_T$-bin in PbPb collisions.
The black and blue curves in the left panel of Fig.~\ref{fig:rgkt} serve to revisit these conclusions.

In the present work, we present Hybrid Model calculations of the Soft Drop angle $R_g$ in heavy-ion collisions that include the effects of Moli\`ere scattering, see the red and green curves in the left panel of Fig.~\ref{fig:rgkt}. This Figure shows Hybrid Model calculations of the ratio of distributions of the scaled Soft Drop angle $R_g/R$ as measured in PbPb collisions to pp collisions for anti-$k_t$ $R = 0.4$ jets with $p_T > 100$ GeV, calculated using grooming parameters $z_{\rm cut}=0.1$ and $\beta=0$. 
We first note that --- consistent with the results of Ref.~\cite{Casalderrey-Solana:2019ubu} --- the inclusion/exclusion of jet wakes has a negligible effect on the calculations shown. 
This confirms that 
since the hadrons which originate from the wake are soft they are largely groomed away by the Soft Drop grooming procedure. This validates our motivation for focusing on these observables, and allows us to focus on the effects of selection bias due to energy loss and Moli\`ere scattering. 
%
We then observe that including Moli\`ere scatterings in our model results in a broadening of the $R_g / R$ distributions in PbPb collisions that is most notable 
towards the periphery of the jet, for $R_g\gtrsim 0.5 R=0.2$. 
In qualitative terms, this
is consistent with what was found in the jet shape observable in the left panel of Fig.~\ref{fig:ffshap}, but here the confounding effects of jet wakes are largely absent.
We see that selection bias pushes the 
$R_g$ distribution down at larger $R_g/R$ and
correspondingly pushes it up at small $R_g/R$, while Moli\`ere scattering acts in the opposite direction and partially counteracts the effects of selection bias due to energy loss.



In the right panel of Fig.~\ref{fig:rgkt}, we present Hybrid Model calculations of PbPb/pp ratios of the distributions of an observable referred to as leading $k_T$~\cite{Mehtar-Tani:2019rrk}, which is closely related to $R_g$ and defined as follows.
$k_T$ is defined to be the splitting scale associated with the C/A reclustering step that maximizes
\begin{equation}
k_T \equiv z(1-z)\, p_T^{\rm parent} \sin\theta,
\end{equation}
where $z$ is the momentum sharing fraction of the splitting, $p_T^{\rm parent}$ is the transverse momentum of the parent parton that splits, and $\theta$ is the opening angle between the two daughter partons after the splitting. Motivated by experimental considerations, we consider only splittings that satisfy $z > z_{\rm cut} = 0.2$, thereby excluding highly asymmetric splittings. (In experimental data, lowering $z_{\rm cut}$ would increase the number of fake subjets arising from upward fluctuations in the QGP background.)
Analogous to what we observed by studying the scaled Soft Drop angle $R_g/R$ in the left panel of Fig.~\ref{fig:rgkt}, the right panel of this Figure shows that the number of larger $k_T$ splittings is reduced by selection bias due to energy loss, and is enhanced when Moli\`ere scatterings are included.
Interestingly, recent measurements of the leading $k_T$ observable by the ALICE Collaboration~\cite{ALICE:2024fip} exhibit sufficient sensitivity to discriminate between the inclusion and exclusion of Moli\`ere scatterings.
This data should therefore definitely be included in any future Bayesian analysis in which experimental measurements of many observables are used to constrain the values of $\kappa_{\rm sc}$, $K$, $g_s$ and $a$.

In both panels of Fig.~\ref{fig:rgkt},
we see the effects of Moli\`ere scattering only partially compensating for the effects of 
selection bias due to energy loss.
This will make it challenging to use measurements of these observables to claim unambiguous
evidence for the presence of Moli\`ere scattering.
Moli\`ere scattering is a physical phenomenon that
results in additional ``prongs'' in jets in PbPb collisions, meaning that more jets should be found with larger values of $R_g/R$ and $k_T$ in
PbPb collisions than in pp collisions.  Unfortunately, our Hybrid Model calculations show that the PbPb/pp ratios are elevated by Moli\`ere scattering, but remain less than one.  This motivates looking at these observables in
samples of jets that are selected in ways that reduce the bias favoring less modified jets.
We will study Hybrid Model calculations of
the $R_g$ distribution in photon-tagged jet samples in subsection~\ref{sec:gamma-jets},
with the goal of reducing the effects of
selection bias.  There we will also investigate jets with a larger radius $R$, where the effects of jet wakes are not completely removed and in fact an interesting interplay between
jet wakes and Moli\`ere scattering is seen.

\begin{figure}
    \centering
     \includegraphics[width=0.49\columnwidth]{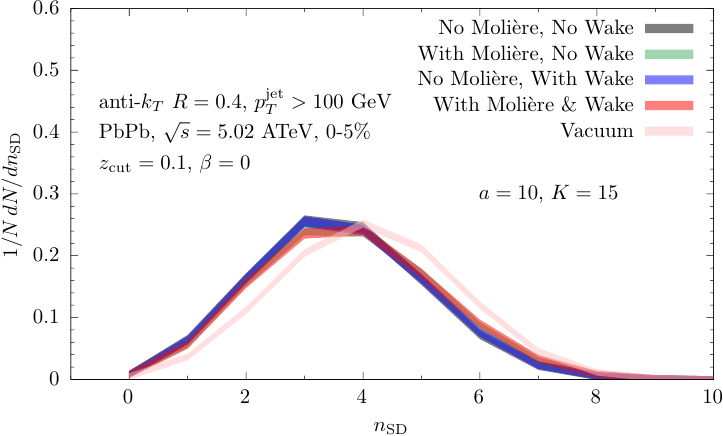}
    \caption{Hybrid Model calculations of the distributions of the number of splittings $n_{\rm SD}$ that satisfy the Soft Drop condition with $z_{\rm cut}=0.1$ and $\beta=0$ within anti-$k_t$ jets of radius $R=0.4$ and $p_T^{\rm jet}>100$~GeV in 0-5\% central PbPb collisions and pp collisions (pink curve) with $\sqrt{s_{\rm NN}}=5.02$~TeV. All distributions are normalized by the number of $R = 0.4$ jets.}
    \label{fig:nsub}
\end{figure}

The first two groomed observables 
that we have examined focus on the properties of a single splitting, or single clustering step, of the shower. 
However, we know that each Moli\`ere scattering adds one parton to the jet shower and yields two scattered partons that may have been scattered by a significant angle. This prompts us to
ask whether Moli\`ere scattering increases the number of subjets within a jet --- a question that we shall focus on in the next subsection.
In the context of this subsection, though, we
take a first look into this idea in Fig.~\ref{fig:nsub} by asking
whether Moli\`ere scattering increases the number of splittings $n_{\rm SD}$ that satisfy the Soft Drop splitting condition, as this is also a way of asking whether the number of semi-hard structures within a jet has been increased (by Moli\`ere scattering) in PbPb collisions relative to pp collisions.
%
%
In Fig.~\ref{fig:nsub} we present Hybrid Model calculations of the probability distribution of finding $n_{\rm SD}$ reclustering steps that satisfy the Soft Drop condition, with $z_{\rm cut}=0.1$ and $\beta=0$. Jets in PbPb collisions are biased towards those that lost the least amount of energy, which tend to have fewer and narrower hard splittings. Thus, we observe a reduction in $n_{\rm SD}$ in the sample of jets
selected in PbPb collisions when compared to that in pp collisions.
%
Once again, jet wakes have almost no effect on this observable since the hadrons originating from jet wakes are too soft to pass the Soft Drop condition. Unfortunately, Moli\`ere scatterings have very little effect on $n_{\rm SD}$, only slightly increasing $n_{\rm SD}$. Hence, we do not believe that measurements of $n_{\rm SD}$ would be sensitive to the presence of Moli\`ere scatterings in PbPb collisions. However, other observables which examine multiple semi-hard structures within a jet may still hold promise for detecting and studying Moli\`ere scattering in heavy-ion collisions. These observables are the focus of the next subsection.


\subsection{Jets within Jets}

In the previous subsection, we saw that groomed substructure is sensitive to the effects of Moli\`ere scattering while remaining relatively insensitive to the effects of jet wakes (provided that the value of $z_{\rm cut}$ is large enough). In the present subsection we explore a different set of substructure observables, which examine the properties of (sub)jets of radius $R_S$ contained within a larger jet of radius $R>R_S$~\cite{Zhang:2015trf}.

\begin{figure}
    \centering
    \includegraphics[width=0.49\columnwidth]{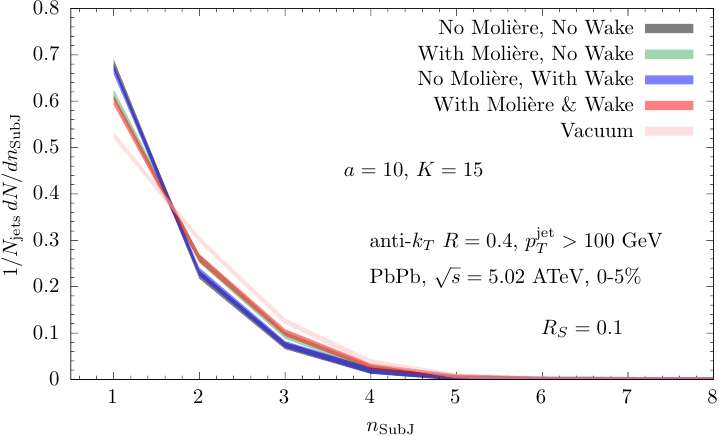}

    \vspace{0.5em}

    \includegraphics[width=0.49\columnwidth]{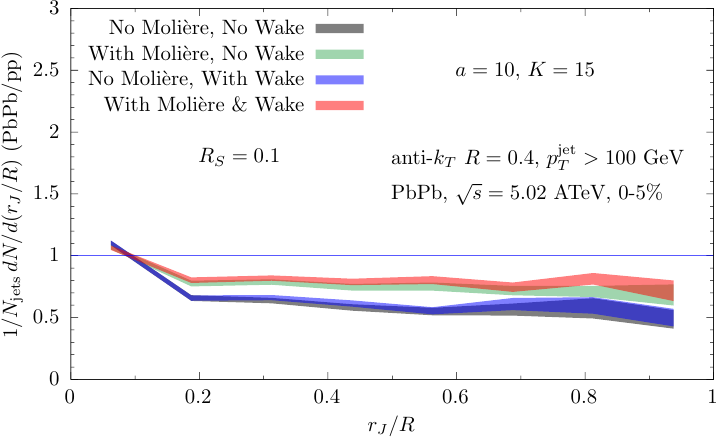}
     \includegraphics[width=0.49\columnwidth]{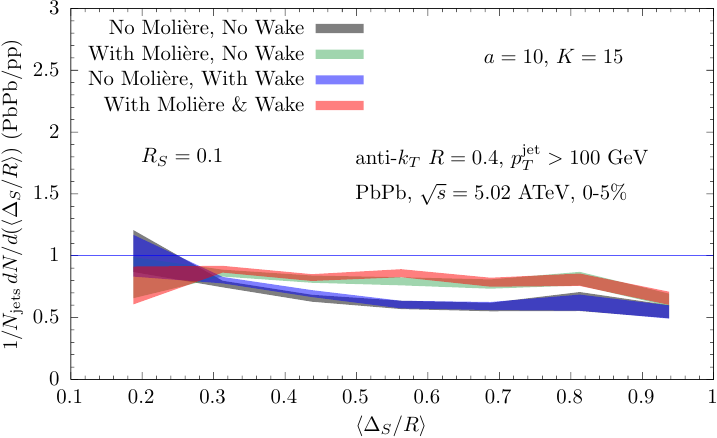}
    \caption{Top: Hybrid Model calculations of the distributions of $n_{\rm subJ}$, the number of subjets  of radius $R_S=0.1$ with $p_T^{\rm Sub}>10$~GeV contained within anti-$k_t$ jets of radius $R=0.4$ and $p_T^{\rm jet}>100$~GeV in 0-5\% central PbPb collisions and pp collisions (pink curve) with $\sqrt{s_{\rm NN}}=5.02$~TeV. Lower panels: Hybrid Model caclulations of the ratios between the distributions of the angular separation $r_J$ between the axis of a subjet with $R_S=0.1$ and the axis of the jet with $R=0.4$ (lower left) and the angular separation $\Delta_S$ between pairs of subjets (lower right) in PbPb collisions and pp collisions.}
    \label{fig:jetdelta}
\end{figure}

The procedure is as follows. First, we reconstruct anti-$k_t$ jets of radius $R$. Then, using only those constituents that that have been clustered into a given jet, we reconstruct anti-$k_t$ jets with a specified smaller radius $R_S<R$. Given the specifics of the anti-$k_t$ algorithm, it will tend to cluster first those particles with larger momenta, rendering it well-suited for finding the semi-hard structures that Moli\`ere scatterings can generate around the original jet axis. 
In the top panel of Fig.~\ref{fig:jetdelta},
we show the results of Hybrid Model calculations  of the distribution of the number of subjets $n_{\rm SubJ}$ of radius $R_S = 0.1$ with $p_T^{\rm Sub}>10$ GeV found within $R = 0.4$ jets with $p_T^{\rm jet}>100$ GeV using this procedure. 
This distribution of the number of jets with $n_{\rm SubJ}$ subjets in pp collisions falls rapidly with increasing $n_{\rm SubJ}$, where having just two subjets is half as likely as having only one subjet.
Since jets in PbPb collisions are biased towards those that lost the least amount of energy and therefore those that contain narrower splittings, we observe an increased number of $R = 0.4$ jets containing only one subjet and a reduction in the number of $R = 0.4$ jets containing multiple subjets, compared to pp collisions. 

We see in the top panel of Fig.~\ref{fig:jetdelta} that
the inclusion of Moli\`ere scatterings enhances the number of $R = 0.4$ jets with $n_{\rm SubJ} > 1$. This enhancement arises because a hard $2 \rightarrow 2$ Moli\`ere scattering can transfer substantial transverse momentum to a thermal parton, producing a semi-hard recoil at a sizable angle with respect to the original jet axis (and can also kick the jet parton that initiated the Moli\`ere scattering with a substantial transverse momentum). 
When either the recoiling parton kicked from the QGP or the scattered jet parton
remains inside the jet with radius $R$ and carries transverse momentum $p_T \gtrsim 10$~GeV, this may result in the reconstruction of one additional subjet of radius $R_S$.  
Since the anti-$k_t$ algorithm preferentially clusters high-$p_T$ particles first, it efficiently resolves these semi-hard structures as independent subjets rather 
than merging them into the leading core. As a consequence, the probability of finding events with $n_{\rm SubJ} > 1$ is enhanced when Moli\`ere scatterings are included.  And indeed, the red and green curves in the top panel of Fig.~\ref{fig:jetdelta} have been pushed toward larger $n_{\rm subJ}$ by Moli\`ere scattering. As is by now familiar, this happens only to a degree that partially compensates for the opposite effect arising from selection bias due to energy loss.

In addition to counting the number of subjets contained within an $R = 0.4$ jet, 
in the lower two panels of Fig.~\ref{fig:jetdelta}
we study the angular distribution of these subjets using two novel observables: (i) the angular separation $r_J$ between the axis of an $R_S = 0.1$ subjet and the axis of the $R = 0.4$ jet in which it is contained, and (ii) the angular separation $\Delta_S$ between two $R_S = 0.1$ subjets within the same $R = 0.4$ jet. While $n_{\rm SubJ}$ counts the multiplicity of resolved subjets, $\langle \Delta_S \rangle$ and $r_J$ quantify how far these subjets are separated from one another and from the central core of the jet. 
The kicks imparted by Moli\`ere scattering should serve to increase both $\Delta_S$ and $r_J$.

The bottom panels of Fig.~\ref{fig:jetdelta} show the PbPb/pp ratios of the normalized distributions $(1/N_{\rm jets})\, dN/d(r_J/R)$ (left) and $(1/N_{\rm jets})\, dN/d(\langle \Delta_S/R \rangle)$ (right), where $r_J$ is the angular separation in $(\eta,\phi)$ space between an $R=0.1$ subjet and the axis of the containing $R=0.4$ jet, and $\langle \Delta_S/R \rangle$ denotes the average angular separation between pairs of $R=0.1$ subjets within an $R = 0.4$ jet. All distributions are normalized by the number of $R = 0.4$ jets.
In the absence of Moli\`ere scatterings, the distributions are suppressed at moderate and large angular separations, reflecting the familiar selection bias toward jets with narrower hard substructures in PbPb collisions. While both observables show little sensitivity to jet wakes, they exhibit visible sensitivity to Moli\`ere scatterings. Hard $2 \rightarrow 2$ scatterings can deflect jet partons, broadening the jet, and can
kick either jet partons or thermal partons out to sizable angles away from the central core of the jet. These recoiling partons and the wakes they excite can be reclustered into the $R=0.4$ jet, and if so may then be reconstructed as an $R_S=0.1$ subjet. This serves to increase both the radial distribution of subjets and the typical angular separation between them, in a way that --- yet again --- partially compensates for the opposite effect coming from selection bias.

Taken together, the results in Fig.~\ref{fig:jetdelta} demonstrate that these jets-within-jets observables provide 
avenues toward the detection and study of Moli\`ere scattering off quasiparticles in QGP
that are complementary to the groomed observables in the previous subsection.
Unfortunately, all of the observables that we have examined so far in this and the previous subsection that focus on modifications to the number and distribution of semi-hard structures within jets 
suffer from significant confounding effects originating from jet selection bias due to energy loss. In the next subsection, we turn to photon-tagged jet observables in which this selection bias can be partially mitigated and, in some cases, even overcome by the effects of Moli\`ere scatterings.

\subsection{Photon-tagged Jet Observables} \label{sec:gamma-jets}

Photon-tagged jet events provide a particularly clean probe of in-medium jet modification. Since the mean free path of photons passing through a droplet of QGP is larger than the size of that droplet formed in heavy-ion collisions, the photon does not interact with the medium or lose energy to it as a passing jet would. 
This means that when we choose our sample of jets to be those jets found
in events in which a photon with
a transverse momentum thus and so,  selecting 
events based upon the photon energy introduces
no bias favoring jets that lose less energy.
Although we shall see that this is not a panacea,
%
this
significantly reduces the selection biases that affect inclusive jet measurements.

\begin{figure*}
    \centering
    \includegraphics[width=0.49\textwidth]{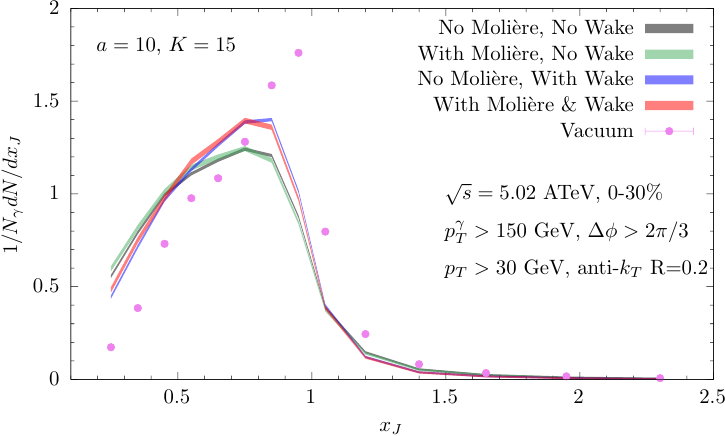}
    \includegraphics[width=0.49\textwidth]{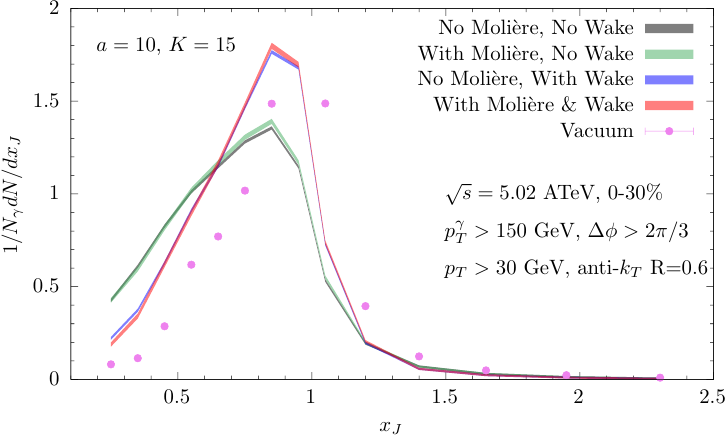}
    \caption{Hybrid Model results for the distributions $(1/N_\gamma)\,dN/dx_J$ of the photon-jet momentum imbalance variable $x_J \equiv p_T^{\rm jet}/p_T^\gamma$ for photon-tagged jets with radii $R = 0.2$ (left) and $R = 0.6$ (right), with $p_T^\gamma>150$~GeV, $\Delta\phi>2\pi/3$, and $p_T^{\rm jet}>30$~GeV in pp (pink points) and 0--30\% central PbPb collisions (colored bands) with $\sqrt{s_{\rm NN}}=5.02$~TeV.}
    \label{fig:xJ}
\end{figure*}

Note that photon-tagged jets in events with
a photon with some $p_T^\gamma$, for example 
the jets in events with a photon with
$p_T^\gamma>150$~GeV, have a rather broad distribution of jet energies even in pp collisions. Of course the photon $p_T$ is balanced by hadronic $p_T$, but this can be distributed among several jets and it includes soft radiation outside the jet cone(s).  Because the jets in PbPb collisions have lost energy, the distribution
of the momentum imbalance variable
$x_J \equiv p_T^{\rm jet} / p_T^\gamma$
will be shifted toward lower $x_J$ in PbPb collisions and will be even broader.
To get a sense of this, and because this distribution will be important to understanding the distributions of observables like the Soft Drop angle $R_g$ that we shall focus on below,
we begin in Fig.~\ref{fig:xJ} by plotting 
the $x_J$ distributions
for photon-tagged jets
selected as follows. We first identify all events with a photon with $p_T^{\gamma} > 150$ GeV and $|\eta^\gamma| < 1.44$. If an event has multiple photons that satisfy this criteria, the photon with the highest $p_T$ is selected as a reference for the photon-tagged jets in the event. We then require that this photon is ``isolated'', which we define as having less than 5 GeV of transverse energy in a cone of $R = 0.4$ around the photon. If this photon is not isolated, then we discard the event altogether. For each selected, isolated photon, we then find and select the highest-$p_T$ jet with $|\eta| < 2$ that is $\Delta \phi > 2 \pi /3$ away from the selected photon. 
In Fig.~\ref{fig:xJ}, we show distributions $(1/N_\gamma)\, dN/dx_J$ of the number of jets within a given bin of $x_J$, normalized by the number of selected photons, for $R=0.2$ (left panel) and $R=0.6$ (right panel) in pp and 0-30\% central PbPb collisions at $\sqrt{s_{\rm NN}}=5.02$~TeV.

In vacuum, the $x_J$ distribution peaks close to unity, reflecting the importance of events where
the photon and a single jet are nearly back-to-back, but the distribution is broad as we have already noted.
In PbPb collisions, jet energy loss shifts the distribution toward smaller values of $x_J$, reflecting our expectation that jet showers which propagate through a droplet of QGP emerge with a reduced transverse momentum relative to that of the photon. 
The inclusion of Moli\`ere scatterings has very little effect on the $x_J$ distributions in Fig.~\ref{fig:xJ}. Instead, the dominant modification comes from the inclusion of jet wakes. For both $R=0.2$ and $R=0.6$, turning on the wake shifts weight in the PbPb distribution toward larger values of $x_J$. 
The jet deposits energy and momentum into the 
droplet of QGP in the form of a wake; at freezeout, the jet wake becomes soft hadrons
with a net momentum reflecting that lost by
the jet; and, some of these soft hadrons
are subsequently clustered into the jet
reconstructed via the anti-$k_t$ algorithm.
This means that if we leave out the 
soft hadrons from jet wakes, the reconstructed jet
$p_T$, and the imbalance $x_J$, are less than they are when we include jet wakes.
Including jet wakes increases the reconstructed jet $p_T$, partially compensating for the energy lost by the hard shower to the medium. 
As expected, this effect is much more pronounced for jets reconstructed with 
the larger jet radius $R=0.6$, as with $R=0.6$ the
jet reconstruction algorithm captures a
substantially larger fraction of the wake than 
is the case if
narrower $R=0.2$ jets are reconstructed. 
The near overlap of the curves with and without Moli\`ere scattering, in both cases including jet wakes, indicates that rare hard elastic scatterings do not appreciably change the overall photon-jet momentum imbalance, even though they can leave clear signatures in more differential substructure observables. To this, we now turn.

\begin{figure*}
    \centering
    \includegraphics[width=0.49\textwidth]{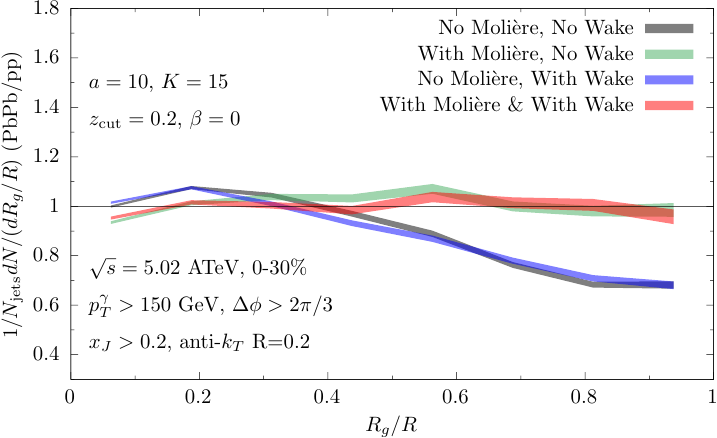}
    \includegraphics[width=0.49\textwidth]{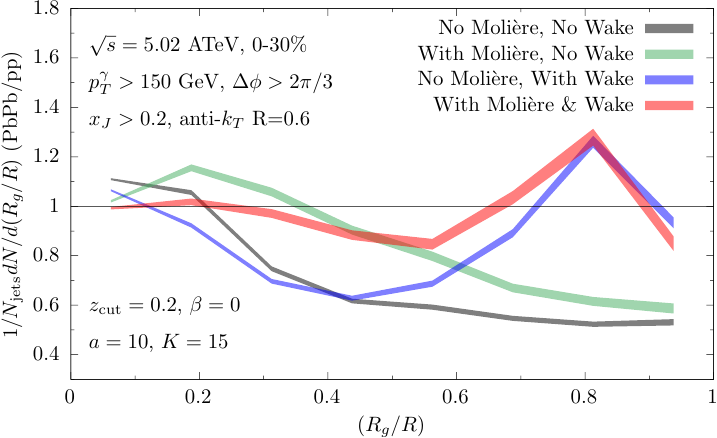}
    \caption{Hybrid Model calculations of the ratios of the distributions of the scaled Soft Drop angle $R_g/R$ for $R=0.2$ (left) and $R=0.6$ (right) jets with $x_J > 0.2$ in photon-jet events with $p_T^\gamma>150$~GeV and $\Delta\phi>2\pi/3$ in 0-30\% central PbPb collisions with $\sqrt{s_{\rm NN}}=5.02$~TeV to same in pp collisions.}
    \label{fig:rgxj0p2}
\end{figure*}
\begin{figure*}
    \centering
    \includegraphics[width=0.49\textwidth]{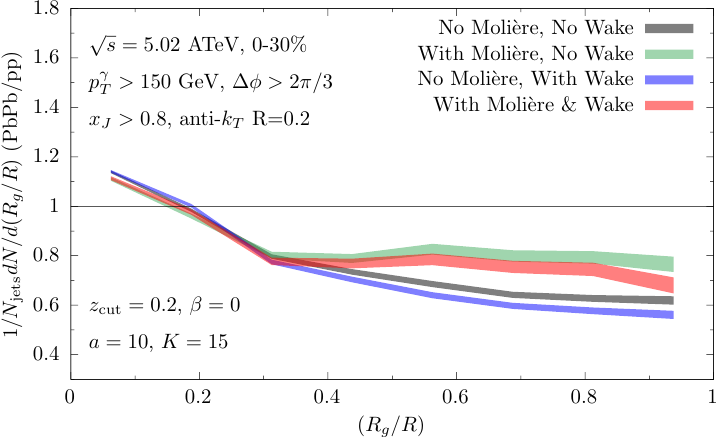}
    \includegraphics[width=0.49\textwidth]{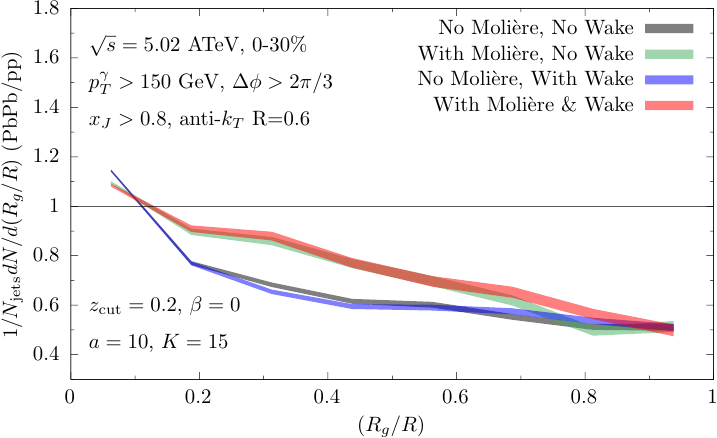}
    \caption{Same as Fig.~\ref{fig:rgxj0p2}, here for jets with $x_J>0.8$: PbPb/pp ratios of the distributions of the scaled Soft Drop angle $R_g/R$ for $R=0.2$ (left) and $R=0.6$ (right) jets with $x_J > 0.8$ in photon-jet events with $p_T^\gamma>150$~GeV and $\Delta\phi>2\pi/3$.}
    \label{fig:rgxj0p8}
\end{figure*}

With the motivations that we described at the end of the previous subsection, we return to analyzing Hybrid Model calculations of the Soft Drop angle $R_g$ --- now in photon-tagged jets. Since selecting events based upon the $p_T^\gamma$ gives us a way to reduce selection biases, and since grooming suppresses soft wide-angle radiation, these observables are especially sensitive to semi-hard structures generated inside the jet by Moli\`ere scatterings. 
Fig.~\ref{fig:rgxj0p2} shows the PbPb/pp ratio of $(1/N_{\rm jets})\, dN/d(R_g/R)$ for photon-tagged jets with $x_J > 0.2$, for $R=0.2$ (left panel) and $R=0.6$ (right panel) jets, with the Soft Drop grooming done using parameters $z_{\rm cut}=0.2$ and $\beta=0$. 
We note that including jets that have lost up to 80\% of their transverse momentum to the medium ($x_J > 0.2$) allows us to significantly, but not completely, eliminate the selection bias due to energy loss.  For example, we see from the left panel of Fig.~\ref{fig:xJ} that there are some $R=0.2$ jets with $x_J<0.2$, many of which are highly modified jets that have lost more than 80\% of their transverse momentum to the medium, and these jets are not included in the sample of jets selected for the $R_g$ analysis in the left panel of Fig.~\ref{fig:rgxj0p2}. 
And indeed, we see a moderate narrowing
of the $R_g$ distribution in the black and blue
curves in this panel, reflecting the 
effects of the remaining selection bias for
$R=0.2$ jets in the absence of Moli\`ere scattering.
It is very pleasing to see, however, that
in this case the bias coming from not selecting the jets that have lost the most energy
is sufficiently modest that including Moli\`ere scattering (red and green curves) completely counteracts it.

We can confirm our understanding of the effects of selection bias by comparing the left panel of Fig.~\ref{fig:rgxj0p2} to the left panel
of Fig.~\ref{fig:rgxj0p8} --- where we have increased the bias by enforcing 
a stricter restriction of $x_J > 0.8$ on the selected jets.
Indeed, the $x_J>0.8$ selection increases the bias favoring jets with smaller $R_g$: we see that even when we include the effects of Moli\`ere scattering the $R_g$ distribution in PbPb collisions remains depressed at larger $R_g$ relative to that in pp collisions. 

Drawing these elements together, we can now
argue that if
measurements in experimental data 
are aligned with what we have seen in the red bands in the left panels of Figs.~\ref{fig:rgxj0p2}, \ref{fig:rgxj0p8} and \ref{fig:xJ}, this would in concert constitute strong evidence for Moli\`ere scattering.
Measuring a PbPb/pp ratio of the Soft Drop angle $R_g$ at (or even above) unity for jets with $x_J>0.2$ in experimental data --- as
for the red band in the left panel of Fig.~\ref{fig:rgxj0p2} --- is of course the key element.  
Confirming the effects of selection bias
by seeing the PbPb/pp ratio drop substantially when only jets with $x_J>0.8$ are included in the sample --- as in the left panel of Fig.~\ref{fig:rgxj0p8} --- is also central to the argument. The third element of the argument is a demonstration
that even with $x_J>0.2$ there must still be some degree of suppression of the PbPb/pp ratio of $R_g$ arising from selection bias: this can
be demonstrated by
measuring the $x_J$ distribution and seeing that --- as in the left panel of Fig.~\ref{fig:xJ} --- selecting $x_J>0.2$ leaves out the most modified jets,  
meaning that some effects of selection bias remain.
That is, seeing the red band in the left panel of Fig.~\ref{fig:xJ} together with comparing the red bands in Fig.~\ref{fig:rgxj0p8} to that in Fig.~\ref{fig:rgxj0p2} tells us that if we could remove {\it all} selection bias by extending the measurement down to $x_J>0$ the PbPb/pp ratio of $R_g$ would be above one --- strong evidence for Moli\`ere scattering.
Or, better to say it more generically, such measurements in concert would constitute strong evidence for hard scattering of jet partons off quasiparticles in the QGP.
This would be an exciting indication that
the particulate structure of QGP can be resolved by energetic partons in jets.

The CMS collaboration has come close to achieving this goal, in the measurements of $R_g$ in photon-tagged jets that they have reported in Ref.~\cite{CMS:2024zjn} for which we had provided Hybrid Model predictions.
They have selected events with $p_T^\gamma>100$~GeV, which means that since reconstructing jets with $p_T$ down to 20 GeV in PbPb collisions is prohibitively challenging
they have only been able to select jets with $x_J>0.8$ and $x_J>0.4$. Their analysis shows clear evidence for greater effects of selection bias due to energy loss in the $x_J>0.8$ sample. For $x_J>0.4$ the PbPb/pp ratio that they have measured is close to unity.  It would be very interesting to 
see their $x_J$ distribution, so as to estimate how large the remaining effects of selection bias are in their $x_J>0.4$ sample, as this could strengthen evidence for Moli\`ere scattering.
We hope, and anticipate, that by going up to $p_T^\gamma>150$~GeV in future higher statistics data sets, as we have done in the Hybrid Model calculations in this paper, the LHC collaborations will be able to
push down to $x_J>0.2$.


We turn next to the Hybrid Model calculations for the PbPb/pp ratio of the Soft Drop angle $R_g$ for the larger radius jets with $R=0.6$ in the right panels of 
Figs.~\ref{fig:rgxj0p2} and \ref{fig:rgxj0p8}.
Although jet wakes have a negligible effect on the modifications of $R_g$ of $R = 0.2$ jets, for jets
with the larger radius $R=0.6$ the impact of jet wakes becomes much more pronounced, in particular at large $R_g\sim 0.5$, meaning $R_g/R\sim 0.8$,
in the sample with $x_J>0.2$ --- see the red and blue bands in the right panel of Fig.~\ref{fig:rgxj0p2}. 
In the absence of jet wakes, we observe a strong suppression in the number of jets with large values of $R_g/R$. 
Moli\`ere scattering increases the number 
of jets with large $R_g$ but the effects of Moli\`ere scattering diminish at the largest $R_g$, suggesting that it is rare for the scattered particles to be deflected away from the core of the jet by more than $\sim 0.5$ radians.
At these largest angles, jet wakes have a very big effect in the right panel of Fig.~\ref{fig:rgxj0p2} --- 
but not in the right panel of Fig.~\ref{fig:rgxj0p8} since when we include only jets with $x_J>0.8$ in the sample we are excluding all jets that have lost any significant energy which means excluding all jets with significant wakes.
%
The effects of jet wakes are amplified for larger-$R$ jets because a greater fraction of the wake remains inside the jet cone, and because the Cambridge--Aachen declustering underlying Soft Drop is intrinsically sensitive to angular structure and is not sensitive to the momenta of jet constituents. 
Note that the behavior seen for $R_g/R<0.3$ in the right panel of Fig.~\ref{fig:rgxj0p2} is quite similar to what we see for $R_g/R<0.9$ in the left panel --- as these correspond to the same range of $R_g$. With $R=0.6$ we can see larger values of $R_g$, where jet wakes make a dominant contribution for $x_J>0.2$.

Comparing Figs.~\ref{fig:rgxj0p2} and~\ref{fig:rgxj0p8} 
makes it apparent that in photon-tagged events
varying both the $x_J$ selection and the 
jet radius $R$ are useful dials, making the interplay between the effects of selection bias due to energy loss, jet wakes, and Moli\`ere scattering beautifully manifest.
Going from $x_J>0.2$ to $x_J>0.8$ dials up the effects of selection bias.  Going from $R=0.2$ jets
to $R=0.6$ jets dials up the effects of jet wakes, in particular in the range $0.4\lesssim R_g \lesssim 0.6$.  And, the effects of Moli\`ere scattering are most prominent in $R=0.2$ jets
with $x_J>0.2$, the left panel of Fig.~\ref{fig:rgxj0p2}.

\begin{figure*}
    \centering
    \includegraphics[width=0.49\textwidth]{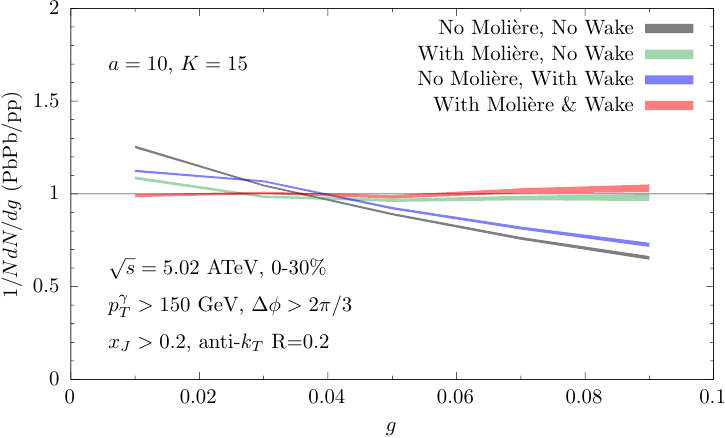}
    \includegraphics[width=0.49\textwidth]{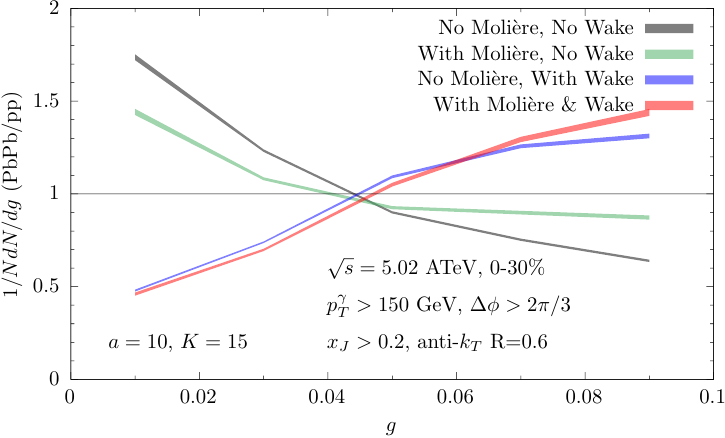}
    \caption{Hybrid Model calculations of the ratios of the distributions of the jet girth $g$ for $R=0.2$ (left) and $R=0.6$ (right) jets with $x_J > 0.2$ in photon-jet events with $p_T^\gamma>150$~GeV and $\Delta\phi>2\pi/3$ in 0-30\% central PbPb collisions with $\sqrt{s_{\rm NN}}=5.02$~TeV to the same in pp collisions.}
    \label{fig:girthxj0p2}
\end{figure*}
\begin{figure*}
    \centering
    \includegraphics[width=0.49\textwidth]{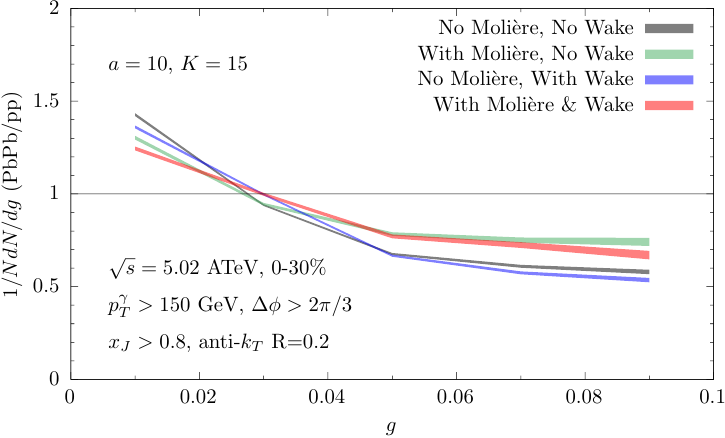}
    \includegraphics[width=0.49\textwidth]{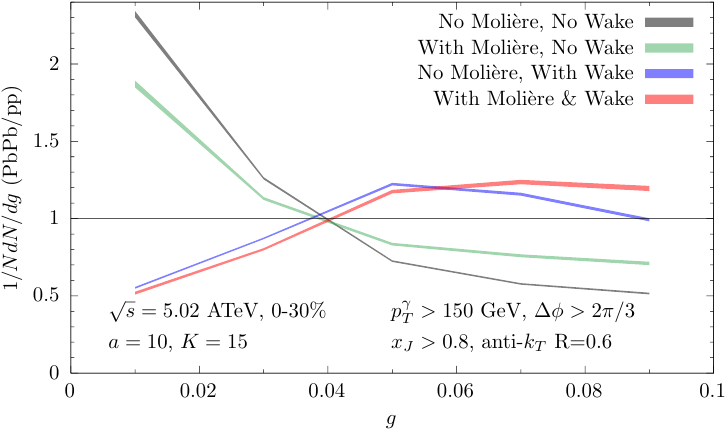}
    \caption{Same as Fig.~\ref{fig:girthxj0p2}, here for jets with $x_J>0.8$: PbPb/pp ratios of the distributions of the jet girth $g$ of $R=0.2$ (left) and $R=0.6$ (right) jets with $x_J > 0.8$ in photon-jet events with $p_T^\gamma>150$~GeV and $\Delta\phi>2\pi/3$.}
    \label{fig:girthxj0p8}
\end{figure*}

A complementary way to quantify jet broadening in photon-tagged events is using the jet girth observable, $g = \int_0^R dr \, r \, \rho(r)$, which is simply the first moment of the jet shape. In practice, it is calculated as $g = \sum_i (p_{T,i}/p_T^{\rm jet})\, r_i$, where $r_i$ denotes the angular distance of jet constituent $i$ from the jet axis in $(\eta, \phi)$ space.
Unlike $R_g$, which isolates a single splitting that passes the Soft Drop condition, the girth is sensitive to the full radial energy profile of the jet and therefore captures both semi-hard prongs and softer medium-induced radiation.
In our calculation of $g$, we include all constituents of each jet without restricting the transverse momenta of the constituents, as appropriate for a calorimetric measurement.
Fig.~\ref{fig:girthxj0p2} shows the PbPb/pp ratio of $(1/N)\, dN/dg$ for photon-tagged jets with $x_J > 0.2$, for $R=0.2$ (left panel) and $R=0.6$ (right panel) jets. Fig.~\ref{fig:girthxj0p8} presents the same jet girth observable in photon-tagged jets, but with the stricter jet selection $x_J>0.8$. 
Comparing these two Figures to 
Figs.~\ref{fig:rgxj0p2} and \ref{fig:rgxj0p8} makes it apparent
that, although the results differ in detail, our Hybrid Model calculations of the jet girth observable are telling the same story that we have 
gleaned from our results for the Soft Drop angle $R_g$.

We begin with the left panel of Fig.~\ref{fig:girthxj0p2}.
For $R=0.2$ jets with $x_J>0.2$, excluding Moli\`ere scatterings leads to a suppression of large-$g$ configurations, consistent with the 
by-now-familiar narrowing of the distribution induced by selection bias and energy transported outside the narrow jet cone. 
Once Moli\`ere scatterings are included, however, the suppression at larger $g$ is essentially completely eliminated, and the PbPb/pp ratio approaches or slightly exceeds unity at large values of $g$. 
This reflects the fact that rare hard Moli\`ere scatterings can redistribute energy to larger angles within the jet cone, increasing the jet's girth.  In the left panel of Fig.~\ref{fig:girthxj0p8} we see 
a suppression in the $g$ distribution at larger $g$ for photon-tagged jets in PbPb collisions regardless of whether Moli\`ere scatterings are included or excluded --- reflecting the stronger effects of selection bias due to energy loss that are imposed when we restrict the jet sample to jets with $x_J>0.8$.  By comparing the red and green bands to the blue and black bands in the left panels of both Figs.~\ref{fig:girthxj0p2} and \ref{fig:girthxj0p8}, we can see that
for jets with $R=0.2$ the increase in jet girth $g$ arising from Moli\`ere scattering dominates over that arising from jet wakes, but it is only for the selection of jets with $x_J>0.2$
in Fig.~\ref{fig:girthxj0p2} that the effect of Moli\`ere scattering suffices to compensate for the opposite effect arising from selection bias.

One may wonder why turning the wake on/off made more of a difference for the jet shape observable in Fig.~\ref{fig:ffshap} than it does for the jet girth observable in the left panels of Figs.~\ref{fig:girthxj0p2} and \ref{fig:girthxj0p8}. 
After all, the jet girth is the first moment of the jet shape $\rho(r)$. 
If we examine Fig.~\ref{fig:ffshap}, though, we notice that including/excluding the wake only makes a difference to the jet shape  for values of $r > 0.2$.  And, the left panels of Figs.~\ref{fig:girthxj0p2} and \ref{fig:girthxj0p8}
depict results for jets with $R=0.2$. We conclude that it is the narrowness of these jets that is responsible for the effects of jet wakes
on the girth distribution being negligible relative to the effects of Moli\`ere scattering.

And indeed, when we look at jets with $R=0.6$ in the right panels of 
Figs.~\ref{fig:girthxj0p2} and \ref{fig:girthxj0p8}
we see a different story. For wider jets with $R=0.6$, in both the $x_J>0.2$ sample and the $x_J>0.8$ sample the effects of jet wakes dominate over the effects of Moli\`ere scattering.
When wakes are included, there is a pronounced enhancement of the PbPb/pp ratio above one for $g \gtrsim 0.04$, since the wake deposits 
energy and momentum in the form of soft hadrons
at sizable angles relative to the jet axis
that remain within the wide $R=0.6$ jet cone.
Including Moli\`ere scatterings on top of this further enhances the jet girth, as hard recoils and their wakes contribute additional weight at large angles, but for $R=0.6$ jets this is a small effect compared to that of the jet wakes.


The one place where our Hybrid Model calculations of the girth observable tell a different story than that told by our calculations of the Soft Drop splitting angle $R_g$ observable is in the right panel of Fig.~\ref{fig:girthxj0p8}. 
Here, unlike in the right panel of Fig.~\ref{fig:rgxj0p8},
even with the stricter requirement $x_J>0.8$ we still observe a clear enhancement of the PbPb/pp ratio above one at $g \gtrsim 0.04$ once jet wakes are included.  This reflects the fact that $R_g$ is a groomed observable, in which the impact of the wake is reduced by grooming away soft jet constituents, whereas the girth includes all constituents weighted by their radial distance from the jet axis.  Even though restricting the selection of jets to $x_J>0.8$ eliminates most jets that have lost energy, enough remain in this wide jet sample that their wakes still have a significant effect on the girth distribution at $g \gtrsim 0.04$.


Just as in the case of the $R_g/R$ observable,
if measurements of the girth of jets with $R=0.2$ selected with $x_J>0.2$ and $x_J>0.8$  in experimental data are aligned with what we have seen in the red bands in the left panels of Figs.~\ref{fig:girthxj0p2}, \ref{fig:girthxj0p8} and \ref{fig:xJ} this would in concert constitute strong evidence for Moli\`ere scattering. 
Measuring a PbPb/pp ratio of the girth $g$ 
around unity for jets with $x_J>0.2$, and confirming that there is still some degree of suppression of this ratio arising from selection bias by measuring the $x_J$ distribution 
and by confirming the effects of selection bias by seeing the PbPb/pp ratio drop substantially when only jets with $x_J>0.8$ are included would
constitute strong evidence for hard scattering of jet partons off quasiparticles in the QGP.

Here, the CMS collaboration is {\it very}
close to achieving this goal in their measurements of the girth $g$ in photon-tagged jets in PbPb and pp collisions~\cite{CMS:2024zjn}.
They have selected events with $p_T^\gamma>100$~GeV and have analyzed
jet samples with $x_J>0.4$ and $x_J>0.8$.
For $x_J>0.4$, four of their five data points for the PbPb/ratio are close to unity, and their results show clear evidence for greater 
effects of selection bias in the $x_J>0.8$ sample.
Although they have not shown their $x_J$ distribution which makes it hard to complete the story as in the preceding paragraph, two further aspects of their analysis and their data make the latter particularly tantalizing. 
First, nine of their ten data points are in good agreement with the Hybrid Model predictions 
that CMS showed in Ref.~\cite{CMS:2024zjn}
--- with Moli\`ere scattering included.
And, second, their one data point that is not described by the Hybrid Model predictions, 
namely the PbPb/pp ratio in their highest girth bin with $0.08<g<0.1$ for jets with $x_J>0.2$,
is above the Hybrid Model calculation by more than three standard deviations and is more than two standard deviations above unity.
This is of course only a single data point, and its large value may represent a statistical fluctuation.  However, the only physical effect that we are aware of that can push the PbPb/pp ratio for girth (or $R_g$) in jets with a cone size as small as $R=0.2$ above unity is
hard scattering of jet partons off quasiparticles in the QGP.  This makes this single data point a tantalizing data point.  The whole suite of CMS measurements of both $g$ and $R_g$ in Ref.~\cite{CMS:2024zjn} motivate: (i) using these data in a future Bayesian analysis of multiple data sets to constrain the values of all four
Hybrid Model parameters, including the two that tell us about Moli\`ere scattering;
(ii) repeating these measurements with higher statistics; and (iii) repeating these measurements 
for jets produced in events with photons with 
$p_T^\gamma>150$~GeV with $x_J>0.2$, as in the Hybrid Model calculations that we have shown here.
The very high CMS data point~\cite{CMS:2024zjn} for the PbPb/pp girth ratio for $0.08<g<0.1$ in jets with $x_J>0.4$ only adds to  these three motivations, as this enhancement, albeit at present only in a single bin of $g$, is consistent with the expectation that hard (Moli\`ere) scattering of jet partons off QGP quasiparticles increases the probability for hard constituents to appear at large angles within the jet.



\section{Concluding Remarks and a Look Ahead}
\label{sec:conclusions}

In this work, we have incorporated perturbative, high momentum-transfer $2 \rightarrow 2$ Moli\`ere scatterings between energetic jet partons and thermal quasiparticles into the Hybrid Model of jet quenching. Prior to this study, the Hybrid Model already included strongly coupled longitudinal energy loss, Gaussian transverse momentum broadening, and the hydrodynamic response of the medium to the passage of a jet in the form of a jet wake. By embedding rare, hard, elastic processes into this framework, we have constructed a description in which soft, strongly coupled interactions and rare, perturbative scatterings coexist within the same dynamical framework.

We began in Section~\ref{sec:moliere} by reviewing perturbative calculations of Moli\`ere scattering for energetic partons traversing a droplet of QGP and generalized earlier results~\cite{DEramo:2018eoy} by tracking both outgoing partons from the $2 \rightarrow 2$ process, rather than only the scattered jet parton. This extension is necessary for a realistic implementation in a jet-quenching framework, since both outgoing scattered partons (the one from the jet and the one from the QGP) continue to propagate through the medium, lose energy to it, excite wakes in the QGP, and may scatter again. In Section~\ref{sec:phase-space-constraints-and-sampling} we identified the region of phase space in which 
$2\rightarrow 2$ elastic scattering
involves high momentum transfer 
making a weakly coupled 
description valid. We then 
constructed the probability distributions for the outgoing parton kinematics subject to the constraints $|t|,|u|>a m_D^2$. 
Enforcing these constraints ensures that we are only including high momentum transfer scattering,
which serves to restrict the allowed ranges of scattering angles and outgoing momenta. 
We provide full details of our calculation of 
Moli\`ere scattering and the resulting analytic expressions for the matrix elements, 
probability distributions, cumulative 
probabilities, and final state kinematics for these scattering processes in four Appendices.

With all of the groundwork in place, in Sec.~\ref{sec:implementing-in-hybrid} we incorporated these processes into the Hybrid Model by computing, at each timestep of the in-medium shower evolution, the probability that a hard scattering occurs and sampling the corresponding subprocess and final-state kinematics from these distributions, allowing rare Molière scatterings to be embedded dynamically 
throughout the Hybrid Model description of
the evolution of a parton shower within 
an expanding cooling droplet of QGP.
We imposed the conditions on the exchanged momentum of a $2 \rightarrow 2$ elastic 
scattering such that $|t|, |u| > a m_D^2$, 
meaning that in adding Moli\`ere scattering, treated perturbatively, to the Hybrid Model we are only adding scattering processes with high momentum transfer. 
The strongly coupled continuous energy loss $dE/dx$  (governed 
by a parameter $\kappa_{\rm sc}$)
and continuous soft Gaussian transverse momentum broadening
(governed by a parameter $K$), incorporated in the Hybrid Model here as in previous work,
serve together to model the strongly coupled dynamics of both elastic and inelastic soft momentum exchange between jet partons and the 
QGP liquid.
In Section~\ref{sec:implementing-in-hybrid}, we chose values $a=10$ and $K=15$ to ensure a smooth matching between the soft, continuous, 
Gaussian-distributed transverse momentum kicks and the hard, perturbative, momentum kicks from Moli\`ere scattering.

A key objective of this study has been to identify jet observables that are sensitive to the presence of Moli\`ere scatterings and, by extension, to the short-distance quasiparticle structure of the quark-gluon plasma. In Section~\ref{sec:results}, we adjusted the longitudinal energy loss parameter $\kappa_{\rm sc}$ so that Hybrid Model results for
inclusive jet $R_{AA}$ and charged 
hadron $R_{\rm AA}$ 
obtained here, with Moli\`ere 
scatterings included, are in good agreement 
with 
Hybrid Model results without Moli\`ere scattering
that were fit to data.
This ensured that any differences observed in jet substructure 
when we turn Moli\`ere scattering off/on
arise from genuine modifications of the internal structure of jets, rather than from changes in overall suppression. 
We investigated such effects in
two classic ungroomed jet observables, the jet shape and jet fragmentation function,
as well as in observables obtained via two different grooming procedures and observables that focus on the number and angular distribution of subjets within jets. We then focussed particular attention on two observables, one groomed and one ungroomed, namely the 
Soft Drop angle $R_g$ and the jet girth $g$,
for jets produced in association with an isolated hard photon in a sample of events selected based upon the $p_T$  of this photon.

We found that Moli\`ere scatterings generate potentially distinctive signatures in a myriad of differential jet substructure observables. In particular, observables that probe angular structure at sufficiently large angles --- such as the Soft Drop angle $R_g$, leading-$k_T$ splittings, subjet multiplicities, and angular separations between subjets --- exhibit sensitivity to the presence of rare, large-angle, $2 \rightarrow 2$ elastic scatterings. 
However, we found that in inclusive jet samples (where the events are selected based upon the $p_T$ of the jet therein)
the effects of Moli\`ere scattering 
are
obscured by the effects of jet selection bias  --- namely that the jets which are selected in any specified $p_T$ bin
are those that have lost the least amount of energy, 
which tend to be those jets with fewer and narrower splittings.  This bias towards jets
that survive their traverse of the QGP droplet
most unscathed serves to push the distributions
of all angular structure observables in PbPb collisions
in a direction that favors narrower structures
and suppresses broader structures relative to what is seen in pp collisions.  Both Moli\`ere scattering and jet wakes have the opposite effect. Jet wakes predominantly modify softer and more diffuse components of the jets and enhance the distribution of hadronic energy out to large angles away from the central core of the jet. Moli\`ere scattering serves in addition to broaden the angular distribution of semi-hard structures as measured via the Soft Drop splitting angle $R_g$ and subjets within jets.

When the selection bias due to parton energy loss is largely mitigated, for example in samples of high-$p_T$-photon-tagged jets selected so as to include
jets down to small values of $x_J\equiv p_T^{\rm jet}/p_T^\gamma$, the broadening of the 
Soft Drop splitting angle $R_g$ and jet girth $g$ 
distributions due to Moli\`ere scattering can potentially overcome the narrowing of these distributions due to jet selection bias,
resulting in a net broadening of the Soft Drop splitting angle $R_g$ and the jet girth $g$. 
Our Hybrid Model calculations show this happening for samples of narrow $R=0.2$ jets with $x_J>0.2$
in events in which the photon has $p_T^\gamma>150$~GeV.
Existing CMS measurements~\cite{CMS:2024zjn} of $R_g$ and $g$ of $R = 0.2$ photon-tagged jets recoiling against photons with $p_T^\gamma > 100$ GeV show strong evidence of the narrowing induced by selection bias in jet samples that satisfy $x_J > 0.8$. The fact that their PbPb/pp ratios for these two observables are close to unity in almost all bins of $R_g/R$ and $g$ in jet samples that satisfy $x_J > 0.4$ is tantalizing, with
this ratio even well above unity in one $g$ bin. 
Together with our results, this
motivates relaxing the jet selection criteria  further to $x_J > 0.2$ in future high-statistics measurements 
with higher photon transverse momenta so that one can see the enhancement in the number
of photon-tagged jets with large $R_g / R$ and large $g$ in PbPb collisions relative to pp collisions --- a definitive consequence of  Moli\`ere scattering.


Our Hybrid Model calculations show that measurements of the scaled Soft Drop angle 
$R_g/R$ and jet girth $g$ in photon-tagged jets
that are either skinny ($R=0.2$) or broad ($R=0.6$)
in samples which either include jets that have lost most of their energy ($x_J>0.2$) or exclude such jets ($x_J>0.8$) can reveal a fascinating 
interplay between the effects of Moli\`ere scattering and jet wakes.
In particular, we found that while measurements of the scaled Soft Drop angle $R_g/R$ and jet girth $g$ in samples of skinny $R=0.2$ photon-tagged jets are especially sensitive to the effects of Moli\`ere scattering, the impact of jet wakes on these observables is comparatively modest. However, for photon-tagged jets with the larger radius $R = 0.6$, contributions from jet wakes become increasingly important, as a greater fraction of the soft, wide-angle medium response remains inside the jet cone and can lead to visible enhancements at large $R_g/R$ and large values of the jet girth $g$. Furthermore, we found that these large-angle enhancements due to jet wakes are further amplified when Moli\`ere scatterings are included, because both outgoing partons produced in a Moli\`ere scattering subsequently propagate through the plasma and each excites its own wake.

Taken together, our results demonstrate that Moli\`ere scatterings leave distinctive and measurable signatures in appropriately chosen jet substructure observables in PbPb collisions. 
This motivates a suite of near-term experimental measurements that can provide definitive 
evidence for weakly coupled hard scattering of jet partons off quasiparticles in the QGP.
Our results also provide strong motivation
for a quantitative confrontation between
results obtained from Hybrid Model calculations including Moli\`ere scattering and experimental data, in particular for a Bayesian  quantification of current constraints on 
the
four Hybrid model
parameters $\kappa_{\rm sc}$, $K$, $g_s$ and $a$
that incorporates extant experimental data
from measurements of (at least) all the observables that we have analyzed in this paper.
This is the key next step needed in order to
obtain a quantitative assessment of the effects of Moli\`ere scattering,
%
%
namely to determine, with a quantitative estimate of the
uncertainty, whether existing measurements 
(and, soon, the near-term measurements that
our results motivate)
reveal the presence of weakly coupled Moli\`ere scatterings and to quantitatively constrain
the associated parameters of our model.
Success in this regard would open the door to using future experimental data to quantify important improvements to our model, including an improved description of the properties and 
momentum distributions of QGP quasiparticles (which we have taken to be non-interacting quarks and gluons with thermal distributions in this exploratory study) and the addition of hard weakly coupled processes that go beyond $2\rightarrow 2$.

In large collision systems such as PbPb, the observable impacts of Moli\`ere scattering compete with the observable impacts of energy loss, in particular via the resulting jet selection bias.
In this context it is important to note that energy loss scales steeply with the in-medium path length $L$ traversed by an energetic parton.
The strongly coupled energy loss that we employ in the Hybrid Model with a rate $dE/dx$ 
given in Eq.~\eqref{eq:elossrate}, a form
based upon holographic calculations in strongly coupled quantum field theory
scales parametrically as $L^3$ for $L \ll x_{\rm stop}$, and then grows rapidly as $L$ approaches 
the distance $x_{\rm stop}$ over which a jet parton thermalizes (unless it splits first).
All effects of elastic scattering, including the probabilities for the high momentum transfer Moli\`ere scattering processes that are our focus in this paper, scale linearly with $L$.
This means that the larger the droplet of QGP, the larger the effects of strongly coupled energy loss, including the effects of the consequent selection bias, will be relative to the effects of elastic
Moli\`ere scattering.
This parametric argument suggests that
light-ion collisions, for example oxygen--oxygen (OO) collisions, could provide a particularly promising arena in which to search for signatures of Moli\`ere scattering off quasiparticles in QGP. 
The smaller characteristic size of QGP droplets in OO collisions reduces the available path length, suppressing the parametrically ${\cal O}(L^3)$ effects of strongly coupled energy loss  more severely than the parametrically ${\cal O}(L)$ 
collisional effects,
thereby amplifying the relative importance of the effects of Moli\`ere scattering processes in experimental observables. 
Recent oxygen--oxygen collisions at the LHC and RHIC, with the earliest measurements of hard probes in these systems having already appeared~\cite{Strangmann2025,ATLAS:2025ooe,CMS:2025bta}, 
present 
an exciting opportunity to use oxygen--oxygen collisions as an arena in which to detect and study Moli\`ere scatterings between energetic jet partons and quasiparticles in the medium.

More broadly, the detection of Moli\`ere scatterings in jet substructure observables would mark a significant advance in our understanding of quark-gluon plasma. It would constitute direct evidence that energetic jet partons resolve quark- and gluon-like quasiparticle at sufficiently short distance scales within QGP, an otherwise strongly coupled liquid. Observing such signatures would represent a concrete manifestation of asymptotic freedom in the strongly coupled matter produced in light- and heavy-ion collisions. And, it would open the door to using experimental measurements of 
jet substructure observables to 
systematically probe the microscopic, particulate, structure of hot quark-gluon plasma --- the primordial liquid that filled the universe in the first microseconds after the Big Bang and that is recreated in these collisions.

\acknowledgments

We gratefully acknowledge helpful conversations with Harry Andrews, Cristian Baldenegro, João Barata, Hannah Bossi, Jasmine Brewer, Quinn Brodsky, Jorge Casalderrey-Solana, Leticia Cunqueiro, Francesco D'Eramo, Jean Du Plessis, Rithya Kunnawalkam Elayavalli, Gian Michele Innocenti, Peter Jacobs, Yen-Jie Lee,  Arthur Lin, Xoan Mayo Lopez, Guilherme Milhano, Matthew Nguyen, Jaime Norman, Molly Park, Gunther Roland, Andrey Sadofyev, Bruno Scheihing-Hitschfeld,  Alba Soto-Ontoso, Rachel Steinhorst, Adam Takacs, Marco van Leeuwen, Xiaojun Yao and Yi Yin.
Research supported in part by the U.S.~Department of Energy, Office of Science, Office of Nuclear Physics under grant Contract Number DE-SC0011090.
DP acknowledges support from the Ram\'on y
Cajal fellowship RYC2023-044989-I. ASK is supported by the National Science Foundation Graduate Research Fellowship Program under Grant No. 2141064.

\appendix

\section{Decomposition of Subprocesses} \label{app:processes}

Recall from Eqs.~\eqref{eq:lin-comb-7} and \eqref{eq:seven-ms} from Sec.~\ref{sec:decomposition}, which we repeat here, that the $2 \rightarrow 2$ elastic scattering amplitudes $|\mathcal{M}^{(n)}({t}, {u})|^2$ and $|\mathcal{M}^{(n)}({u}, {t})|^2$ can be decomposed into a linear combination of 7 terms $m_i$
\begin{equation}
    \frac{1}{g_s^4} \left| \mathcal{M}^{(n)}({t}, {u}) \right|^2 
    = \sum_i c_i^{(n)} m_i({t}, {u}), \qquad \frac{1}{g_s^4} \left| \mathcal{M}^{(\tilde{n})}({u}, {t}) \right|^2 
    = \sum_i \tilde{c}_i^{(\tilde{n})} m_i({t}, {u}),
\end{equation}
where the 7 terms $m_i$ in question are
\begin{equation}
\begin{split}
m_1 = \left( \frac{{s}}{{t}} \right)^2, \quad
m_2 = -\left( \frac{{s}}{{t}} \right), \quad
m_3 = 1, \quad
m_4 = -\left( \frac{{t}}{{s}} \right), \quad
m_5 = \left( \frac{{t}}{{s}} \right)^2, \\
m_6 = -\left( \frac{{t}}{{s} + {t}} \right) 
= \frac{{t}}{{u}}, \quad
m_7 = \left( \frac{{t}}{{s} + {t}} \right)^2 
= \left( \frac{{t}}{{u}} \right)^2.
\end{split}
\end{equation}
Using Table~\ref{tab:QCDprocesses}, we determine that each $c_i^{(n)}$ is given by the entry in the $n$-th row and $i$-th column of the matrix
\begin{equation}
\mathbf{C} = \begin{pmatrix}
C_1 &\hspace{10pt} C_2 - C_1 &\hspace{10pt} C_1+ C_2 &\hspace{10pt} 0 &\hspace{10pt} 0 &\hspace{10pt} C_1 + C_2 &\hspace{10pt} C_1 \\
C_1 &\hspace{10pt} C_2 - C_1 &\hspace{10pt} C_1+ C_2 &\hspace{10pt} 0 &\hspace{10pt} 0 &\hspace{10pt} C_1 + C_2 &\hspace{10pt} C_1 \\
C_1 &\hspace{10pt} -(C_1 + C_2) &\hspace{10pt} C_1 + 2C_2 &\hspace{10pt} -(C_1 + C_2) &\hspace{10pt} C_1 &\hspace{10pt} 0 &\hspace{10pt} 0 \\
C_1 &\hspace{10pt} -C_1 &\hspace{10pt} C_1/2 &\hspace{10pt} 0 &\hspace{10pt} 0 &\hspace{10pt} 0 &\hspace{10pt} 0 \\
C_1 &\hspace{10pt} -C_1 &\hspace{10pt} C_1/2 &\hspace{10pt} 0 &\hspace{10pt} 0 &\hspace{10pt} 0 &\hspace{10pt} 0 \\
C_1 &\hspace{10pt} -C_1 &\hspace{10pt} C_1/2 &\hspace{10pt} 0 &\hspace{10pt} 0 &\hspace{10pt} 0 &\hspace{10pt} 0 \\
0 &\hspace{10pt} 0 &\hspace{10pt} C_1/2 &\hspace{10pt} -C_1 &\hspace{10pt} C_1 &\hspace{10pt} 0 &\hspace{10pt} 0 \\
0 &\hspace{10pt} C_3 &\hspace{10pt} -(C_3 + C_4) &\hspace{10pt} 2C_4 &\hspace{10pt} -2C_4 &\hspace{10pt} C_3 &\hspace{10pt} 0 \\
2C_4 &\hspace{10pt} -2C_4 &\hspace{10pt} 2C_3 + C_4 &\hspace{10pt} -C_3 &\hspace{10pt} 0 &\hspace{10pt} C_3 &\hspace{10pt} 0 \\
2C_4 &\hspace{10pt} -2C_4 &\hspace{10pt} 2C_3 + C_4 &\hspace{10pt} -C_3 &\hspace{10pt} 0 &\hspace{10pt} C_3 &\hspace{10pt} 0 \\
2C_5 &\hspace{10pt} -2C_5 &\hspace{10pt} 4C_5 &\hspace{10pt} -2C_5 &\hspace{10pt} 2C_5 &\hspace{10pt} 2C_5 &\hspace{10pt} 2C_5
\end{pmatrix},
\end{equation}
where
\begin{subequations}\label{eq:something}
    \begin{align}
    C_1&=16(d_FC_F)^2/d_A\,,\\
    C_2&=2C_3-C_4\,,\\
     C_3&=8d_FC_F^2\,,\\
    C_4&=8d_FC_FC_A\,,\\
    C_5&=8d_AC_A^2.
    \end{align}
\end{subequations}
Swapping ${t}$ and ${u}$, one can calculate that each $\tilde{c}_i^{(\tilde{n})}$ is given by the entry in the $\tilde{n}$-th row and $i$-th column of the matrix
\begin{equation}
\mathbf{\tilde{C}} = \begin{pmatrix}
C_1 &\hspace{10pt} C_2 - C_1 &\hspace{10pt} C_1+ C_2 &\hspace{10pt} 0 &\hspace{10pt} 0 &\hspace{10pt} C_1 + C_2 &\hspace{10pt} C_1 \\
C_1 &\hspace{10pt} C_2 - C_1 &\hspace{10pt} C_1+ C_2 &\hspace{10pt} 0 &\hspace{10pt} 0 &\hspace{10pt} C_1 + C_2 &\hspace{10pt} C_1 \\
0 &\hspace{10pt} 0 &\hspace{10pt} C_1 &\hspace{10pt} C_2 - C_1 &\hspace{10pt} C_1 &\hspace{10pt} C_1 - C_2 &\hspace{10pt} C1 \\
0 &\hspace{10pt} 0 &\hspace{10pt} C_1/2 &\hspace{10pt} 0 &\hspace{10pt} 0 &\hspace{10pt} C_1 &\hspace{10pt} C_1 \\
0 &\hspace{10pt} 0 &\hspace{10pt} C_1/2 &\hspace{10pt} 0 &\hspace{10pt} 0 &\hspace{10pt} C_1 &\hspace{10pt} C_1 \\
0 &\hspace{10pt} 0 &\hspace{10pt} C_1/2 &\hspace{10pt} 0 &\hspace{10pt} 0 &\hspace{10pt} C_1 &\hspace{10pt} C_1 \\
0 &\hspace{10pt} 0 &\hspace{10pt} C_1/2 &\hspace{10pt} -C_1 &\hspace{10pt} C_1 &\hspace{10pt} 0 &\hspace{10pt} 0 \\
0 &\hspace{10pt} C_3 &\hspace{10pt} -(C_3 + C_4) &\hspace{10pt} 2C_4 &\hspace{10pt} -2C_4 &\hspace{10pt} C_3 &\hspace{10pt} 0 \\
0 &\hspace{10pt} C_3 &\hspace{10pt} C_4 &\hspace{10pt} C_3 &\hspace{10pt} 0 &\hspace{10pt} 2C_4 &\hspace{10pt} 2C_4 \\
0 &\hspace{10pt} C_3 &\hspace{10pt} C_4 &\hspace{10pt} C_3 &\hspace{10pt} 0 &\hspace{10pt} 2C_4 &\hspace{10pt} 2C_4 \\
2C_5 &\hspace{10pt} -2C_5 &\hspace{10pt} 4C_5 &\hspace{10pt} -2C_5 &\hspace{10pt} 2C_5 &\hspace{10pt} 2C_5 &\hspace{10pt} 2C_5
\end{pmatrix}\ .
\end{equation}

\section{Evaluating $u$ at $k_T = k_T^{\rm min}$} \label{app:uatktmin}
In this Appendix, we derive Eq.~\eqref{eq:uKmin}, which gives the value of $u$ at the kinematic endpoint $k_T = k_T^{\min}$ where the momentum $k_T$  of the thermal parton participating in the elastic scattering takes on its minimum possible value, given the values of $p_{\rm in}$, $\theta$, and $p$. From Eq.~\eqref{eq:rewriting-stu}, the Mandelstam variable $u$ can be written as
\begin{equation}
u = t\,(C_u - D\cos\phi),
\end{equation}
with
\begin{equation}
2q^2 C \equiv (p_{\rm in}+p)(k_T+k_\chi)+q^2,
\qquad
C_u \equiv C-1,
\end{equation}
and
\begin{equation}
2q^2 D \equiv \sqrt{(4p_{\rm in}p+t)(4k_Tk_\chi+t)}.
\end{equation}
Since $k_T^{\min}=(q-\omega)/2$ and, $k_\chi = k_T + \omega$, at $k_T = k_T^{\rm min}$ we have
\begin{equation}
k_\chi\bigg|_{k_T=k_T^{\min}}=\frac{q+\omega}{2},
\qquad
4k^{\rm min}_Tk_\chi\bigg|_{k_T=k_T^{\min}} = q^2-\omega^2.
\end{equation}
Using $t=\omega^2-q^2$, it follows that $4k_Tk_\chi+t = 0$ when $k_T=k_T^{\rm min}$,
and therefore
\begin{equation}
D\big|_{k_T=k_T^{\min}}=0.
\end{equation}

As a result, $u$ becomes independent of $\phi$ at this endpoint: $u(k_T=k_T^{\min}) = t\,C_u$. To evaluate $C_u$, note that at $k_T=k_T^{\min}$ we have $k_T+k_\chi = q$.
Substituting into the definition of $C$ gives $2q^2 C = (p_{\rm in}+p)q + q^2
= q(p_{\rm in}+p+q)$, so that
\begin{equation}
C\bigg|_{k_T=k_T^{\min}} = \frac{p_{\rm in}+p+q}{2q},
\qquad
C_u\bigg|_{k_T=k_T^{\min}} = \frac{p_{\rm in}+p-q}{2q}.
\end{equation}
Using $\omega = p_{\rm in}-p$, the minimum momentum of the thermal parton in the medium can be written as $k_T^{\min}=(q-\omega)/2= (q-p_{\rm in}+p)/2$,
which implies $p-k_T^{\min}= (p_{\rm in}+p-q)/2$.
Therefore,
\begin{equation}
C_u\big|_{k_T=k_T^{\min}} = \frac{p-k_T^{\min}}{q}.
\end{equation}
Combining these results yields
\begin{equation}
u(k_T=k_T^{\min}) = t\,\frac{(p-k_T^{\min})}{q},
\end{equation}
and hence
\begin{equation}
\left|u(k_T=k_T^{\min})\right|
= \frac{|t|\,(p-k_T^{\min})}{q},
\end{equation}
which is Eq.~\eqref{eq:uKmin}.

\section{Kinematics of Elastic $2 \rightarrow 2$  Scattering}
\label{app:kinematics}

In this Appendix, we outline a derivation of the kinematics necessary for a full description of each scattering process, once one has 
chosen values of the four 
variables $(x,\tilde k_\chi^{\rm min},\tilde k, \phi)$ according to the appropriate probability
distribution as described in
Section \ref{sec:sampling}. 
We are considering an incoming massless parton with four–momentum $p_{\text{in}}^\mu=(p_{\text{in}},0,0,p_{\text{in}})$ that scatters elastically off a thermal parton carrying four-momentum $k^\mu  = (k_T, \bm{k}_T)$. 
After the collision, the 
two resulting partons have
momenta $p^\mu=(E,\bm p)$ and $k_\chi^\mu=(E_\chi,\bm k_\chi)$.  A generic $2 \rightarrow 2$ process with four massless particles would thus appear to have $4\times4=16$ real variables.
However, the on-shell conditions for these massless partons remove four independent variables. Overall momentum conservation removes another four. Knowledge of $\vec{p}_{\rm in}$ fixes three variables.  And, an overall rotation of the system about the $\hat z$ axis can be fixed
arbitrarily.
%
Thus, the scattering process can be specified by four variables.  As described in Sections~\ref{sec:moliere} and \ref{sec:phase-space-constraints-and-sampling}, the straightforward choice for these four variables is:
the outgoing momentum $p = |\bm p|$ of one of the partons in the final state, the angle
$\theta$ of this outgoing parton relative to the direction of the incident jet parton,
the transverse momentum exchange $k_T = |\bm k_T|$, and the angle $\phi$ between the two planes identified by the pair of vectors $(\bm{p}, \bm{p} - \bm{p}_{\rm in})$ and $( \bm{p} - \bm{p}_{\rm in}, \bm{k}_T )$.

In this Appendix, we shall describe how the complete kinematics of a $2 \rightarrow 2$ elastic scattering is determined from knowledge of the four variables $p$, $\theta$, $k_T$, and $\phi$. Before we do that, though, we describe how $p$, $\theta$, $k_T$, and $\phi$ themselves are determined from the four variables $x$, $\tilde k_\chi^{\rm min}$, $\tilde k$, and $\phi$ that we sample using the procedure outlined in Sec.~\ref{sec:sampling}. First, we state the obvious: $\phi$ is one of the sampled variables and is unchanged. Second, using Eq.~\eqref{eq:kchi-min} and the facts that $|t| = am_D^2 / x$ and $k_T = k_T^{\rm min} + \tilde{k} T$, we identify
\begin{equation}
    k_T = \frac{|t|}{4 \tilde{k}_\chi^{\rm min} T} + \tilde{k} T = \frac{am_D^2}{4 \, x \, \tilde{k}_\chi^{\rm min} T} + \tilde{k} T.
\end{equation}
Third, using Eq.~\eqref{eq:omega-and-q} and the fact that $p = p_{\rm in} - \omega$, we write
\begin{equation}
    p = p_{\rm in} - \tilde{k}_\chi^{\rm min}T + \frac{|t|}{4\tilde{k}_\chi^{\rm min}T} = p_{\rm in} - \tilde{k}_\chi^{\rm min}T + \frac{am_D^2}{4 \, x \, \tilde{k}_\chi^{\rm min}T}.
\end{equation}
Finally, since $|t|=2p_{\rm in}p(1-\cos\theta)$, we identify
\begin{equation}
    \theta = \arccos\!\left(1-\frac{|t|}{2p_{\rm in}p}\right).
\end{equation}



In the remainder of this Appendix, we will show how, via energy-momentum conservation, knowledge of the four variables $p$, $\theta$, $k_T$, and $\phi$ servies to specify  the complete kinematics of a $2 \rightarrow 2$ elastic scattering process. We work in the frame where $\bm p_{\text{in}}=p_{\text{in}}\,\hat z$ and place $\bm p$ in the $(x,z)$ plane, meaning that the angle $\theta$ between $\bm p_{\text{in}}$ 
and $\bm p$
is an angle in this plane: $\bm p=(p\cos\theta) \hat z+(p\sin\theta) \hat x$. Conservation of momentum then  relates the three components of the recoil momentum ${\bf k}_\chi$ to the three components of the momentum ${\bf k}_T$ of the struck thermal particle:
\begin{align}
k_{\chi x} & = k_{Tx}-p\sin\theta \label{eq:kchixkTx}\\
k_{\chi y} & =k_{Ty} \label{eq:kchiykTy}\\
k_{\chi z} & =k_{Tz}-p\cos\theta+p_{\rm in}\ . \label{eq:kchixkTz}\\
\end{align}
Conservation of energy can then can then be cast as a relationship between $k_{Tz}$ and $k_{Tx}$:
\begin{equation}
k_T(p-p_{\rm in})+k_{Tz}p_{\rm in}+pp_{\rm in}-p(k_{Tz}+p_{\rm in})\cos{\theta}-k_{Tx} p\sin{\theta}=0\ .
\label{eq:consvenergyconstraint}
\end{equation}
We can obtain a second relationship between
$k_{Tz}$ and $k_{Tx}$ by noting that, by definition,
\begin{equation}
\cos(\phi)=\frac{(\bm p\times \bm q)\cdot (\bm k_T\times \bm q)}{|\bm p\times \bm q||\bm k_T \times \bm q|},
\label{eq:cosphi}
\end{equation}
where $\bm q = \bm p - \bm p_{\rm in}$,
and writing 
Eq.~\eqref{eq:cosphi} in the form
\begin{equation}
k_{Tx} p_{\rm in} - k_{Tx} p \cos\theta + k_{Tz} p \sin\theta = \cos\phi \sqrt{k_T^2q^2 - (k_{Tz}p_{\rm in} - k_{Tz}p \cos\theta-k_{Tx} p\sin\theta)^2}.
\label{eq:phidefnconstraint}
\end{equation}
%
%
%
%
%
With one minor caveat, the two equations~\eqref{eq:consvenergyconstraint} and~\eqref{eq:phidefnconstraint}
together with knowledge of $k_T\equiv | {\bf k}_T |$ specifies $k_{Tx}$, $k_{Ty}$ and $k_{Tz}$, which then means that via Eqs.~\eqref{eq:kchixkTx}, \eqref{eq:kchiykTy} and \eqref{eq:kchixkTz} we have also specified $k_{\chi x}$,
$k_{\chi y}$ and $k_{\chi z}$. 
The minor caveat is that
since neither Eq.~\eqref{eq:consvenergyconstraint} nor Eq.~\eqref{eq:phidefnconstraint} depends on $k_{Ty}$, the sign of $k_{Ty}$ is not specified,
but we can choose this sign arbitrarily.
%
Explicit solution of Eqs.~\eqref{eq:consvenergyconstraint} and~\eqref{eq:phidefnconstraint} then yields:
	\begin{equation}
k_{Tx}=2\cos{\phi}\left(p_{\rm in}-p\cos\theta\right)\frac{pp_{\rm in}}{q^2}\sin\frac{\theta}{2}\sqrt{\frac{k_Tk_\chi}{pp_{\rm in}}-\sin^2{\frac{\theta}{2}}}
+\frac{p\sin\theta}{2}\left(1+\frac{k_T^2-k_\chi^2}{q^2}\right)
\end{equation}

	\begin{equation}
k_{Ty}=2\sin\frac{\theta}{2}\sin\phi\,\frac{pp_{\rm in}}{q}\sqrt{\frac{k_Tk_\chi}{pp_{\rm in}}-\sin^2{\frac{\theta}{2}}}
\end{equation}

	\begin{equation}
k_{Tz}=\frac{1}{2}(-p_{\rm in}+p\cos\theta)\left(\frac{k_T^2-k_\chi^2}{q^2}+1\right)+2\cos{\phi}\frac{p^2p_{\rm in}}{q^2}\sin\theta\sin\frac{\theta}{2}\sqrt{\frac{k_Tk_\chi}{pp_{\rm in}}-\sin^2{\frac{\theta}{2}}}\ .
\end{equation}
This completes the specification of the momenta of all four particles involved in the $2\rightarrow 2$ scattering process.



\section{Integrals} \label{app:integrals}

Section~\ref{sec:sampling} outlines the procedure for choosing the kinematic variables $\phi, \tilde{k}, \tilde{k}_\chi^{\rm min}$, and $x$ from 
the appropriate probability distributions via
computing
the conditional distribution functions $P_i$, where $i$ runs from 1 to 7, given in Eqs.~\eqref{eq:x-basis}, \eqref{eq:kchimin-basis}, \eqref{eq:k-basis} and \eqref{eq:phi-basis}, and the associated cumulative distribution functions. 
The procedure requires us to perform successive integrals over each of the kinematic variables. 
The total probability that a Moli\`ere scattering occurs in a timestep $\Delta t$ requires one performing one additional integral over $x$,
as in Eq.~\eqref{eq:four-integrals-to-be-done} and then adding the probabilities with appropriate weights as discussed below that equation.

The purpose of this Appendix is to describe the explicit calculation of these integrals. The arguments of these integrals are $m_i^{D, B}({t}, {u}) |\mathcal{J}|P_u$, where $\mathcal{P}_u = \Theta(|u| - am_D^2)$ enforces the constraint $|u| > am_D^2$ from Eq.~\eqref{eq:constraint2}, $|\mathcal{J}|$ is the Jacobian resulting from the change of variables given in Eq.~\eqref{eq:changeofvariables}, and the $m_i^{D, B}$ are defined in Eq.~\eqref{eq:midb}. There, each $m_i^{D, B}$ is written in terms of functions $m_i$ which depend on the Mandelstam variables:
\begin{equation}\label{initterms}
    \begin{split}
    m_1=\frac{s^2}{t^2}=(C-D\cos\phi)^2\,,
 \quad
    m_2=-\frac{s}{t}=C-D\cos\phi\,,
\quad
m_3=1\,,
\\
\quad
m_4=-\frac{t}{s}=(C-D\cos\phi)^{-1}\,,
\quad
m_5=\frac{t^2}{s^2}=(C-D\cos\phi)^{-2}\,,
\\
\quad
m_6=\frac{t}{u}=(C_u-D\cos\phi)^{-1}\,,
\quad
m_7=\frac{t^2}{u^2}=(C_u-D\cos\phi)^{-2}\, ,
    \end{split}
\end{equation}
with $C$, $D$ and $C_u$ as defined in Eqs.~\eqref{eq:c}, \eqref{eq:cu}, and \eqref{eq:d}. As discussed in Section~\ref{sec:phasespace}, the constraint $|u| > am_D^2$ can be cast in the form \eqref{eq:curly-Pu}, which we repeat here:
\begin{equation}
{\cal P}_u=\Theta(|u|-am_D^2)=\Theta(C_u-x-D\cos\phi)\ .
\label{eq:curly-Pu-App}
\end{equation}
It is useful to decompose this Heaviside function into three regimes of the variable $\tilde{k}$, as follows. There are three kinematic regimes for which $C_u - x - D \cos \phi > 0$ can potentially be satisfied:
\begin{enumerate} [label=(\roman*)]
    \item If $(C_u - x) / D \ge 1$, the inequality is satisfied for all $\phi$ and hence ${\cal P}_u=1$.
    \item If $-1< (C_u - x) / D <1$, the inequality is satisfied only for $\phi\in(\phi_u,2\pi-\phi_u)$ where
    \begin{equation}
    \phi_u \equiv \arccos\!\left(\frac{C_u-x}{D}\right)\in(0,\pi)\,,
    \end{equation}
    and the constraint ${\cal P}_u$ in Eq.~\eqref{eq:curly-Pu-App} is equivalent to
    \begin{equation}
    {\cal P}_u= \Theta(\phi-\phi_u)\Theta(2\pi - \phi_u-\phi)\ .
    \end{equation}
    \item If $ (C_u - x) / D \le -1$, the inequality is never satisfied, meaning that ${\cal P}_u=0$.
\end{enumerate}
The boundaries between (i) and (ii) and between (ii) and (iii) occur when 
$D^2-(C_u-x)^2=0$, which defines two solutions $\tilde{k}^{\pm} \equiv \frac{1}{T} \left(\sqrt{k_u(x+1)}\pm\sqrt{(k_u+q)x}\right)^2$, where we have defined 
$k_u \equiv p - k_T^{\rm min}$, meaning that $\tilde k_u=\tilde k + \tilde p -\tilde k_T$.

When $\tilde{k} > \tilde{k}^+$, ${\cal P}_u = 1$. When $\tilde{k}^- < \tilde{k} < \tilde{k}^+$, $P_u = 1$ if and only if $\cos \phi < (C_u - x)/D$. And if $\tilde{k} < \tilde{k}^-$, then one might expect ${\cal P}_u = 0$. However, this third case is more nuanced than that because it is only valid when $D > 0$. If $D = 0$, then the quantity $(C_u - x)/D$ is ill-defined. $D = 0$ when $\tilde{k} = 0$, which occurs when $k_T = k_T^{\rm min}$. Eq.~\eqref{eq:uKmin} implies that $u(\tilde{k} = 0) = |t| k_u / q$. Thus, $|u(\tilde{k} = 0)| > a m_D^2$ is satisfied if and only if $k_u > qx$. So, when $0 < \tilde{k} < \tilde{k}^-$, ${\cal P}_u = 0$ and when $\tilde{k} = 0$, ${\cal P}_u = \Theta(\tilde{k}_u - \tilde{q}x)$.  (Note that $\tilde k_u<\tilde q x$ for $0 < \tilde{k} < \tilde{k}^-$, whereas when $\tilde k=0$ it is possible that $\tilde k_u>\tilde q x$. This means that in fact ${\cal P}_u = \Theta(\tilde{k}_u - \tilde{q}x)$ for all $\tilde k<\tilde k^-$.)
In light of all of these considerations, 
we see that the constraint $|u|>am_D^2$, namely Eq.~\eqref{eq:curly-Pu-App}, takes different forms in different regimes of $\tilde{k}$ and can be written explicitly in terms of our variables as a sum of Heaviside $\Theta$ functions like so: 
\begin{equation}\label{eq:scriptP}
{\cal P}_u=\Theta(\tilde{k}>\tilde{k}^+)+\Theta\left(\frac{C_u-x}{D}>\cos\phi\right)\Theta(\tilde{k}^+>\tilde{k}>\tilde{k}^-)+\Theta(\tilde{k}_u-\tilde{q}x)\Theta(\tilde{k}^->\tilde{k})\ .
\end{equation}
There are three terms in Eq.~\eqref{eq:scriptP}, which correspond to the three different regimes of the momentum variable $k$. 
We will refer to these three terms as ${\cal P}_{u1},{\cal P}_{u2},{\cal P}_{u3}$ in order from highest to lowest $k$. Thus, ${\cal P}_u={\cal P}_{u1}+{\cal P}_{u2}+{\cal P}_{u3}$.


We can further expand the Heaviside function $\Theta(\tilde{k}_u-\tilde{q}x)=\Theta(|u(k=0)|-am_D^2)$ that arises in ${\cal P}_{u3}$ as
\begin{equation}\label{sqxdef}
\Theta\left(\tilde{k}_u-\tilde{q}x\right)=\Theta\left(k_{\chi}^{\rm min+}>k_{\chi}^{\rm min}>k_{\chi}^{\rm min^-}\right)\Theta\left(\frac{p_{\rm in}^2}{am_D^2}-1>x\right)\Theta\left(p_{\rm in}^2-am_D^2\right),
\end{equation}
where
\begin{equation}
k_{\chi}^{\rm min \pm} \equiv \frac{p_{\rm in}\pm \sqrt{p_{\rm in}^2-am_D^2(1+x)}}{2(1+x)}\ .
\end{equation}

\subsection{\texorpdfstring{$\phi$}{phi} Integrals } \label{app:phi-integrals}

We begin with the conditional probability distributions $P_i(\phi)$ in Eq.~\eqref{eq:phi-basis}, and the cumulative probability distributions obtained by integrating these: $\int_o^\phi d \phi' P_i(\phi')$.
The necessary integrals over $\phi$ are all of the form 
\begin{equation} \label{eq:phicdf}
    A\int \frac{d\phi}{2\pi}(C-D\cos\phi)^n {
    \cal P}_u,
\end{equation}
where $A$ is a placeholder for all factors which do not depend on $\phi$. In this subsection where we focus on doing the $\phi$ integrals, we will omit $A$ from our expressions. We define the following function of $\phi$ and $z$, which will be useful throughout:
\begin{equation}\label{eq:varphi-defn}
    \varphi(\phi,z)=2\pi\left \lfloor \frac{\phi}{2\pi}+\frac{1}{2} \right \rfloor +2{\rm arctan}\left(\sqrt{\frac{z+1}{z-1}}\tan(\frac{\phi}{2})\right)\ .
\end{equation}
Then,
\begin{subequations}\label{eq:phihelp}
    \begin{align}
    \int (C-D\cos(\phi))^{-2} \frac{d\phi}{2\pi} &=\frac{C}{(C^2-D^2)^{3/2}}\bigg(\frac{\varphi(\phi,C/D)}{2\pi}+\frac{D}{2\pi C}\frac{\sqrt{C^2-D^2}\sin(\phi)}{C-D\cos(\phi)}\bigg),
    \\
   \int (C-D\cos(\phi))^{-1} \frac{d\phi}{2\pi} &=\frac{1}{\sqrt{C^2-D^2}}\frac{\varphi(\phi,C/D)}{2\pi},
   \\
   \int \frac{d\phi}{2\pi} &=\frac{\phi}{2\pi},
    \\
	\int C-D\cos(\phi) \frac{d\phi}{2\pi}&=\frac{C\phi-D\sin(\phi)}{2\pi},
    \\
    \int (C-D\cos(\phi))^2 \frac{d\phi}{2\pi}&=\frac{C^2\phi-2CD\sin(\phi)+D^2(\phi/2+\sin(2\phi)/4)}{2\pi}.
    \end{align}
\end{subequations}
Since ${\cal P}_{u1}$ and ${\cal P}_{u3}$ are independent of $\phi$,
\begin{equation}
 \int_0^{\phi}\frac{d\phi'}{2\pi} m_i({\cal P}_{u1}+{\cal P}_{u3})=({\cal P}_{u1}+{\cal P}_{u3})\int_0^{\phi}\frac{d\phi'}{2\pi}m_i
\end{equation}
Defining $\phi_u \equiv \arccos[(C_u-x)/D] \in (0, \pi)$, we have
\begin{equation}
 \int_0^{\phi}\frac{d\phi'}{2\pi} m_i{\cal P}_{u2}=\Theta(\tilde{k}^+>\tilde{k}>\tilde{k}^-)\left(\int_{\phi_u}^{\phi}\Theta(\phi>\phi_u)+\int_{\phi}^{2\pi-\phi_u}\Theta(\phi>2\pi-\phi_u)\right)\frac{d\phi'}{2\pi}m_i
\end{equation}
From the definitions of $C$, $D$, and $C_u$ 
in terms of the kinematic variables, we obtain
\begin{subequations}
 \begin{align}
 \sqrt{C\pm D}=&\frac{\sqrt{(\tilde{k}_u+\tilde{q})(\tilde{k}+\tilde{q})}}{\tilde{q}}\pm\frac{\sqrt{\tilde{k}_u \tilde{k}}}{\tilde{q}};~~~~~~\sqrt{C^2-D^2}=\frac{\tilde{k}+\tilde{k}_u+\tilde{q}}{\tilde{q}}\\
\sqrt{C_u\pm D}=&\left|\frac{\sqrt{\tilde{k}_u(\tilde{k}+\tilde{q})}}{\tilde{q}}\pm\frac{\sqrt{(\tilde{k}_u+\tilde{q})\tilde{k}}}{\tilde{q}}\right|;~~~~\sqrt{C_u^2-D^2}=\frac{|\tilde{k}-\tilde{k}_u|}{\tilde{q}}.
\end{align}
\end{subequations}
Finally, if we define ${{\cal P}}_{u13} \equiv {\cal P}_{u1} + {\cal P}_{u3} = \Theta(k>k^-)+\Theta(k<k^-)\Theta(\tilde{k}_u-\tilde{q}x)$, 
then we may use Eqs.~\eqref{eq:phihelp} to determine every integral of $m_i {\cal P}_u$ over $\phi$,  obtaining 
\begin{subequations}
\begin{align}
 &\int_0^{2\pi}\frac{d\phi'}{2\pi} m_1{\cal P}_u={{\cal P}}_{u13}(C^2+\frac{D^2}{2})-\Theta(\tilde{k}^+>\tilde{k}>\tilde{k}^-)\int_{0}^{\phi_u}\frac{d\phi'}{\pi}(C-D\cos(\phi'))^2,
    \\
 &\int_0^{2\pi}\frac{d\phi'}{2\pi} m_2{\cal P}_u={{\cal P}}_{u13}C-\Theta(\tilde{k}^+>\tilde{k}>\tilde{k}^-)\int_{0}^{\phi_u}\frac{d\phi'}{\pi}(C-D\cos(\phi')),
   \\
 &\int_0^{2\pi}\frac{d\phi'}{2\pi} m_3{\cal P}_u={{\cal P}}_{u13}-\Theta(\tilde{k}^+>\tilde{k}>\tilde{k}^-)\frac{\phi_u}{\pi},
    \\
  &\int_0^{2\pi}\frac{d\phi'}{2\pi} m_4{\cal P}_u={{\cal P}}_{u13}\frac{1}{\sqrt{C^2 - D^2}}-\Theta(\tilde{k}^+>\tilde{k}>\tilde{k}^-)\int_{0}^{\phi_u}\frac{d\phi'}{\pi}(C-D\cos(\phi'))^{-1},
    \\
 &\int_0^{2\pi}\frac{d\phi'}{2\pi} m_5{\cal P}_u={{\cal P}}_{u13}\frac{C}{(C^2 - D^2)^{3/2}}-\Theta(\tilde{k}^+>\tilde{k}>\tilde{k}^-)\int_{0}^{\phi_u}\frac{d\phi'}{\pi}(C-D\cos(\phi'))^{-2},
    \\
 &\int_0^{2\pi}\frac{d\phi'}{2\pi} m_6{\cal P}_u={{\cal P}}_{u13}\frac{1}{\sqrt{C_u^2 - D^2}}-\Theta(\tilde{k}^+>\tilde{k}>\tilde{k}^-)\int_{0}^{\phi_u}\frac{d\phi'}{\pi}(C_u-D\cos(\phi'))^{-1},
    \\
 &\int_0^{2\pi}\frac{d\phi'}{2\pi} m_7{\cal P}_u={{\cal P}}_{u13}\frac{C_u}{(C_u^2 - D^2)^{3/2}}-\Theta(\tilde{k}^+>\tilde{k}>\tilde{k}^-)\int_{0}^{\phi_u}\frac{d\phi'}{\pi}(C_u-D\cos(\phi'))^{-2}.
\end{align}
\label{eq:C11}
\end{subequations} 

The prefactor $A$ in Eq.~\eqref{eq:phicdf} arises from two sources:  
(i) the Jacobian associated with the change of integration variables  
$(p,\theta,k_T,\phi)\to(x,\tilde{k}_{\min}^{\chi},\tilde{k},\phi)$, and  
(ii) the thermal occupation factors $n_D(k_T)[1\pm n_B(k_\chi)]$. In Sec.~\ref{sec:phasespace}, the phase-space measure transforms as
\begin{equation}
\int dp\,d\theta\,dk_T\,d\phi\;P_t P_u
\;\longrightarrow\;
\int_0^1 dx \int_0^{p_{\rm in}} d\tilde{k}_{\min}^{\chi}
\int_0^\infty d\tilde{k}\int_0^{2\pi} d\phi\; P_u\,|J| \, .
\end{equation}
As per Eq.~\eqref{eq:jacobianprefactor}, the Jacobian $|\mathcal{J}|$ is proportional to $\eta/(\tilde{k}_T^{\min}\,x)$, with $\eta^{-1} \equiv (4\pi)^3 \tilde{p}_{\rm in}^2$. Furthermore, the integrand contains the product
\begin{equation}
n_D(k_T)\,[1\pm n_B(k_\chi)], 
\qquad k_\chi = k_T + \omega ,
\end{equation}
with
\begin{equation}
n(k) = \frac{1}{e^{k/T}\mp 1} .
\end{equation}
This can be rewritten algebraically as
\begin{align}
n_D(k_T)\,[1\pm n_B(k_T+\omega)]
&= \frac{1}{e^{k_T/T}\mp_D 1}
\left(1 \pm_B \frac{1}{e^{(k_T+\omega)/T}\mp_B 1}\right) \nonumber\\[4pt]
&= \frac{e^{k_T/T}}
{\big(e^{k_T/T}\mp_D 1\big)\big(e^{k_T/T}\mp_B e^{-\omega/T}\big)} .
\end{align}
Thus, the Jacobian and the thermal weights yield the prefactor
\begin{equation}
A = \eta\,\frac{\tilde{k}_T^{\min}}{x}\;
\frac{e^{k_T/T}}
{\big(e^{k_T/T}\mp_D 1\big)\big(e^{k_T/T}\mp_B e^{-\omega/T}\big)}.
\end{equation}
Therefore, upon restoring the explicit expression for the factor of $A$ from Eq.~\eqref{eq:phicdf} 
means that all the integrals \eqref{eq:C11} take the form
\begin{equation}
\mathcal{I}_\phi^{(i)} \equiv \eta\,\frac{\tilde{k}_T^{\min}}{x}\;
\frac{e^{k_T/T}}
{\big(e^{k_T/T}\mp_D 1\big)\big(e^{k_T/T}\mp_B e^{-\omega/T}\big)}
\left(\int_0^{2\pi}\frac{d\phi'}{2\pi}\, m_i\, P_u\right).
\label{eq:phi-integral-expression}
\end{equation}
Given values for $\tilde k$, $\tilde{k}_\chi^{\rm min}$, and $x$, we can then employ these seven integrals \eqref{eq:phi-integral-expression} (one for each value of $i$) to 
choose the value of $\phi$ according to the appropriate conditional probability distribution using the method
described in Section~\ref{sec:sampling}.

\subsection{\texorpdfstring{$\tilde{k}$}~ Integrals} \label{app:k-integrals}

Next, in order to compute the cumulative distribution for the conditional probability
$P(\tilde k)$ in Eq.~\eqref{eq:k-basis} 
or in order to compute the conditional probability
$P(\tilde k_\chi^{\rm min})$ 
in Eq.~\eqref{eq:kchimin-basis}, we need to write the expression~\eqref{eq:phi-integral-expression} in terms of $\tilde{k}=\tilde{k}_T-\tilde{k}_T^{\rm min}$ and then integrate over $\tilde{k}$, which is equivalent to integrating $\mathcal{I}_\phi^{(i)}(\tilde k_T)$ from Eq.~\eqref{eq:phi-integral-expression} over $\tilde{k}_T$ starting from $\tilde{k}_T^{\rm min}$:
\begin{equation}
\int_0^\infty d\tilde{k}\, 
\mathcal{I}_\phi^{(i)}(\tilde k+\tilde k_T^{\rm min})
=
\int_{\tilde{k}_T^{\rm min}}^\infty d\tilde{k}_T\, \mathcal{I}_\phi^{(i)}(\tilde k_T)\ .
\end{equation}
To this we now turn.
%
Upon
defining
\begin{equation}
	\delta\equiv\pm_De^{(-q+\omega)/(2T)}=\pm_D e^{-k_{T}^{\rm min}/T} \,
\end{equation}
\begin{equation}
	\beta\equiv\pm_Be^{(-q-\omega)/(2T)}=\pm_B e^{-k_{\chi}^{\rm min}/T}.
\end{equation}
we obtain
\begin{subequations}
\label{eq:geometricexpansion}
\begin{align}
    \int_{\tilde{k}_T^{\rm min}}^\infty d\tilde{k}_T \, \mathcal{I}_\phi^{(i)} &= \frac{\eta e^{-\tilde{k}_{T}^{\rm min}}\tilde{k}_T^{\rm min}}{x}\int_0^\infty d\tilde{k}\frac{e^{\tilde{k}}}{(e^{\tilde{k}}-\delta)(e^{\tilde{k}}-\beta)}\bigg(\int_0^{2\pi}\frac{d\phi}{2\pi}m_i{\cal P}_u\bigg)\\
 &=\frac{\eta e^{-\tilde{k}_{T}^{\rm min}}\tilde{k}_T^{\rm min}}{x}\sum_{n=1}^{\infty}\frac{\delta^n-\beta^n}{\delta-\beta}\int_0^\infty d\tilde{k} e^{-n\tilde{k}}\bigg(\int_0^{2\pi}\frac{d\phi}{2\pi}m_i{\cal P}_u\bigg)\ .
 \label{eq:k-main}
 \end{align}
\end{subequations}

\noindent Many of the required integrals take the form 
\begin{equation}
	\int d\tilde{k}\frac{e^{\tilde{k}}(\tilde{k}+\tilde{k}_u+\tilde{q})^j}{(e^{\tilde{k}}-\delta)(e^{\tilde{k}}-\beta)}=\frac{-1}{\delta-\beta}\sum_{n=1}^{\infty}\frac{(\delta e^{\tilde{k}_u+\tilde{q}})^n-(\beta e^{\tilde{k}_u+\tilde{q}})^n}{n^{j+1}}\Gamma(j+1,n(\tilde{k}+\tilde{k}_u+\tilde{q})).
\end{equation}
If we were able to integrate over the whole range, many of these would simplify upon using the polylogarithm identity 
\begin{equation}
\int_{0}^{\infty}d\tilde{k}\frac{e^{\tilde{k}}k^n}{(e^{\tilde{k}}-\delta)(e^{\tilde{k}}-\beta)}=n!\frac{Li_{n+1}(\delta)-Li_{n+1}(\beta)}{\delta-\beta}\ .
\end{equation}
What makes this more complicated is the ${\cal P}_u$ on the right-hand side of 
Eq.~\eqref{eq:k-main}, as
the decomposition ${\cal P}_u = {\cal P}_{u1} + {\cal P}_{u2} + {\cal P}_{u3}$ in Eq.~\eqref{eq:scriptP}  results in 
different expressions in each of the ranges $\tilde{k} < \tilde{k}^-$, $\tilde{k}^- < \tilde{k} < \tilde{k}^+$, and $\tilde{k}^+ < \tilde{k}$, which
we must handle in turn.

We begin by considering 
the range $\tilde{k}^- < \tilde{k} < \tilde{k}^+$, where the integrals in Eq.~\eqref{eq:k-main} are of the form
\begin{equation}
\int_{\tilde{k}^-}^{\tilde{k}} d\tilde{k} e^{-n\tilde{k}}\int_{0}^{\phi_u(\tilde{k})}\frac{d\phi'}{2\pi}(C-D\cos(\phi'))^m,
\end{equation}
where $n$ and $m$ are integers.
%
%
%
The boundaries of the range in $\tilde{k}$ that we are considering are given by the expressions $\tilde k^\pm  = \frac{1}{T} \left(\sqrt{k_u(x+1)}\;\pm\;\sqrt{(k_u+q)\,x}\right)^2$, which can be expanded so as to yield
\begin{equation}
    \tilde k^\pm = \left(\tilde k_u + x(2\tilde k_u+\tilde q)\right) \;\pm\; 2\sqrt{\tilde k_u(\tilde k_u+\tilde q)x(x+1)}.
    \label{eq:kpm_expanded}
\end{equation}
We then define the midpoint $\tilde k_0$ and halfwidth $\Delta_k$ of the interval
$[\tilde k^-,\tilde k^+]$ by
\begin{equation}
\tilde k_0 \equiv \frac{\tilde k^+ + \tilde k^-}{2},\qquad
\Delta_k \equiv \frac{\tilde k^+ - \tilde k^-}{2}.
\label{eq:def_k0_Delta}
\end{equation}
We may then identify
\begin{align}
\tilde k_0
&= \tilde k_u + x(2\tilde k_u+\tilde q),
\label{eq:k0_result}
\\
\Delta_k
&= 2\sqrt{\tilde k_u(\tilde k_u+\tilde q)x(x+1)},
\label{eq:Delta_result}
\end{align}
and rewrite Eq.~\eqref{eq:kpm_expanded} as
\begin{equation}
\tilde k^\pm = \tilde k_0 \pm \Delta_k.
\label{eq:kpm_center_radius}
\end{equation}
Having defined $\Delta_k$ and $\tilde{k}_0$, we can also define an angle $\psi$ by
\begin{equation}
\Delta_k\sin\psi \equiv \sqrt{(\tilde k^+ - \tilde k)(\tilde k - \tilde k^-)},
\qquad
\Delta_k\cos\psi \equiv \tilde k - \tilde k_0\,.
\label{eq:def_psi}
\end{equation}
upon noting that 
\begin{align}
(\Delta_k\sin\psi)^2+(\Delta_k\cos\psi)^2
&=(\tilde k^+ - \tilde k)(\tilde k - \tilde k^-)+(\tilde k-\tilde k_0)^2 \\
&=(\tilde k_0 + \Delta_k - \tilde k)(\tilde k - \tilde k_0 + \Delta_k)+(\tilde k-\tilde k_0)^2 \\
&=\left[\Delta_k^2-(\tilde k-\tilde k_0)^2\right]+(\tilde k-\tilde k_0)^2\\
&=\Delta_k^2.
\end{align}

The following identities, computed using the definitions of $\tilde{k}^\pm$ and $k_u$ from above and $q$ from Eq.~\eqref{eq:qdefn}, will  prove useful:
\begin{subequations}
\begin{align}
\sqrt{\tilde{k}^+\tilde{k}^-}=&\Big|\tilde{k}_u-x\tilde{q}\Big|\\
\sqrt{(\tilde{k}^++\tilde{q})(\tilde{k}^-+\tilde{q})}=&\tilde{k}_u+(x+1)\tilde{q}\\
\sqrt{\tilde{k}^{\pm}+\tilde{q}}=&\sqrt{(\tilde{k}_u+\tilde{q})(x+1)}\pm\sqrt{\tilde{k}_ux}\\
\sqrt{\tilde{k}^{\pm}}=&\Bigg|\sqrt{\tilde{k}_u(x+1)}\pm\sqrt{(\tilde{k}_u+\tilde{q})x}\Bigg|
\end{align}\\
\end{subequations}
Furthermore, we can rewrite $\phi_u = \arccos[(C_u-x)/D]$, identify $D\sin(\phi_u)$, and calculate $d\phi_u / d\psi$ as
\begin{subequations}
\begin{align}
\phi_u=&{\rm arctan}\left(\frac{\tilde{q}x-\tilde{k}_u}{\tilde{k}^+}\tan\frac{\psi}{2}\right)+{\rm arctan}\left(\frac{\tilde{q}x+\tilde{k}_u+\tilde{q}}{\tilde{k}^++\tilde{q}}\tan\frac{\psi}{2}\right)\\
D\sin(\phi_u)=&\sqrt{D^2-(C_u-x)^2}=\frac{1}{\tilde{q}}\sqrt{(\tilde{k}-\tilde{k}^-)(\tilde{k}^+-\tilde{k})}=\frac{\Delta_k}{\tilde{q}}\sin\psi \label{eq:b22b}\\
\frac{d\phi_u}{d\psi}=&\frac{d\phi_u}{d\tilde{k}}\frac{d\tilde{k}}{d\psi}=\frac{1}{2}\left(\frac{x\tilde{q}-\tilde{k}_u}{\tilde{k}}+\frac{x\tilde{q}+\tilde{k}_u+\tilde{q}}{\tilde{k}+\tilde{q}}\right). \label{eq:b22c}
\end{align}
\end{subequations}


Using Eq.~\eqref{eq:b22b}, we note that at $\tilde{k} = \tilde{k}^{\pm}$, $D^2 - (C_u - x)^2 = 0$. That is, $(C_u - x)/D = \pm 1$ at $\tilde{k} = \tilde{k}^{\pm}$. So, $\phi_u = {\rm arccos}(\pm 1)$ at $\tilde{k} = \tilde{k}^\pm$. Thus, $\phi_u|_{\tilde{k}=\tilde{k}^+}=0$. We also note from Eq.~\eqref{eq:scriptP} that when $\tilde{k} = \tilde{k}^-$, we require that $\tilde{q}x > \tilde{k}_u$. Thus,
%
\begin{eqnarray}
\phi_u|_{\tilde{k}=\tilde{k}^-}&=&\arccos \left( {\rm {\rm sgn} } \left(\sqrt{\tilde{k}_u(1+x)}-\sqrt{(\tilde{k}_u+\tilde{q})x}\right) \right)=\arccos \left( {\rm {\rm sgn} } \left(\tilde{k}_u-\tilde{q}x \right) \right)\nonumber\\
&=&\pi\Theta\left(\tilde{q}x-\tilde{k}_u\right)\ .
\end{eqnarray}
We also evaluate
\begin{subequations}
\begin{align}
\varphi(\phi_u,C/D)|_{\tilde{k}=\tilde{k}^+}&= 0 \quad & \varphi(\phi_u,C_u/D)|_{\tilde{k}=\tilde{k}^+}&=0\\
\varphi(\phi_u,C/D)|_{\tilde{k}=\tilde{k}^-}&= \pi\Theta(\tilde{q}x-\tilde{k}_u) \quad & \varphi(\phi_u,C_u/D)|_{\tilde{k}=\tilde{k}^-}&=\pi\Theta(\tilde{q}x-\tilde{k}_u)
\end{align}
\end{subequations}
Upon defining 
\begin{equation}
\Phi \equiv \varphi(\phi_u,C/D)\ ;\quad\Phi_u\equiv \varphi(\phi_u,C_u/D)=\Phi_u^0 {\rm sgn} (\tilde{k}-\tilde{k}_u)\ ,
\end{equation}
in terms of the function $\varphi$ that we defined in Eq.~\eqref{eq:varphi-defn}, we can derive the following expressions that we will also need for computing the integral of $\mathcal{I}$ over $\tilde{k}$ in Eq.~\eqref{eq:geometricexpansion}:
\begin{subequations}
\label{eq:dphidpsi}
\begin{align}
\Phi=&\psi+{\rm arctan}\left(\frac{\tilde{q}x-\tilde{k}_u}{\tilde{k}^+}\tan\frac{\psi}{2}\right)-{\rm arctan}\left(\frac{\tilde{q}x+\tilde{k}_u+\tilde{q}}{\tilde{k}^++\tilde{q}}\tan\frac{\psi}{2}\right),\\
\Phi_u=&(2\psi-\Phi){\rm sgn} (\tilde{k}-\tilde{k}_u)+2\pi\Theta(\tilde{k}_u-\tilde{k}),\\
\frac{d\Phi}{d\psi}=&\frac{d\Phi}{d\tilde{k}}\frac{d\tilde{k}}{d\psi}=\frac{1}{2}\left(2+\frac{x\tilde{q}-\tilde{k}_u}{\tilde{k}}-\frac{x\tilde{q}+\tilde{k}_u+\tilde{q}}{\tilde{k}+\tilde{q}}\right),\\
\frac{d\Phi_u^0}{d\psi}=&\frac{d\Phi_u^0}{d\tilde{k}}\frac{d\tilde{k}}{d\psi}=\frac{1}{2}\left(2-\frac{x\tilde{q}-\tilde{k}_u}{\tilde{k}}+\frac{x\tilde{q}+\tilde{k}_u+\tilde{q}}{\tilde{k}+\tilde{q}}\right).
\end{align}
\end{subequations}

We have almost completed laying the groundwork needed for the evaluation of the 
integral in Eq.~\eqref{eq:geometricexpansion}
for each of the 
seven integrands $m_i \mathcal{P}_u$.
First, though, we need the following integrals, each of which can be obtained via integration by parts after recognizing that $d\tilde{k} = - \Delta_k \sin\psi \, d\psi$:
\begin{subequations}
    \begin{align}
&\int dke^{-n\tilde{k}}\frac{\phi_u}{\pi}=-\frac{e^{-n\tilde{k}}\phi_u}{n\pi}+\frac{1}{n}\int\frac{d\psi}{\pi}\frac{d\phi_u}{d\psi}e^{-n\tilde{k}}
\\
&\int d\tilde{k}e^{-n\tilde{k}}\int_0^{\phi_u}
\frac{d\phi'}{\pi(C-D\cos\phi')}
\\ \nonumber
&\qquad\qquad=\tilde{q}e^{n(\tilde{k}_u+\tilde{q})}\left(-E_1\left(n(\tilde{k}+\tilde{k}_u+\tilde{q})\right)\frac{\Phi}{\pi} + \int\frac{d\psi}{\pi}\frac{d\Phi}{d\psi}E_1\left(n(\tilde{k}+\tilde{k}_u+\tilde{q})\right)\right)
\\
&\int d\tilde{k}e^{-n\tilde{k}}\int_{0}^{\phi_u}
\frac{d\phi'}{\pi(C_u-D\cos\phi')}
\\ \nonumber
&\qquad\qquad=\tilde{q}e^{-n\tilde{k}_u}\left(-E_1\left(n(\tilde{k}-\tilde{k}_u)\right)\frac{\Phi_u}{\pi}+\int\frac{d\psi}{\pi}\frac{d\Phi_u^0}{d\psi}E_1\left(n(\tilde{k}-\tilde{k}_u)\right)\right),
\\ 
&\int d\tilde{k}e^{-n\tilde{k}}\int_0^{\phi_u}(C-D\cos\phi')\frac{d\phi'}{\pi}
\\ \nonumber
&\qquad\qquad=-\frac{\phi_u}{\pi} \left(\frac{h}{\tilde{q}^2n^2}\Gamma(2,n\tilde{k})+\frac{f}{n\tilde{q}}e^{-n\tilde{k}}\right)\\\nonumber
&\qquad\qquad~~~~+\int\frac{d\psi}{\pi}\left(\frac{d\phi_u}{d\psi}\left(\frac{h}{\tilde{q}^2n^2}\Gamma(2,n\tilde{k})+\frac{f}{n\tilde{q}}e^{-n\tilde{k}}\right)+\frac{\Delta_k^2\sin^2\psi}{\tilde{q}}e^{-n\tilde{k}}\right)
\\
&\int d\tilde{k}e^{-n\tilde{k}}\int_0^{\phi_u}(C-D\cos\phi')^2\frac{d\phi'}{\pi}\\\nonumber
&\qquad\qquad=\frac{\phi_u}{\pi}\left(\frac{h^2+2f \tilde{k}_u}{\tilde{q}^4n^3}\Gamma(3,n\tilde{k})+\frac{2f}{\tilde{q}^3n^2}(h + \tilde{k}_u )\Gamma(2,n\tilde{k})+\frac{f^2}{\tilde{q}^2n}e^{-n\tilde{k}}\right)\\\nonumber
&\qquad\qquad~~~~+\frac{1}{\tilde{q}^2n}\int\frac{d\psi}{\pi}\frac{d\phi_u}{d\psi}\left(\frac{h^2+2f \tilde{k}_u}{\tilde{q}^2n^2}\Gamma(3,n\tilde{k})+\frac{2f}{\tilde{q}n}(h + \tilde{k}_u)\Gamma(2,n\tilde{k})+f^2e^{-n\tilde{k}}\right)\\\nonumber
&\qquad\qquad~~~~+\int\frac{d\psi}{\pi}\frac{\Delta_k^2}{2\tilde{q}}\sin^2\psi \, e^{-n\tilde{k}}\left(3\left(\frac{f}{\tilde{q}} + \frac{h}{\tilde{q}^2} \tilde{k} \right)+1+x\right),
\\
&\int d\tilde{k} e^{-n\tilde{k}}\int_0^{\phi_u}
\frac{d\phi'}{\pi(C-D\cos\phi')^{2}}\\\nonumber 
&\qquad\qquad=-\tilde{q}\frac{\Phi}{\pi}e^{n(\tilde{k}_u+\tilde{q})}\left[hn\Gamma\left(-1,n(\tilde{k}+\tilde{k}_u+\tilde{q})\right)-2f \tilde{k}_u n^2\Gamma\left(-2,n(\tilde{k}+\tilde{k}_u+\tilde{q})\right)\right]\\\nonumber
&\qquad\qquad~~~~+\tilde{q}e^{n(\tilde{k}_u+\tilde{q})}\int\frac{d\psi}{\pi}\frac{d\Phi}{d\psi}\left[hn\Gamma\left(-1,n(\tilde{k}+\tilde{k}_u+\tilde{q})\right)-2f \tilde{k}_u n^2\Gamma\left(-2,n(\tilde{k}+\tilde{k}_u+\tilde{q})\right)\right]\\\nonumber
&\qquad\qquad~~~~-\tilde{q}\int\frac{d\psi}{\pi}e^{-n\tilde{k}}\frac{\Delta_k^2\sin^2\psi}{(1+x)(\tilde{k}+\tilde{k}_u+\tilde{q})^2}
\\ 
&\int d\tilde{k}e^{-n\tilde{k}}\int_0^{\phi_u}
\frac{d\phi'}{\pi(C_u-D\cos\phi')^{2}}\\\nonumber
&\qquad\qquad=-\tilde{q}\frac{\Phi_u}{\pi}e^{-n\tilde{k}_u}\left[hn\Gamma\left(-1,n(\tilde{k}-\tilde{k}_u)\right)+2f \tilde{k}_u n^2\Gamma\left(-2,n(\tilde{k}-\tilde{k}_u)\right)\right]\\\nonumber
&\qquad\qquad~~~~+\tilde{q}e^{-n\tilde{k}_u}\int\frac{d\psi}{\pi}\frac{d\Phi_u^0}{d\psi}\left[hn\Gamma\left(-1,n(\tilde{k}-\tilde{k}_u)\right)+2f \tilde{k}_u n^2\Gamma\left(-2,n(\tilde{k}-\tilde{k}_u)\right)\right] \\\nonumber
&\qquad\qquad~~~~-\tilde{q}\int\frac{d\psi}{\pi}e^{-n\tilde{k}}\frac{\Delta_k^2\sin^2\psi}{x(\tilde{k}-\tilde{k}_u)^2},
\end{align}
\end{subequations}
where $h = 2 \tilde{k}_u + \tilde{q}$, $f = \tilde{k}_u + \tilde{q}$, $E_n(x) \equiv \int_1^\infty dt\, e^{-xt}/t^n$ is the generalized exponential integral, and $\Gamma(s,x) \equiv \int_{x}^{\infty} dt \, t^{\,s-1} e^{-t}$ is the upper incomplete gamma function.

%
%
%
We note, again, a simplification when integrating over the full range. For example, the integral of $m_3 \mathcal{P}_u$ evaluates to
%

\begin{align}
&\int_0^{\infty}d\tilde{k}e^{-n\tilde{k}}\int_0^{2\pi}\frac{d\phi'}{2\pi} m_3{\cal P}_u= \left(\int_0^{\tilde{k}^-}+\int_{\tilde{k}^-}^{\tilde{k}^+}+\int_{\tilde{k}^+}^{\infty}\right)d\tilde{k}\, e^{-n\tilde{k}} \int_0^{2\pi}\frac{d\phi'}{2\pi} m_3{\cal P}_u\\\nonumber
&~~~=\left(\int_0^{\tilde{k}^-}+\int_{\tilde{k}^-}^{\tilde{k}^+}+\int_{\tilde{k}^+}^{\infty}\right)d\tilde{k} \, e^{-n\tilde{\tilde{k}}}\left({\cal P}_{u13}-\Theta\left(\tilde{k}^-<\tilde{k}<\tilde{k}^+\right)\frac{\phi_u}{\pi}\right)\\\nonumber
&~~~=\left(\int_0^{\tilde{k}^-}\Theta\left(\tilde{k}_u-\tilde{q}x\right)+\int_{\tilde{k}^+}^{\infty}\right)d\tilde{k}\, e^{-n\tilde{k}}-\int_{\tilde{k}^-}^{\tilde{k}^+}d\tilde{k} \, e^{-n\tilde{k}}\frac{\phi_u}{\pi}\\\nonumber
&~~~=\frac{1}{n}\Theta\left(\tilde{k}_u-\tilde{q}x\right)-\frac{1}{n}\int_{\pi}^{0}\frac{d\psi}{\pi}\frac{d\phi_u}{d\psi}e^{-n\tilde{k}}\\\nonumber
&~~~=\frac{1}{n}\Theta\left(\tilde{k}_u-\tilde{q}x\right)+\frac{1}{n}\int_{0}^{\pi}\frac{d\psi}{\pi}\frac{d\phi_u}{d\psi}e^{-n\tilde{k}},
\end{align}
where in the fourth equality, we have employed integration by parts and a change of integration variable to $\psi$. 
%
%
%

Finally, we obtain explicit expressions for the seven integrals over $\tilde k$ that we need in Section~\ref{sec:phase-space-constraints-and-sampling} in order
to compute the cumulative distribution for the conditional probability
$P(\tilde k)$ in Eq.~\eqref{eq:k-basis} 
or the conditional probability
$P(\tilde k_\chi^{\rm min})$ 
in Eq.~\eqref{eq:kchimin-basis}:
%
\begin{subequations}
    \begin{align}
&\int_0^{\infty}d\tilde{k} \, e^{-n\tilde{k}}\int_0^{2\pi}\frac{d\phi'}{2\pi} m_1{\cal P}_u\\\nonumber
&\qquad=\Theta(\tilde{k}_u-\tilde{q}x)\left(\frac{2(h^2+2f\tilde{k}_u)}{\tilde{q}^4n^3}+\frac{2f}{\tilde{q}^3n^2}(h+\tilde{k}_u)+\frac{f^2}{\tilde{q}^2n}\right)\\\nonumber
&\qquad~~~+\frac{1}{\tilde{q}^2n}\int_0^{\pi}\frac{d\psi}{\pi}\frac{d\phi_u}{d\psi}\left(\frac{h^2+2f \tilde{k}_u}{\tilde{q}^2n^2}\Gamma(3,n\tilde{k})+\frac{2f}{\tilde{q}n}(h + \tilde{k}_u)\Gamma(2,n\tilde{k})+f^2e^{-n\tilde{k}}\right)\\\nonumber
&\qquad~~~+\frac{1}{2\tilde{q}}\int_0^{\pi}\frac{d\psi}{\pi}\Delta_k^2\sin^2\psi \, e^{-n\tilde{k}}\left[3\left(\frac{f}{\tilde{q}} + \frac{h}{\tilde{q}^2} \tilde{k} \right)+1+x\right]\\
&\int_0^{\infty}d\tilde{k} \, e^{-n\tilde{k}}\int_0^{2\pi}\frac{d\phi'}{2\pi} m_2{\cal P}_u\\\nonumber
&\qquad=\Theta(\tilde{k}_u-\tilde{q}x)\left(\frac{h}{\tilde{q}^2n^2}+\frac{f}{n\tilde{q}}\right) \\ \nonumber
&\qquad~~~+\int_0^{\pi}\frac{d\psi}{\pi}\left[\frac{d\phi_u}{d\psi}\left(\frac{h}{\tilde{q}^2n^2}\Gamma(2,n\tilde{k})+\frac{f}{n\tilde{q}}e^{-n\tilde{k}}\right)+\frac{\Delta_k^2\sin^2\psi}{\tilde{q}}e^{-n\tilde{k}}\right] \\
&\int_0^{\infty}d\tilde{k} \, e^{-n\tilde{k}}\int_0^{2\pi}\frac{d\phi'}{2\pi} m_3{\cal P}_u=\frac{1}{n}\Theta\left(\tilde{k}_u-\tilde{q}x\right)+\frac{1}{n}\int_0^{\pi}\frac{d\psi}{\pi}\frac{d\phi_u}{d\psi}e^{-n\tilde{k}}\\
&\int_0^{\infty}d\tilde{k} \, e^{-n\tilde{k}}\int_0^{2\pi}\frac{d\phi'}{2\pi} m_4{\cal P}_u
\\ \nonumber
&\qquad=\tilde{q}e^{n(\tilde{k}_u+\tilde{q})}\left[E_1\left(n(\tilde{k}_u+\tilde{q})\right)\Theta\left(\tilde{k}_u-\tilde{q}x\right)+\int_0^{\pi}\frac{d\psi}{\pi}\frac{d\Phi}{d\psi}E_1\left(n(\tilde{k}+\tilde{k}_u+\tilde{q})\right)\right]\\
&\int_0^{\infty}d\tilde{k} \, e^{-n\tilde{k}}\int_0^{2\pi}\frac{d\phi'}{2\pi} m_5{\cal P}_u \\\nonumber
&\qquad=\Theta(\tilde{k}_u-\tilde{q}x)\,\frac{\tilde{q}\,e^{n(\tilde{k}_u + \tilde{q})}}{(\tilde{k}_u + \tilde{q})} \left[h\,E_{2}\left(n(\tilde{k}_u + \tilde{q})\right) - 2\tilde{k}_u\,E_{3}\left(n(\tilde{k}_u + \tilde{q})\right) \right] \\\nonumber
&\qquad~~~+\tilde{q}e^{n(\tilde{k}_u+\tilde{q})}\int_0^{\pi}\frac{d\psi}{\pi}\frac{d\Phi}{d\psi}\left[hn\Gamma\left(-1,n(\tilde{k}+\tilde{k}_u+\tilde{q})\right)-2f \tilde{k}_u n^2\Gamma\left(-2,n(\tilde{k}+\tilde{k}_u+\tilde{q})\right)\right]\\\nonumber
&\qquad~~~-\tilde{q}\int_0^{\pi}\frac{d\psi}{\pi}e^{-n\tilde{k}}\frac{\Delta_k^2\sin^2\psi}{(1+x)(\tilde{k}+\tilde{k}_u+\tilde{q})^2} \, , \\ 
&\int_0^{\infty}d\tilde{k} \, e^{-n\tilde{k}}\int_0^{2\pi}\frac{d\phi'}{2\pi} m_6{\cal P}_u
\\ \nonumber
&\qquad=\tilde{q}e^{-n\tilde{k}_u}\left[-E_1\left(-n\tilde{k}_u\right)\Theta\left(\tilde{k}_u-\tilde{q}x\right)+\int_0^{\pi}\frac{d\psi}{\pi}\frac{d\Phi_u^0}{d\psi}E_1\left(n(\tilde{k}-\tilde{k}_u)\right)\right]\ , \\
&\int_0^{\infty}d\tilde{k} \, e^{-n\tilde{k}}\int_0^{2\pi}\frac{d\phi'}{2\pi} m_7 {\cal P}_u \\\nonumber
&\qquad=\Theta(\tilde{k}_u-\tilde{q}x)\,\tilde{q}\left[ 3+\frac{2\tilde{q}}{\tilde{k}_u}+n(\tilde{q}+\tilde{k}_u)+e^{-n\tilde{k}_u}E_1(-n\tilde{k}_u)\left(n(\tilde{q}+2\tilde{k}_u)+n^2\tilde{k}_u(\tilde{q}+\tilde{k}_u)\right) \right] \\\nonumber
&\qquad~~~+\tilde{q}e^{-n\tilde{k}_u}\int_0^{\pi}\frac{d\psi}{\pi}\frac{d\Phi_u^0}{d\psi}\left[hn\Gamma\left(-1,n(\tilde{k}-\tilde{k}_u)\right)+2f \tilde{k}_u n^2\Gamma\left(-2,n(\tilde{k}-\tilde{k}_u)\right)\right] \\\nonumber
&\qquad~~~-\tilde{q}\int_0^{\pi}\frac{d\psi}{\pi}e^{-n\tilde{k}}\frac{\Delta_k^2\sin^2\psi}{x(\tilde{k}-\tilde{k}_u)^2}.
\end{align}
\end{subequations}
The integrals over $\psi$ in the expressions above can be evaluated using Eq.~\eqref{eq:b22c}, Eqs.~\eqref{eq:dphidpsi}, and upon noting the following expansions
\begin{subequations}
    \begin{align}
e^{-n\Delta_k\cos\psi}=&\sum_{r=-\infty}^{\infty}I_r(n\Delta_k)(-1)^r \cos(r\psi),\\
\frac{\sqrt{a^2-1}}{a+\cos\psi}=&1+2\sum_{n=1}^\infty \left(\sqrt{a^2-1}-a\right)^n\cos(n\psi),
    \end{align}
\end{subequations}
which admit integrals 
\begin{subequations}
\begin{align}
\frac{1}{\pi}\int_0^{\pi}d\psi \, e^{-n\Delta_k\cos\psi}=&I_0(n\Delta_k),\\
\frac{1}{\pi}\int_0^{\pi}d\psi \, \frac{\sqrt{a^2-1}}{a+\cos\psi}e^{-n\Delta_k\cos\psi} = &I_0(n\Delta_k)+2\sum_{r=1}^\infty(-1)^r \left(\sqrt{a^2-1}-a\right)^rI_r(n\Delta_k)\,,
\end{align}
\end{subequations}
where $I_r$ is the $r$'th modified Bessel function of the first kind.

\subsection{$k_{\chi}^{\rm min}$ and $x$ 
Integrals} \label{app:numerical-integrals}

Next, in order to compute the cumulative distribution for the conditional probability
$P(\tilde k_\chi^{\rm min})$ in 
Eq.~\eqref{eq:kchimin-basis} 
which we reproduce here
\begin{equation} \label{eq:kchimin-basis-again}
    P_i(\tilde{k}_{\chi}^{\rm min}|x) \propto \int_0^{\infty} d\tilde{k}\int_0^{2\pi} d\phi \, m_i^{D,B}({t}, {u}) |\mathcal{J}| {\cal P}_u \ ,
\end{equation} 
or in order to compute the conditional probability
$P(x)$ 
in Eq.~\eqref{eq:kchimin-basis}, reproduced here
\begin{equation} \label{eq:x-basis-again}
    P_i(x | ab \leftrightarrow cd) \propto \int_0^{p_{\rm in}}d\tilde{k}_{\chi}^{\rm min}\int_0^{\infty} d\tilde{k}\int_0^{2\pi} d\phi \, m_i^{D,B}({t}, {u}) |\mathcal{J}| {\cal P}_u\ ,
\end{equation}
we need to integrate the expressions that we have obtained in Appendix~\ref{app:k-integrals} over $\tilde k_\chi^{\rm min}$.
And then the last step, needed in order
to obtain the cumulative distribution for the
conditional probability $P(x)$ and the total probability for elastic scattering, is to 
integrate over $x$.  We perform these integrals numerically.



We perform the  integral over $k_{\chi}^{\min}$ using GSL's QAGS algorithm~\cite{GSL,Galassi:2019czg, Piessens1983}. 
%
Direct numerical evaluation of the $x$ integral after numerically integrating over $\tilde k_\chi^{\rm min}$ would be computationally expensive,
and so we use a precomputed, finely grained table of cumulative integrals in $x$, binned in $p_{\rm in}/T$. We cover $p_{\rm in}/T \in [0,1500]$ with $N_{p_{\rm in}}=100$ uniformly-sized bins. For an arbitrary $p_{\rm in}/T$ at runtime we linearly interpolate between the two nearest bins. For each $p_{\rm in}/T$ bin and each $(n,i,D,B)$ channel we tabulate the seven cumulative integrals
\begin{equation}\label{eq:Phi-i}
    \Phi_i(x) \;\equiv\; \int_{0}^{x}\! \mathrm{d}x'\; P_i(x'\,|\,ab\leftrightarrow cd),
\qquad x\in[0,1],
\end{equation}
for a uniformly spaced array of $N_x = 100$ bins in $x \in [0, 1]$. Finally, recalling that $n$ denotes the exponent in the convergent geometric sum in Eq.~\eqref{eq:geometricexpansion}, 
for numerical evaluation we must truncate the sum at some finite $n$, as it is not possible to evaluate the requisite integrals for infinitely many values of $n$. 
We have performed the numerical integrations of the integrals needed to evaluate the sum of the first 17 terms, $1\leq n \leq 17$,
in the convergent geometric sum~\eqref{eq:geometricexpansion}.
For an arbitrary value of $x$ sampled at runtime from $\Phi_i(x)$, we calculate $P_i(\tilde{k}_{\chi}^{\rm min}|x)$ in Eq.~\eqref{eq:kchimin-basis-again} by linearly interpolating between the value of $P_i(\tilde{k}_{\chi}^{\rm min}|x)$ computed for the two nearest bins in $x$.  

Finally, last but not least, the total probability of elastic scattering is obtained by evaluating the seven $\Phi_i(x)$ in Eq.~\eqref{eq:Phi-i} at $x=1$ to obtain the 
seven $P_i$ of Eq.~\eqref{eq:four-integrals-to-be-done}, and then evaluating the appropriate linear combination as described below Eq.~\eqref{eq:four-integrals-to-be-done}.


\bibliography{bibliography}

\providecommand{\href}[2]{#2}\begingroup\raggedright\begin{thebibliography}{100}

\bibitem{PHENIX:2004vcz}
{\bf PHENIX} Collaboration, K.~Adcox et~al., {\it {Formation of dense partonic
  matter in relativistic nucleus-nucleus collisions at RHIC: Experimental
  evaluation by the PHENIX collaboration}},  {\em Nucl. Phys. A} {\bf 757}
  (2005) 184--283, [\href{http://arxiv.org/abs/nucl-ex/0410003}{{\tt
  nucl-ex/0410003}}].

\bibitem{BRAHMS:2004adc}
{\bf BRAHMS} Collaboration, I.~Arsene et~al., {\it {Quark gluon plasma and
  color glass condensate at RHIC? The Perspective from the BRAHMS experiment}},
   {\em Nucl. Phys. A} {\bf 757} (2005) 1--27,
  [\href{http://arxiv.org/abs/nucl-ex/0410020}{{\tt nucl-ex/0410020}}].

\bibitem{PHOBOS:2004zne}
{\bf PHOBOS} Collaboration, B.~B. Back et~al., {\it {The PHOBOS perspective on
  discoveries at RHIC}},  {\em Nucl. Phys. A} {\bf 757} (2005) 28--101,
  [\href{http://arxiv.org/abs/nucl-ex/0410022}{{\tt nucl-ex/0410022}}].

\bibitem{STAR:2005gfr}
{\bf STAR} Collaboration, J.~Adams et~al., {\it {Experimental and theoretical
  challenges in the search for the quark gluon plasma: The STAR Collaboration's
  critical assessment of the evidence from RHIC collisions}},  {\em Nucl. Phys.
  A} {\bf 757} (2005) 102--183,
  [\href{http://arxiv.org/abs/nucl-ex/0501009}{{\tt nucl-ex/0501009}}].

\bibitem{Gyulassy:2004zy}
M.~Gyulassy and L.~McLerran, {\it {New forms of QCD matter discovered at
  RHIC}},  {\em Nucl. Phys. A} {\bf 750} (2005) 30--63,
  [\href{http://arxiv.org/abs/nucl-th/0405013}{{\tt nucl-th/0405013}}].

\bibitem{dEnterria:2009xfs}
D.~d'Enterria, {\it {Jet quenching}},  {\em Landolt-Bornstein} {\bf 23} (2010)
  471, [\href{http://arxiv.org/abs/0902.2011}{{\tt arXiv:0902.2011}}].

\bibitem{Wiedemann:2009sh}
U.~A. Wiedemann, {\it {Jet Quenching in Heavy Ion Collisions}},
  \href{http://arxiv.org/abs/0908.2306}{{\tt arXiv:0908.2306}}.

\bibitem{Majumder:2010qh}
A.~Majumder and M.~Van~Leeuwen, {\it {The Theory and Phenomenology of
  Perturbative QCD Based Jet Quenching}},  {\em Prog. Part. Nucl. Phys.} {\bf
  66} (2011) 41--92, [\href{http://arxiv.org/abs/1002.2206}{{\tt
  arXiv:1002.2206}}].

\bibitem{Jacak:2012dx}
B.~V. Jacak and B.~Muller, {\it {The exploration of hot nuclear matter}},  {\em
  Science} {\bf 337} (2012) 310--314.

\bibitem{Muller:2012zq}
B.~Muller, J.~Schukraft, and B.~Wyslouch, {\it {First Results from Pb+Pb
  collisions at the LHC}},  {\em Ann. Rev. Nucl. Part. Sci.} {\bf 62} (2012)
  361--386, [\href{http://arxiv.org/abs/1202.3233}{{\tt arXiv:1202.3233}}].

\bibitem{Mehtar-Tani:2013pia}
Y.~Mehtar-Tani, J.~G. Milhano, and K.~Tywoniuk, {\it {Jet physics in heavy-ion
  collisions}},  {\em Int. J. Mod. Phys. A} {\bf 28} (2013) 1340013,
  [\href{http://arxiv.org/abs/1302.2579}{{\tt arXiv:1302.2579}}].

\bibitem{Connors:2017ptx}
M.~Connors, C.~Nattrass, R.~Reed, and S.~Salur, {\it {Jet measurements in heavy
  ion physics}},  {\em Rev. Mod. Phys.} {\bf 90} (2018) 025005,
  [\href{http://arxiv.org/abs/1705.01974}{{\tt arXiv:1705.01974}}].

\bibitem{Busza:2018rrf}
W.~Busza, K.~Rajagopal, and W.~van~der Schee, {\it {Heavy Ion Collisions: The
  Big Picture, and the Big Questions}},  {\em Ann. Rev. Nucl. Part. Sci.} {\bf
  68} (2018) 339--376, [\href{http://arxiv.org/abs/1802.04801}{{\tt
  arXiv:1802.04801}}].

\bibitem{Cao:2020wlm}
S.~Cao and X.-N. Wang, {\it {Jet quenching and medium response in high-energy
  heavy-ion collisions: a review}},  {\em Rept. Prog. Phys.} {\bf 84} (2021),
  no.~2 024301, [\href{http://arxiv.org/abs/2002.04028}{{\tt
  arXiv:2002.04028}}].

\bibitem{Cunqueiro:2021wls}
L.~Cunqueiro and A.~M. Sickles, {\it {Studying the QGP with Jets at the LHC and
  RHIC}},  {\em Prog. Part. Nucl. Phys.} {\bf 124} (2022) 103940,
  [\href{http://arxiv.org/abs/2110.14490}{{\tt arXiv:2110.14490}}].

\bibitem{Apolinario:2022vzg}
L.~Apolin\'ario, Y.-J. Lee, and M.~Winn, {\it {Heavy quarks and jets as probes
  of the QGP}},  {\em Prog. Part. Nucl. Phys.} {\bf 127} (2022) 103990,
  [\href{http://arxiv.org/abs/2203.16352}{{\tt arXiv:2203.16352}}].

\bibitem{Wang:2025lct}
X.-N. Wang and U.~A. Wiedemann, {\it {QGP@50: More than Four Decades of Jet
  Quenching}},  8, 2025.
\newblock \href{http://arxiv.org/abs/2508.18794}{{\tt arXiv:2508.18794}}.

\bibitem{Liu:2006ug}
H.~Liu, K.~Rajagopal, and U.~A. Wiedemann, {\it {Calculating the jet quenching
  parameter from AdS/CFT}},  {\em Phys. Rev. Lett.} {\bf 97} (2006) 182301,
  [\href{http://arxiv.org/abs/hep-ph/0605178}{{\tt hep-ph/0605178}}].

\bibitem{Liu:2006he}
H.~Liu, K.~Rajagopal, and U.~A. Wiedemann, {\it {Wilson loops in heavy ion
  collisions and their calculation in AdS/CFT}},  {\em JHEP} {\bf 03} (2007)
  066, [\href{http://arxiv.org/abs/hep-ph/0612168}{{\tt hep-ph/0612168}}].

\bibitem{DEramo:2010wup}
F.~D'Eramo, H.~Liu, and K.~Rajagopal, {\it {Transverse Momentum Broadening and
  the Jet Quenching Parameter, Redux}},  {\em Phys. Rev. D} {\bf 84} (2011)
  065015, [\href{http://arxiv.org/abs/1006.1367}{{\tt arXiv:1006.1367}}].

\bibitem{DEramo:2012uzl}
F.~D'Eramo, M.~Lekaveckas, H.~Liu, and K.~Rajagopal, {\it {Momentum Broadening
  in Weakly Coupled Quark-Gluon Plasma (with a view to finding the
  quasiparticles within liquid quark-gluon plasma)}},  {\em JHEP} {\bf 05}
  (2013) 031, [\href{http://arxiv.org/abs/1211.1922}{{\tt arXiv:1211.1922}}].

\bibitem{Kurkela:2014tla}
A.~Kurkela and U.~A. Wiedemann, {\it {Picturing perturbative parton cascades in
  QCD matter}},  {\em Phys. Lett. B} {\bf 740} (2015) 172--178,
  [\href{http://arxiv.org/abs/1407.0293}{{\tt arXiv:1407.0293}}].

\bibitem{Moliere:1947zza}
G.~Moli\`ere, {\it {Theorie der Streuung schneller geladener Teilchen I.
  Einzelstreuung am abgeschirmten Coulomb-Feld}},  {\em Zeitschrift für
  Naturforschung A} {\bf 2} (1947) 133.

\bibitem{Moliere:1948zz}
G.~Moli\`ere, {\it {Theorie der Streuung schneller geladener Teilchen II.
  Mehrfach- und Vielfachstreuung}},  {\em Zeitschrift für Naturforschung A}
  {\bf 3} (1948) 78--97.

\bibitem{Moliere:1955zz}
G.~Moli\`ere, {\it {Theorie der Streuung schneller geladener Teilchen III. Die
  Vielfachstreuung von Bahnspuren unter Berücksichtigung der statistichen
  Kopplung}},  {\em Zeitschrift für Naturforschung A} {\bf 10} (1955) 177.

\bibitem{Casalderrey-Solana:2014bpa}
J.~Casalderrey-Solana, D.~C. Gulhan, J.~G. Milhano, D.~Pablos, and
  K.~Rajagopal, {\it {A Hybrid Strong/Weak Coupling Approach to Jet
  Quenching}},  {\em JHEP} {\bf 10} (2014) 019,
  [\href{http://arxiv.org/abs/1405.3864}{{\tt arXiv:1405.3864}}]. [Erratum:
  JHEP 09, 175 (2015)].

\bibitem{Casalderrey-Solana:2015vaa}
J.~Casalderrey-Solana, D.~C. Gulhan, J.~G. Milhano, D.~Pablos, and
  K.~Rajagopal, {\it {Predictions for Boson-Jet Observables and Fragmentation
  Function Ratios from a Hybrid Strong/Weak Coupling Model for Jet Quenching}},
   {\em JHEP} {\bf 03} (2016) 053, [\href{http://arxiv.org/abs/1508.00815}{{\tt
  arXiv:1508.00815}}].

\bibitem{Casalderrey-Solana:2016jvj}
J.~Casalderrey-Solana, D.~Gulhan, G.~Milhano, D.~Pablos, and K.~Rajagopal, {\it
  {Angular Structure of Jet Quenching Within a Hybrid Strong/Weak Coupling
  Model}},  {\em JHEP} {\bf 03} (2017) 135,
  [\href{http://arxiv.org/abs/1609.05842}{{\tt arXiv:1609.05842}}].

\bibitem{Hulcher:2017cpt}
Z.~Hulcher, D.~Pablos, and K.~Rajagopal, {\it {Resolution Effects in the Hybrid
  Strong/Weak Coupling Model}},  {\em JHEP} {\bf 03} (2018) 010,
  [\href{http://arxiv.org/abs/1707.05245}{{\tt arXiv:1707.05245}}].

\bibitem{Casalderrey-Solana:2018wrw}
J.~Casalderrey-Solana, Z.~Hulcher, G.~Milhano, D.~Pablos, and K.~Rajagopal,
  {\it {Simultaneous description of hadron and jet suppression in heavy-ion
  collisions}},  {\em Phys. Rev. C} {\bf 99} (2019), no.~5 051901,
  [\href{http://arxiv.org/abs/1808.07386}{{\tt arXiv:1808.07386}}].

\bibitem{Casalderrey-Solana:2019ubu}
J.~Casalderrey-Solana, G.~Milhano, D.~Pablos, and K.~Rajagopal, {\it
  {Modification of Jet Substructure in Heavy Ion Collisions as a Probe of the
  Resolution Length of Quark-Gluon Plasma}},  {\em JHEP} {\bf 01} (2020) 044,
  [\href{http://arxiv.org/abs/1907.11248}{{\tt arXiv:1907.11248}}].

\bibitem{Hulcher:2022kmn}
Z.~Hulcher, D.~Pablos, and K.~Rajagopal, {\it {Sensitivity of Jet Observables
  to the Presence of Quasi-particles in QGP}},  {\em Acta Phys. Polon. Supp.}
  {\bf 16} (2023), no.~1 1--A57, [\href{http://arxiv.org/abs/2208.13593}{{\tt
  arXiv:2208.13593}}].

\bibitem{Bossi:2024qho}
H.~Bossi, A.~S. Kudinoor, I.~Moult, D.~Pablos, A.~Rai, and K.~Rajagopal, {\it
  {Imaging the wakes of jets with energy-energy-energy correlators}},  {\em
  JHEP} {\bf 12} (2024) 073, [\href{http://arxiv.org/abs/2407.13818}{{\tt
  arXiv:2407.13818}}].

\bibitem{Kudinoor:2025ilx}
A.~S. Kudinoor, D.~Pablos, and K.~Rajagopal, {\it {Visualizing how the
  structure of large-radius jets shapes their wakes}},  {\em JHEP} {\bf 01}
  (2026) 020, [\href{http://arxiv.org/abs/2501.18683}{{\tt arXiv:2501.18683}}].

\bibitem{Kudinoor:2025gao}
A.~S. Kudinoor, D.~Pablos, and K.~Rajagopal, {\it {Constraining the Resolution
  Length of Quark-Gluon Plasma with New Jet Substructure Measurements}},
  \href{http://arxiv.org/abs/2509.08881}{{\tt arXiv:2509.08881}}.

\bibitem{Beraudo:2025nvq}
A.~Beraudo, J.~F. Du~Plessis, D.~Pablos, and K.~Rajagopal, {\it {Heavy Quark
  Energy Loss in the Hybrid Model}},
  \href{http://arxiv.org/abs/2510.24847}{{\tt arXiv:2510.24847}}.

\bibitem{Zapp:2008gi}
K.~Zapp, G.~Ingelman, J.~Rathsman, J.~Stachel, and U.~A. Wiedemann, {\it {A
  Monte Carlo Model for 'Jet Quenching'}},  {\em Eur. Phys. J. C} {\bf 60}
  (2009) 617--632, [\href{http://arxiv.org/abs/0804.3568}{{\tt
  arXiv:0804.3568}}].

\bibitem{Zapp:2013vla}
K.~C. Zapp, {\it {JEWEL 2.0.0: directions for use}},  {\em Eur. Phys. J. C}
  {\bf 74} (2014), no.~2 2762, [\href{http://arxiv.org/abs/1311.0048}{{\tt
  arXiv:1311.0048}}].

\bibitem{He:2015pra}
Y.~He, T.~Luo, X.-N. Wang, and Y.~Zhu, {\it {Linear Boltzmann Transport for Jet
  Propagation in the Quark-Gluon Plasma: Elastic Processes and Medium Recoil}},
   {\em Phys. Rev. C} {\bf 91} (2015) 054908,
  [\href{http://arxiv.org/abs/1503.03313}{{\tt arXiv:1503.03313}}]. [Erratum:
  Phys.Rev.C 97, 019902 (2018)].

\bibitem{Milhano:2017nzm}
G.~Milhano, U.~A. Wiedemann, and K.~C. Zapp, {\it {Sensitivity of jet
  substructure to jet-induced medium response}},  {\em Phys. Lett. B} {\bf 779}
  (2018) 409--413, [\href{http://arxiv.org/abs/1707.04142}{{\tt
  arXiv:1707.04142}}].

\bibitem{Park:2018acg}
C.~Park, S.~Jeon, and C.~Gale, {\it {Jet modification with medium recoil in
  quark-gluon plasma}},  {\em Nucl. Phys. A} {\bf 982} (2019) 643--646,
  [\href{http://arxiv.org/abs/1807.06550}{{\tt arXiv:1807.06550}}].

\bibitem{Ke:2020clc}
W.~Ke and X.-N. Wang, {\it {QGP modification to single inclusive jets in a
  calibrated transport model}},  {\em JHEP} {\bf 05} (2021) 041,
  [\href{http://arxiv.org/abs/2010.13680}{{\tt arXiv:2010.13680}}].

\bibitem{Dai:2020rlu}
T.~Dai, J.-F. Paquet, D.~Teaney, and S.~A. Bass, {\it {Parton energy loss in a
  hard-soft factorized approach}},  {\em Phys. Rev. C} {\bf 105} (2022), no.~3
  034905, [\href{http://arxiv.org/abs/2012.03441}{{\tt arXiv:2012.03441}}].

\bibitem{Tachibana:2025rcx}
Y.~Tachibana et~al., {\it {Effect of recoils on soft-drop-groomed observables
  in {\ensuremath{\gamma}}-tagged jets in a multistage approach}},  {\em Phys.
  Rev. C} {\bf 113} (2026), no.~3 034910,
  [\href{http://arxiv.org/abs/2503.23693}{{\tt arXiv:2503.23693}}].

\bibitem{Jing:2025bwi}
P.~Jing, Y.~Dang, Y.~He, S.~Cao, L.~Yi, and X.-N. Wang, {\it {Emergence of
  thermal recoil jets in high-energy heavy-ion collisions}},
  \href{http://arxiv.org/abs/2512.12715}{{\tt arXiv:2512.12715}}.

\bibitem{Soudi:2025lei}
I.~Soudi and A.~Takacs, {\it {Deriving a parton shower for jet thermalization
  in QCD plasmas}},  \href{http://arxiv.org/abs/2510.25837}{{\tt
  arXiv:2510.25837}}.

\bibitem{Boguslavski:2025ylx}
K.~Boguslavski, F.~Lindenbauer, A.~Mazeliauskas, A.~Takacs, and F.~Zhou, {\it
  {Minijet thermalization and jet transport coefficients in QCD kinetic
  theory}},  \href{http://arxiv.org/abs/2510.25669}{{\tt arXiv:2510.25669}}.

\bibitem{Chesler:2014jva}
P.~M. Chesler and K.~Rajagopal, {\it {Jet quenching in strongly coupled
  plasma}},  {\em Phys. Rev. D} {\bf 90} (2014), no.~2 025033,
  [\href{http://arxiv.org/abs/1402.6756}{{\tt arXiv:1402.6756}}].

\bibitem{Chesler:2015nqz}
P.~M. Chesler and K.~Rajagopal, {\it {On the Evolution of Jet Energy and
  Opening Angle in Strongly Coupled Plasma}},  {\em JHEP} {\bf 05} (2016) 098,
  [\href{http://arxiv.org/abs/1511.07567}{{\tt arXiv:1511.07567}}].

\bibitem{Casalderrey-Solana:2020rsj}
J.~Casalderrey-Solana, J.~G. Milhano, D.~Pablos, K.~Rajagopal, and X.~Yao, {\it
  {Jet Wake from Linearized Hydrodynamics}},  {\em JHEP} {\bf 05} (2021) 230,
  [\href{http://arxiv.org/abs/2010.01140}{{\tt arXiv:2010.01140}}].

\bibitem{DEramo:2018eoy}
F.~D'Eramo, K.~Rajagopal, and Y.~Yin, {\it {Moli\`ere scattering in quark-gluon
  plasma: finding point-like scatterers in a liquid}},  {\em JHEP} {\bf 01}
  (2019) 172, [\href{http://arxiv.org/abs/1808.03250}{{\tt arXiv:1808.03250}}].

\bibitem{Pablos:2024muu}
D.~Pablos and S.~Sanjurjo, {\it {Color coherence effects in dipole-quark
  scattering in the soft limit}},  {\em Phys. Rev. D} {\bf 110} (2024), no.~11
  L111502, [\href{http://arxiv.org/abs/2406.08550}{{\tt arXiv:2406.08550}}].

\bibitem{CMS:2024zjn}
{\bf CMS} Collaboration, A.~Hayrapetyan et~al., {\it {Girth and groomed radius
  of jets recoiling against isolated photons in lead-lead and proton-proton
  collisions at $\sqrt{s_{\rm NN}}=5.02$~TeV}},  {\em Phys. Lett. B} {\bf 861}
  (2025) 139088, [\href{http://arxiv.org/abs/2405.02737}{{\tt
  arXiv:2405.02737}}].

\bibitem{Arnold:2002zm}
P.~B. Arnold, G.~D. Moore, and L.~G. Yaffe, {\it {Effective kinetic theory for
  high temperature gauge theories}},  {\em JHEP} {\bf 01} (2003) 030,
  [\href{http://arxiv.org/abs/hep-ph/0209353}{{\tt hep-ph/0209353}}].

\bibitem{Sjostrand:2014zea}
T.~Sj\"ostrand, S.~Ask, J.~R. Christiansen, R.~Corke, N.~Desai, P.~Ilten,
  S.~Mrenna, S.~Prestel, C.~O. Rasmussen, and P.~Z. Skands, {\it {An
  introduction to PYTHIA 8.2}},  {\em Comput. Phys. Commun.} {\bf 191} (2015)
  159--177, [\href{http://arxiv.org/abs/1410.3012}{{\tt arXiv:1410.3012}}].

\bibitem{Eskola:2009uj}
K.~J. Eskola, H.~Paukkunen, and C.~A. Salgado, {\it {EPS09: A New Generation of
  NLO and LO Nuclear Parton Distribution Functions}},  {\em JHEP} {\bf 04}
  (2009) 065, [\href{http://arxiv.org/abs/0902.4154}{{\tt arXiv:0902.4154}}].

\bibitem{Gubser:2008as}
S.~S. Gubser, D.~R. Gulotta, S.~S. Pufu, and F.~D. Rocha, {\it {Gluon energy
  loss in the gauge-string duality}},  {\em JHEP} {\bf 10} (2008) 052,
  [\href{http://arxiv.org/abs/0803.1470}{{\tt arXiv:0803.1470}}].

\bibitem{PhysRevD.10.186}
F.~Cooper and G.~Frye, {\it Single-particle distribution in the hydrodynamic
  and statistical thermodynamic models of multiparticle production},  {\em
  Phys. Rev. D} {\bf 10} (Jul, 1974) 186--189.

\bibitem{CMS:2025dua}
{\bf CMS} Collaboration, V.~Chekhovsky et~al., {\it {Evidence of medium
  response to hard probes using correlations of Z bosons with hadrons in heavy
  ion collisions}},  \href{http://arxiv.org/abs/2507.09307}{{\tt
  arXiv:2507.09307}}.

\bibitem{Caucal:2021lgf}
P.~Caucal and Y.~Mehtar-Tani, {\it {Anomalous diffusion in QCD matter}},  {\em
  Phys. Rev. D} {\bf 106} (2022), no.~5 L051501,
  [\href{http://arxiv.org/abs/2109.12041}{{\tt arXiv:2109.12041}}].

\bibitem{Caucal:2022fhc}
P.~Caucal and Y.~Mehtar-Tani, {\it {Universality aspects of quantum corrections
  to transverse momentum broadening in QCD media}},  {\em JHEP} {\bf 09} (2022)
  023, [\href{http://arxiv.org/abs/2203.09407}{{\tt arXiv:2203.09407}}].

\bibitem{Barata:2020rdn}
J.~Barata, Y.~Mehtar-Tani, A.~Soto-Ontoso, and K.~Tywoniuk, {\it {Revisiting
  transverse momentum broadening in dense QCD media}},  {\em Phys. Rev. D} {\bf
  104} (2021), no.~5 054047, [\href{http://arxiv.org/abs/2009.13667}{{\tt
  arXiv:2009.13667}}].

\bibitem{Barata:2021wuf}
J.~Barata, Y.~Mehtar-Tani, A.~Soto-Ontoso, and K.~Tywoniuk, {\it
  {Medium-induced radiative kernel with the Improved Opacity Expansion}},  {\em
  JHEP} {\bf 09} (2021) 153, [\href{http://arxiv.org/abs/2106.07402}{{\tt
  arXiv:2106.07402}}].

\bibitem{Armesto:2004pt}
N.~Armesto, C.~A. Salgado, and U.~A. Wiedemann, {\it {Measuring the collective
  flow with jets}},  {\em Phys. Rev. Lett.} {\bf 93} (2004) 242301,
  [\href{http://arxiv.org/abs/hep-ph/0405301}{{\tt hep-ph/0405301}}].

\bibitem{He:2020iow}
Y.~He, L.-G. Pang, and X.-N. Wang, {\it {Gradient Tomography of Jet Quenching
  in Heavy-Ion Collisions}},  {\em Phys. Rev. Lett.} {\bf 125} (2020), no.~12
  122301, [\href{http://arxiv.org/abs/2001.08273}{{\tt arXiv:2001.08273}}].

\bibitem{Barata:2022utc}
J.~Barata, A.~V. Sadofyev, and X.-N. Wang, {\it {Quantum partonic transport in
  QCD matter}},  {\em Phys. Rev. D} {\bf 107} (2023), no.~5 L051503,
  [\href{http://arxiv.org/abs/2210.06519}{{\tt arXiv:2210.06519}}].

\bibitem{Fu:2022idl}
Y.~Fu, J.~Casalderrey-Solana, and X.-N. Wang, {\it {Asymmetric transverse
  momentum broadening in an inhomogeneous medium}},  {\em Phys. Rev. D} {\bf
  107} (2023), no.~5 054038, [\href{http://arxiv.org/abs/2204.05323}{{\tt
  arXiv:2204.05323}}].

\bibitem{Sadofyev:2021ohn}
A.~V. Sadofyev, M.~D. Sievert, and I.~Vitev, {\it {Ab~initio coupling of jets
  to collective flow in the opacity expansion approach}},  {\em Phys. Rev. D}
  {\bf 104} (2021), no.~9 094044, [\href{http://arxiv.org/abs/2104.09513}{{\tt
  arXiv:2104.09513}}].

\bibitem{Barata:2022krd}
J.~Barata, A.~V. Sadofyev, and C.~A. Salgado, {\it {Jet broadening in dense
  inhomogeneous matter}},  {\em Phys. Rev. D} {\bf 105} (2022), no.~11 114010,
  [\href{http://arxiv.org/abs/2202.08847}{{\tt arXiv:2202.08847}}].

\bibitem{Barata:2023zqg}
J.~Barata, J.~G. Milhano, and A.~V. Sadofyev, {\it {Picturing QCD jets in
  anisotropic matter: from jet shapes to energy energy correlators}},  {\em
  Eur. Phys. J. C} {\bf 84} (2024), no.~2 174,
  [\href{http://arxiv.org/abs/2308.01294}{{\tt arXiv:2308.01294}}].

\bibitem{Kuzmin:2023hko}
M.~V. Kuzmin, X.~Mayo~L{\'o}pez, J.~Reiten, and A.~V. Sadofyev, {\it {Jet
  quenching in anisotropic flowing matter}},  {\em Phys. Rev. D} {\bf 109}
  (2024), no.~1 014036, [\href{http://arxiv.org/abs/2309.00683}{{\tt
  arXiv:2309.00683}}].

\bibitem{Kuzmin:2024smy}
M.~V. Kuzmin and X.~Mayo~L{\'o}pez, {\it {Gluon radiation inside a flowing
  medium}},  \href{http://arxiv.org/abs/2406.14628}{{\tt arXiv:2406.14628}}.

\bibitem{Bahder:2024jpa}
J.~Bahder, H.~Rahman, M.~D. Sievert, and I.~Vitev, {\it {Signatures of jet
  drift in quark-gluon plasma hard-probe observables}},  {\em Phys. Rev. Res.}
  {\bf 8} (2026), no.~1 L012016, [\href{http://arxiv.org/abs/2412.05474}{{\tt
  arXiv:2412.05474}}].

\bibitem{Armesto:2009fj}
N.~Armesto, L.~Cunqueiro, and C.~A. Salgado, {\it {Q-PYTHIA: A Medium-modified
  implementation of final state radiation}},  {\em Eur. Phys. J. C} {\bf 63}
  (2009) 679--690, [\href{http://arxiv.org/abs/0907.1014}{{\tt
  arXiv:0907.1014}}].

\bibitem{JETSCAPE:2017eso}
{\bf JETSCAPE} Collaboration, S.~Cao et~al., {\it {Multistage Monte-Carlo
  simulation of jet modification in a static medium}},  {\em Phys. Rev. C} {\bf
  96} (2017), no.~2 024909, [\href{http://arxiv.org/abs/1705.00050}{{\tt
  arXiv:1705.00050}}].

\bibitem{JETSCAPE:2019udz}
{\bf JETSCAPE} Collaboration, A.~Kumar et~al., {\it {JETSCAPE framework: $p+p$
  results}},  {\em Phys. Rev. C} {\bf 102} (2020), no.~5 054906,
  [\href{http://arxiv.org/abs/1910.05481}{{\tt arXiv:1910.05481}}].

\bibitem{JETSCAPE:2024nkj}
{\bf JETSCAPE} Collaboration, C.~Sirimanna et~al., {\it {Hard-photon-triggered
  jets in p-p and A-A collisions}},  {\em Phys. Rev. C} {\bf 111} (2025), no.~6
  064911, [\href{http://arxiv.org/abs/2412.19738}{{\tt arXiv:2412.19738}}].

\bibitem{Cacciari:2008gp}
M.~Cacciari, G.~P. Salam, and G.~Soyez, {\it {The anti-$k_t$ jet clustering
  algorithm}},  {\em JHEP} {\bf 04} (2008) 063,
  [\href{http://arxiv.org/abs/0802.1189}{{\tt arXiv:0802.1189}}].

\bibitem{Cacciari:2011ma}
M.~Cacciari, G.~P. Salam, and G.~Soyez, {\it {FastJet User Manual}},  {\em Eur.
  Phys. J. C} {\bf 72} (2012) 1896, [\href{http://arxiv.org/abs/1111.6097}{{\tt
  arXiv:1111.6097}}].

\bibitem{Milhano:2015mng}
J.~G. Milhano and K.~C. Zapp, {\it {Origins of the di-jet asymmetry in heavy
  ion collisions}},  {\em Eur. Phys. J. C} {\bf 76} (2016), no.~5 288,
  [\href{http://arxiv.org/abs/1512.08107}{{\tt arXiv:1512.08107}}].

\bibitem{Rajagopal:2016uip}
K.~Rajagopal, A.~V. Sadofyev, and W.~van~der Schee, {\it {Evolution of the jet
  opening angle distribution in holographic plasma}},  {\em Phys. Rev. Lett.}
  {\bf 116} (2016), no.~21 211603, [\href{http://arxiv.org/abs/1602.04187}{{\tt
  arXiv:1602.04187}}].

\bibitem{Brewer:2017fqy}
J.~Brewer, K.~Rajagopal, A.~Sadofyev, and W.~Van Der~Schee, {\it {Evolution of
  the Mean Jet Shape and Dijet Asymmetry Distribution of an Ensemble of
  Holographic Jets in Strongly Coupled Plasma}},  {\em JHEP} {\bf 02} (2018)
  015, [\href{http://arxiv.org/abs/1710.03237}{{\tt arXiv:1710.03237}}].

\bibitem{Mehtar-Tani:2017web}
Y.~Mehtar-Tani and K.~Tywoniuk, {\it {Sudakov suppression of jets in QCD
  media}},  {\em Phys. Rev. D} {\bf 98} (2018), no.~5 051501,
  [\href{http://arxiv.org/abs/1707.07361}{{\tt arXiv:1707.07361}}].

\bibitem{Caucal:2019uvr}
P.~Caucal, E.~Iancu, and G.~Soyez, {\it {Deciphering the $z_g$ distribution in
  ultrarelativistic heavy ion collisions}},  {\em JHEP} {\bf 10} (2019) 273,
  [\href{http://arxiv.org/abs/1907.04866}{{\tt arXiv:1907.04866}}].

\bibitem{Du:2020pmp}
Y.-L. Du, D.~Pablos, and K.~Tywoniuk, {\it {Deep learning jet modifications in
  heavy-ion collisions}},  {\em JHEP} {\bf 21} (2020) 206,
  [\href{http://arxiv.org/abs/2012.07797}{{\tt arXiv:2012.07797}}].

\bibitem{Caucal:2021cfb}
P.~Caucal, A.~Soto-Ontoso, and A.~Takacs, {\it {Dynamically groomed jet radius
  in heavy-ion collisions}},  {\em Phys. Rev. D} {\bf 105} (2022), no.~11
  114046, [\href{http://arxiv.org/abs/2111.14768}{{\tt arXiv:2111.14768}}].

\bibitem{Brewer:2021hmh}
J.~Brewer, Q.~Brodsky, and K.~Rajagopal, {\it {Disentangling jet modification
  in jet simulations and in Z+jet data}},  {\em JHEP} {\bf 02} (2022) 175,
  [\href{http://arxiv.org/abs/2110.13159}{{\tt arXiv:2110.13159}}].

\bibitem{Pablos:2022mrx}
D.~Pablos and A.~Soto-Ontoso, {\it {Pushing forward jet substructure
  measurements in heavy-ion collisions}},  {\em Phys. Rev. D} {\bf 107} (2023),
  no.~9 094003, [\href{http://arxiv.org/abs/2210.07901}{{\tt
  arXiv:2210.07901}}].

\bibitem{Dokshitzer:1997in}
Y.~L. Dokshitzer, G.~D. Leder, S.~Moretti, and B.~R. Webber, {\it {Better jet
  clustering algorithms}},  {\em JHEP} {\bf 08} (1997) 001,
  [\href{http://arxiv.org/abs/hep-ph/9707323}{{\tt hep-ph/9707323}}].

\bibitem{Wobisch:1998wt}
M.~Wobisch and T.~Wengler, {\it {Hadronization corrections to jet
  cross-sections in deep inelastic scattering}},  in {\em {Workshop on Monte
  Carlo Generators for HERA Physics (Plenary Starting Meeting)}}, pp.~270--279,
  4, 1998.
\newblock \href{http://arxiv.org/abs/hep-ph/9907280}{{\tt hep-ph/9907280}}.

\bibitem{Larkoski:2014wba}
A.~J. Larkoski, S.~Marzani, G.~Soyez, and J.~Thaler, {\it {Soft Drop}},  {\em
  JHEP} {\bf 05} (2014) 146, [\href{http://arxiv.org/abs/1402.2657}{{\tt
  arXiv:1402.2657}}].

\bibitem{CMS:2017qlm}
{\bf CMS} Collaboration, A.~M. Sirunyan et~al., {\it {Measurement of the
  Splitting Function in $pp$ and Pb-Pb Collisions at $\sqrt{s_{_{\mathrm{NN}}}}
  =$ 5.02 TeV}},  {\em Phys. Rev. Lett.} {\bf 120} (2018), no.~14 142302,
  [\href{http://arxiv.org/abs/1708.09429}{{\tt arXiv:1708.09429}}].

\bibitem{ATLAS:2019mgf}
{\bf ATLAS} Collaboration, G.~Aad et~al., {\it {Measurement of soft-drop jet
  observables in $pp$ collisions with the ATLAS detector at $\sqrt {s}$ =13
  TeV}},  {\em Phys. Rev. D} {\bf 101} (2020), no.~5 052007,
  [\href{http://arxiv.org/abs/1912.09837}{{\tt arXiv:1912.09837}}].

\bibitem{CMS:2018fof}
{\bf CMS} Collaboration, A.~M. Sirunyan et~al., {\it {Measurement of the
  groomed jet mass in PbPb and pp collisions at $ \sqrt{s_{\mathrm{NN}}}=5.02 $
  TeV}},  {\em JHEP} {\bf 10} (2018) 161,
  [\href{http://arxiv.org/abs/1805.05145}{{\tt arXiv:1805.05145}}].

\bibitem{ALICE:2019ykw}
{\bf ALICE} Collaboration, S.~Acharya et~al., {\it {Exploration of jet
  substructure using iterative declustering in pp and Pb{\textendash}Pb
  collisions at LHC energies}},  {\em Phys. Lett. B} {\bf 802} (2020) 135227,
  [\href{http://arxiv.org/abs/1905.02512}{{\tt arXiv:1905.02512}}].

\bibitem{ALargeIonColliderExperiment:2021mqf}
{\bf A Large Ion Collider Experiment, ALICE} Collaboration, S.~Acharya et~al.,
  {\it {Measurement of the groomed jet radius and momentum splitting fraction
  in pp and Pb$-$Pb collisions at $\sqrt{s_{NN}} = 5.02$ TeV}},  {\em Phys.
  Rev. Lett.} {\bf 128} (2022), no.~10 102001,
  [\href{http://arxiv.org/abs/2107.12984}{{\tt arXiv:2107.12984}}].

\bibitem{ATLAS:2022vii}
{\bf ATLAS} Collaboration, G.~Aad et~al., {\it {Measurement of
  substructure-dependent jet suppression in Pb+Pb collisions at 5.02 TeV with
  the ATLAS detector}},  {\em Phys. Rev. C} {\bf 107} (2023), no.~5 054909,
  [\href{http://arxiv.org/abs/2211.11470}{{\tt arXiv:2211.11470}}].

\bibitem{ATLAS:2025svn}
{\bf ATLAS} Collaboration, G.~Aad et~al., {\it {Measurement of
  substructure-dependent suppression of large-radius jets with charged
  particles in Pb+Pb collisions with ATLAS}},  {\em Phys. Lett. B} {\bf 871}
  (2025) 139929, [\href{http://arxiv.org/abs/2504.04805}{{\tt
  arXiv:2504.04805}}].

\bibitem{Andrews:2019hvc}
H.~A. Andrews, {\em {Exploring the phase space of medium induced QCD radiation
  with jets in ALICE at the LHC}}.
\newblock PhD thesis, Birmingham U., 2019.

\bibitem{Apolinario:2026hff}
L.~Apolin{\'a}rio, D.~Costa, and A.~Soto-Ontoso, {\it {Bottom-up approach to
  describe groomed jet data in heavy-ion collisions}},
  \href{http://arxiv.org/abs/2601.13310}{{\tt arXiv:2601.13310}}.

\bibitem{Mehtar-Tani:2019rrk}
Y.~Mehtar-Tani, A.~Soto-Ontoso, and K.~Tywoniuk, {\it {Dynamical grooming of
  QCD jets}},  {\em Phys. Rev. D} {\bf 101} (2020), no.~3 034004,
  [\href{http://arxiv.org/abs/1911.00375}{{\tt arXiv:1911.00375}}].

\bibitem{ALICE:2024fip}
{\bf ALICE} Collaboration, S.~Acharya et~al., {\it {Search for quasi-particle
  scattering in the quark-gluon plasma with jet splittings in pp and Pb$-$Pb
  collisions at $\sqrt{s_{\rm NN}}$ = 5.02 TeV}},  {\em Phys. Rev. Lett.} {\bf
  135} (2025), no.~3 031901, [\href{http://arxiv.org/abs/2409.12837}{{\tt
  arXiv:2409.12837}}].

\bibitem{Zhang:2015trf}
X.~Zhang, L.~Apolin{\'a}rio, J.~G. Milhano, and M.~P{\l}osko{\'n}, {\it
  {Sub-jet structure as a discriminating quenching probe}},  {\em Nucl. Phys.
  A} {\bf 956} (2016) 597--600, [\href{http://arxiv.org/abs/1512.09255}{{\tt
  arXiv:1512.09255}}].

\bibitem{Strangmann2025}
{\bf ALICE} Collaboration, N.~Strangmann, ``Nuclear modification of $\pi^0$
  production in {O-O} collisions with {ALICE}.'' Talk at the VIII-th
  International Conference on the Initial Stages of High-Energy Nuclear
  Collisions (Initial Stages 2025), Sept., 2025.

\bibitem{ATLAS:2025ooe}
{\bf ATLAS} Collaboration, {\it {Measurement of dijet transverse momentum
  balance in O+O and $pp$ collisions at 5.36 TeV with the ATLAS detector}},
  2025.
\newblock ATLAS-CONF-2025-010.

\bibitem{CMS:2025bta}
{\bf CMS} Collaboration, A.~Hayrapetyan et~al., {\it {Discovery of suppressed
  charged-particle production in ultrarelativistic oxygen-oxygen collisions}},
  \href{http://arxiv.org/abs/2510.09864}{{\tt arXiv:2510.09864}}.

\bibitem{GSL}
M.~Galassi, J.~Davies, J.~Theiler, B.~Gough, G.~Jungman, M.~Booth, and
  F.~Rossi, {\em GNU Scientific Library Reference Manual}.
\newblock Free Software Foundation, 3rd~ed., 2009.
\newblock Version 1.12.

\bibitem{Galassi:2019czg}
M.~C. Galassi, J.~Davies, J.~Theiler, B.~Gough, G.~Jungman, P.~Alken, M.~Booth,
  F.~Rossi, and R.~Ulerich, {\em {GNU Scientific Library}}.
\newblock Network Theory, Ltd., 8, 2019.

\bibitem{Piessens1983}
R.~Piessens, E.~de~Doncker-Kapenga, C.~W. Überhuber, and D.~K. Kahaner, {\em
  QUADPACK: A Subroutine Package for Automatic Integration}.
\newblock Springer, Berlin, 1983.

\end{thebibliography}\endgroup
\bibliographystyle{JHEP}

\end{document}